%% file: race.tex
\pdfoutput=1

\documentclass[acmsmall,nonacm]{acmart}

\input{preamble.tex}

\AtBeginDocument{%
	\providecommand\BibTeX{{%
			\normalfont B\kern-0.5em{\scshape i\kern-0.25em b}\kern-0.8em\TeX}}}

\setcopyright{none}

\acmYear{xxxx}
\acmDOI{}

\begin{document}


\title[SymmSpMV with RACE]{A Recursive Algebraic Coloring Technique for Hardware-Efficient Symmetric Sparse Matrix-Vector Multiplication}

\author{CHRISTIE  ALAPPAT}
\email{christie.alappat@fau.de}
\affiliation{
	\institution{Department of Computer Science, Friedrich-Alexander-Universit\"at Erlangen-N\"urnberg}}

\author{GEORG HAGER}
\email{georg.hager@fau.de}
\affiliation{
	\institution{Erlangen Regional Computing Center, Friedrich-Alexander-Universit\"at Erlangen-N\"urnberg}}

\author{OLAF SCHENK}
\email{olaf.schenk@usi.ch}
\affiliation{
	\institution{Institute of Computational Science, Universit\`{a} della Svizzera italiana}}

\author{JONAS THIES}
\email{jonas.thies@dlr.de}
\affiliation{
	\institution{Simulation and Software Technology, German Aerospace Center}}

\author{ACHIM BASERMANN}
\email{achim.basermann@dlr.de}
\affiliation{
	\institution{Simulation and Software Technology, German Aerospace Center}}

\author{ALAN R. BISHOP}
\email{arb@lanl.gov}
\affiliation{\institution{Theory, Simulation and Computation, Los Alamos National Laboratory}}

\author{HOLGER FEHSKE}
\email{fehske@physik.uni-greifswald.de}
\affiliation{\institution{Institute of Physics, University of Greifswald}}

\author{GERHARD WELLEIN}
\email{gerhard.wellein@fau.de}
\affiliation{
	\institution{Department of Computer Science, Friedrich-Alexander-Universit\"at Erlangen-N\"urnberg}}


\renewcommand{\shortauthors}{Christie Alappat et al.}

\input{abstract.tex}

%
%


%
%



\maketitle


\section{Introduction and Related Work}
\label{Sec:related_work} 
\input{related_work.tex}

  
\subsection*{Contribution and Outline}
\label{Sec:contribution}
\input{contribution.tex}

\section{Hardware and Software Environment}
\label{Sec:test_bed}
\input{test_bed_and_others.tex}

\section{Kernels}
\label{Sec:test_kernels}
\input{kernels.tex}

\subsection{Analysis of the \acrshort{SymmSpMV} kernel using parallel coloring schemes for the Spin-26 matrix}
\label{Sec:motivation}
\input{motivation.tex}

\section{Recursive Algebraic Coloring Engine (RACE)}
\label{Sec:race}
\input{race_method.tex}

\section{Parameter study}
\label{Sec:param_study}
\input{parameter_study.tex}

\section{Performance Evaluation of SymmSpMV using RACE}
\label{Sec:expt}
\input{experiment_and_results.tex}

\section{Conclusion and Outlook}
\input{conclusion.tex}

\label{Sec:conclusion}


\begin{acks}
The project is funded by the German DFG priority programme 1648
``Software for Exascale Computing (SPPEXA)'' and the Swiss National
Science Foundation (SNF) under the projects ``Dual-Phase Steels -- From
Micro to Macro Properties (EXASTEEL-2)'' (DFG, SNF) and ``Equipping
Sparse Solvers for Exascale (ESSEX-II)'' (DFG). The authors wish to thank
Andreas Alvermann for providing access to his ScaMaC library, Thomas Gruber for supporting our \LIKWID measurements and Moritz Kreutzer for helpful discussions.
\end{acks}
\clearpage
\printglossaries
\clearpage

\bibliographystyle{ACM-Reference-Format}
\bibliography{references}



\medskip





\clearpage
\appendix

\section{Algorithms}\label{Sec:algo}
\begin{algorithm}[htp]
	\caption{Construction of levels}
	\label{alg:BFS}
	\begin{algorithmic}[1]
		\STATE $integer::root = n$ {\textcolor{gray}{\hspace{1em} \% Choose starting node}}
		\STATE $bool::marked\_all = false$ {\textcolor{gray}{\hspace{1em} \% Stopping criteria}}
		\STATE $integer::N = nrows(graph)$
		\STATE $integer::distFromRoot[N] = \{-1\}$
		\STATE $integer::curr\_children[] = \{\}$
		\STATE $curr\_children.push\_back(root)$;
		\STATE $integer::currLvl = 0$
		\WHILE{$!marked\_all$}
		\STATE $marked\_all$ = true
		\STATE $integer::nxt\_children[] = \{\}$ 
		\FOR{$i=1:size(curr\_children)$}
		\IF{$distFromRoot[curr\_children[i]]==-1$}
		\STATE $distFromRoot[curr\_children[i]]=currLvl$
		\FOR{$j$ in $graph[curr\_children[i]].children$}
		\IF{$distFromRoot[j]==-1$}
		\STATE $nxt\_children.push\_back(j)$
		\ENDIF
		\ENDFOR		
		\ENDIF
		\ENDFOR	
		\STATE $curr\_children = nxt\_children$
		\STATE $currLvl = currLvl + 1$
		\ENDWHILE
	\end{algorithmic}
\end{algorithm}

\begin{algorithm}[htbp]
	\caption{Load Balancing for two sweep, \DTWO, two colors} 
	\label{alg:LB}
	\begin{algorithmic}[1]
		\IF{\cref{subsec:LB}}
		\STATE{$integer::nthreads$ = \acrshort{nthreads}}
		\STATE {$integer::len$ = $2*nthreads$ \textcolor{gray}{\hspace{1em} \% number of \levelGroups}}
		\STATE{$integer::worker[len] = {1}$}
		\STATE{$integer::T\_ptr[len+1] = linspace(0, \acrshort{totalLvl}, len) $ \textcolor{gray}{\hspace{0em} \% \levelGroup pointer}}
		\ELSE
		\STATE{$integer::nthreads = n_t(T_{s-1}(i))$ \textcolor{gray}{\% $i$ is the index of \levelGroup in stage $s-1$}}
		\STATE{\textcolor{gray}{\hspace{15em}\% where recursion is applied.}}
		\STATE {$integer::len$ = $2*nthreads$ \textcolor{gray}{\hspace{1.5em} \% number of \levelGroups}}
		\STATE{$integer::worker[len] = [n_t(T_{s}(0)), ... ,n_t(T_{s}(len-1))] = b$}
		\STATE{$integer::T\_ptr[len+1] = linspace(0, \acrshort{totalLvl}, len) $}
		\ENDIF	
		\STATE{$bool::exit = false$}
		\STATE{$integer::T\_size[len], absRankIdx[len], rankIdx[len], currRank$}
		\STATE{$double::mean\_r, mean\_b, diff[len], var, newVar$}
		\WHILE{$!(exit)$} 
		\STATE\label{lb_line5} {$T\_size[:]$ = update($T\_ptr[:]$) \textcolor{gray}{\hspace{0em} \% $T\_size$ contains nrows in each \levelGroup}}
		\STATE {$integer::T\_size\_worker[:]$ = $T\_size[:]/worker[:]$} 
		\STATE {$mean\_r$ = sum($T\_size\_worker[0:2:len-1]$) / $nthreads$}
		\textcolor{gray}{\%mean of red color}
		\STATE {$mean\_b$ = sum($T\_size\_worker[1:2:len-1]$) / $nthreads$}
		\textcolor{gray}{\%mean of blue color}
		\STATE {$diff[0:2:len-1]$ = $T\_size\_worker[0:2:len-1] .- mean\_r$}
		\STATE {$diff[1:2:len-1]$ = $T\_size\_worker[1:2:len-1] .- mean\_b$}
		\STATE\label{lb_line10} {$var$ = dot\_product($diff,diff$)/len} \textcolor{gray}{\% overall variance}
		\STATE {$absRankIdx$ = argsort(-abs($diff$))  \textcolor{gray}{\% ranking according to absolute deviation}}
		\STATE {$rankIdx$ = argsort($diff$)} \textcolor{gray}{\% ranking according to signed deviation}
		\STATE {$currRank = 0, newVar = var$}
		\STATE {$integer::old\_T\_ptr[len+1]$ = $T\_ptr[:], acquireIdx, giveIdx$}
		\WHILE{$newVar \geq var$}
		\STATE {$T\_ptr$ = $old\_T\_ptr$}
		\STATE {$bool::fail$=true}
		\IF{ $diff[absRankIdx[currRank]] < 0$ }
		\FOR{$el$ in $rankIdx[(len-1):-1:0]$}
		\IF{$(T\_Ptr[el+1] - T\_ptr[el]) > 2$}
		\STATE {$acquireIdx$ = el}
		\STATE {$fail$=false}
		\STATE {$break$}
		\ENDIF
		\ENDFOR
		\STATE {shift($T\_ptr, acquireIdx, currRank$) \textcolor{gray}{\% shifts $T\_ptr$ by 1  from  $acquireIdx$}} 
		\STATE {\textcolor{gray} {\hspace{6.5em} \% to $currRank$ if $currIdx < acquireIdx$ else shift by -1}}
		\ELSIF{ $(T\_ptr[currRank+1]-T\_ptr[currRank]) > 2$ }
		\STATE {$giveIdx = rankIdx[0]$}
		\STATE {$fail$=false}
		\STATE {shift($T\_ptr,currRank,giveIdx$)}
		\ENDIF
		\algstore{lbAlg}
	\end{algorithmic}
\end{algorithm}

\begin{algorithm}
	\begin{algorithmic}	
		\algrestore{lbAlg}
		\IF{!$fail$}
		\STATE {$newVar$ = calculate\_variance($T\_ptr$) \textcolor{gray}{\% as seen in \textcolor{darkgray}{Line 17} to \textcolor{darkgray}{Line 23}}}
		\ENDIF
		\IF{$ (currRank == (len-1)) \text{  } \&\&  \text{  } (newVar \geq var) $}
		\STATE {$T\_Ptr = old\_T\_ptr$}
		\STATE {$exit$ = true}
		\STATE {$break$}
		\ENDIF
		\STATE {$currRank += 1$}
		\ENDWHILE
		\ENDWHILE
	\end{algorithmic}
\end{algorithm} 


\end{document}

%% file: preamble.tex
\usepackage{graphicx,epstopdf} 
\usepackage[caption=false]{subfig} 
\usepackage{braket,amsfonts,amsopn} 
\usepackage{centernot}
 \usepackage{multirow}
\usepackage{xspace} 
\usepackage{comment}
\usepackage{siunitx} 
\usepackage{amssymb}
\usepackage[normalem]{ulem}
\usepackage[acronym, nonumberlist]{glossaries} 
\usepackage{tikz}
\glossarystyle{list}
\makeglossaries
\newcommand\mathmacro[1][A]{\ensuremath{{#1}}}

\newacronym{nrows}{\mathmacro[N_\mathrm r]}{number of rows}
\newacronym{nrowsEff}{\mathmacro[N_\mathrm{r}^\mathrm{eff}]}{effective number of rows}
\newacronym{nnz}{\mathmacro[N_\mathrm{nz}]}{total number of nonzeros}
\newacronym{NNZR}{\mathmacro[N_\mathrm{nzr}]}{average number of nonzeros per row}
\newacronym{SymmNNZR}{\mathmacro[N_\mathrm{nzr}^\mathrm{symm}]}{average number of nonzeros per row in symmetric matrix}
\newacronym{totalLvl}{\mathmacro[N_\ell]}{total levels in the graph}
\newacronym{nthreads}{\mathmacro[N_\mathrm t]}{number of threads}
\newacronym{threadEff}{\mathmacro[N_\mathrm{t}^\mathrm{eff}]}{effective number of threads}
\newacronym{L_i}{\mathmacro[L(i)]}{$i$th level}
\newacronym{T_i}{\mathmacro[T(i)]}{$i$th level group}
\newacronym{L_si}{\mathmacro[L_s(i)]}{$i$th level at stage $s$}
\newacronym{T_si}{\mathmacro[T_s(i)]}{$i$th level group at stage $s$}
\newacronym{s}{\mathmacro[s]}{stage number of recursion}
\newacronym{b_s}{\mathmacro[b_\mathrm s]}{single socket bandwidth}
\newacronym{eta}{\mathmacro[\eta]}{theoretical parallel efficiency}

\newacronym{RACE}{RACE}{recursive algebraic coloring engine}
\newacronym{RSB}{RSB}{recursive sparse blocks}
\newacronym{SpMV}{SpMV}{sparse matrix-vector multiplication}
\newacronym{SymmSpMV}{SymmSpMV}{symmetric sparse matrix-vector multiplication}
\newacronym{SpMTV}{SpMTV}{sparse matrix transposed vector multiplication}
\newacronym{GS}{GS}{Gauss-Seidel iterative solver}
\newacronym{SymmGS}{SymmGS}{symmetric Gauss-Seidel iterative solver}
\newacronym{KACZ}{KACZ}{Kaczmarz iterative solver}
\newacronym{SymmKACZ}{SymmKACZ}{symmetric Kaczmarz iterative solver}
\newacronym{CRS}{CRS}{compressed row storage}
\newacronym{MC}{MC}{multicoloring}
\newacronym{ABMC}{ABMC}{algebraic block multicoloring}
\newacronym{LLC}{LLC}{last level cache}
\newacronym{LCC}{LCC}{low core count}
\newacronym{HCC}{HCC}{high core count}
\newacronym{RCM}{RCM}{reverse Cuthill McKee}
\newacronym{BFS}{BFS}{breadth-first search}
\newacronym{MKL}{Intel MKL}{Intel math kernel library}
\newacronym{SNC}{SNC}{sub-NUMA clustering}

\definecolor{amber}{rgb}{1.0, 0.49, 0.0}
\definecolor{carmine}{rgb}{0.59, 0.0, 0.09}

\ifpdf
\hypersetup{
  pdftitle={RACE},
  pdfauthor={}
}
\fi

\def\GW{\color{blue}}
\def\CA{\color{red}}

\newcommand{\abs}{\mbox{abs}}
\newcommand{\eos}{~.}
\newcommand{\cma}{~,}
\newcommand{\DONE}{distance-$1$\xspace}
\newcommand{\DTWO}{distance-$2$\xspace}
\newcommand{\DK}{distance-$k$\xspace}

\newcommand{\etal}{et al.\xspace}
\newcommand{\Intel}{Intel\xspace}
 
\newcommand{\IVB}{Ivy Bridge EP\xspace}

\newcommand{\SKX}{Skylake SP\xspace}

\newcommand{\LIKWID}{LIKWID\xspace}
\newcommand{\likwidBench}{\texttt{likwid-bench}\xspace}
\newcommand{\likwidPerfctr}{\texttt{likwid-perfctr}\xspace}
\newcommand{\ie}{i.e.,\xspace}
\newcommand{\cf}{cf. \xspace}

\newcommand{\FLOP}{\mbox{flops}\xspace}
\newcommand{\BYTE}{\mbox{bytes}\xspace}

\newcommand{\MB}{\mbox{MB}\xspace}
\newcommand{\KB}{\mbox{kB}\xspace}
\newcommand{\GB}{\mbox{GB}\xspace}
\newcommand{\TB}{\mbox{TB}\xspace}
\newcommand{\GBS}{\mbox{GB/s}}
\newcommand{\MiB}{\mbox{MiB}\xspace}
\newcommand{\KiB}{\mbox{KiB}\xspace}
\newcommand{\GiB}{\mbox{GiB}\xspace}

\newcommand{\GHZ}{\mbox{GHz}\xspace}
\newcommand{\SIMD}{SIMD\xspace}
\newcommand{\CPU}{CPU\xspace}

\newcommand{\Stex}{Stencil example\xspace}
\newcommand{\stex}{stencil example\xspace}
\newcommand{\Inorder}{In order\xspace}
\newcommand{\inorder}{in order\xspace}
\newcommand{\levelPtr}{{\tt level\_ptr}\xspace}
\newcommand{\atleast}{at least\xspace}
\newcommand{\level}{level\xspace}
\newcommand{\levels}{levels\xspace}
\newcommand{\levelGroup}{level group\xspace}
\newcommand{\levelGroups}{level groups\xspace}

\newcommand{\eg}{e.g.,\xspace}
\newcommand{\aka}{a.k.a.\xspace}
\newcommand{\subgraph}{subgraph\xspace}
\newcommand{\subgraphs}{subgraphs\xspace}
\newcommand{\levelTree}{{\tt level\_tree}\xspace}

\newcommand{\effPar}{\emph{effective parallelism}\xspace}
\newcommand{\effRow}{\emph{effective row count}\xspace}

\newcommand{\fracUnit}[2]{\Big[\frac{\mbox{#1}}{\mbox{#2}}\Big]}

\newcommand{\roofline}{roof\/line\xspace}
\newcommand{\METIS}{METIS\xspace}
\newcommand{\COLPACK}{COLPACK\xspace}
\newcommand{\SPMP}{Intel SpMP\xspace}
\newcommand{\GF}{\mbox{GF/s}\xspace}

\newcommand{\CAcomm}[1]{{\color{orange}{#1}\color{black}}}
\newcommand{\sref}[1]{{\textnormal{\protect\subref{#1}}}}

\usepackage{lipsum}
\usepackage{amsfonts}
\usepackage{graphicx}
\usepackage{epstopdf}
\usepackage{cleveref}
\usepackage{booktabs}
\usepackage{algorithm}
\usepackage{algcompatible}

\ifpdf
  \DeclareGraphicsExtensions{.eps,.pdf,.png,.jpg}
\else
  \DeclareGraphicsExtensions{.eps}
\fi

\usepackage{amsopn}

\usepackage{graphicx}						

\usepackage{amsfonts}						
\usepackage{amsmath}
\usepackage{amssymb}
\usepackage{mathrsfs}

%% file: abstract.tex
\begin{abstract}

The \acrfull{SymmSpMV} is an important building block for many
numerical linear algebra kernel operations or graph traversal
applications. Parallelizing \acrshort{SymmSpMV} on today's multicore
platforms with up to 100 cores is difficult due to the need to manage
conflicting updates on the result vector. Coloring approaches can be
used to solve this problem without data duplication, but existing
coloring algorithms do not take load balancing and deep memory
hierarchies into account, hampering scalability and full-chip
performance. In this work, we propose the \acrfull{RACE}, a novel
coloring algorithm and open-source library implementation, which
eliminates the shortcomings of previous coloring methods in
terms of hardware efficiency and parallelization overhead. We describe
the level construction, \DK coloring, and load balancing steps in
\acrshort{RACE}, use it to parallelize \acrshort{SymmSpMV}, and
compare its performance on 31 sparse matrices with other
state-of-the-art coloring techniques and Intel MKL on two modern
multicore processors.  \acrshort{RACE} outperforms all other
approaches substantially and behaves in accordance with the \roofline
model. Outliers are discussed and analyzed in detail.
While we focus on \acrshort{SymmSpMV} in this paper, 
our algorithm and software is applicable to any sparse matrix operation 
with data dependencies that can be resolved by distance-k coloring.

%

\end{abstract}

%% file: related_work.tex
The efficient solution of linear systems or eigenvalue problems
involving large sparse matrices has been an active research field in
parallel and high performance computing for many decades. Well-known,
traditional application areas include quantum physics, quantum
chemistry or engineering. In recent years, new fields such as social
graph analysis ~\cite{Simpson:2018:BGP:3218176.3218232} or spectral
clustering in the context of learning
algorithms \cite{vonLuxburg2007,JMLR:v17:16-109} have further
increased the need for hardware-efficient, parallel sparse solvers
and/or efficient matrix-free solvers. Assuming sufficiently large
problems, the solvers are typically based on iterative subspace
methods and may include advanced preconditioning techniques. In many
methods, two components, \acrfull{SpMV} and coloring techniques, are
crucial for hardware efficiency and parallel scalability. Typically,
these two components are considered to be orthogonal, \ie hardware
efficiency for \Acrshort{SpMV} is mainly related to data formats and
local structures while coloring is used to address dependencies in the
enclosing iteration scheme.  Interestingly, the hardware-efficient
parallelization of symmetric \Acrshort{SpMV} has not attracted a lot
of attention over the years, though symmetry is widespread in the
application fields.

The \Acrshort{SpMV} operation is an essential building block 
in a number of applications such as algebraic multigrid methods, 
sparse iterative solvers, shortest path algorithms, breadth first search algorithms, 
and Markov cluster algorithms, and therefore it is an integral part 
of numerous scientific algorithms.
In the past decades, much research has been
focusing on designing new data structures, efficient algorithms, and
parallelization  techniques for the \acrshort{SpMV} operation. Its performance is typically limited by main memory bandwidth. On
cache-based architectures, the main factors that influence performance
are spatial access locality to the matrix data and temporal locality when
reusing the elements of the vectors involved. To address this problem, over the
last two decades a plethora of  partitioning techniques
and data structures to improve \acrshort{SpMV} on cache-based
architectures have been suggested, including 
cache-oblivious methods using hypergraph partitioning. One
of the first studies on temporal locality optimizations was done
by Toledo~\cite{Toledo:1997:IMP:279511.279532}, who investigated
Cuthill--McKee (CM) ordering
techniques on three-dimensional finite-element test matrices when
used in combination with blocking into small dense blocks. Various
authors~\cite{Williams:2009:OSM:1513001.1513318,doi:10.1177/1094342004041296}
used advanced data storage formats and techniques such as register and cache blocking  for  
\acrshort{SpMV} by splitting the matrix into several smaller $p \times
q$ sparse submatrices and presented an analytic cache-aware model to
determine the optimal block size. These algorithms are, e.g.,
included in OSKI~\cite{1742-6596-16-1-071}, which is a collection of
low-level primitives of tuned sparse kernels for modern cache-based
superscalar machines. Kreutzer \etal~\cite{Moritz_sell} and 
Xing \etal~\cite{Liu:2013:ESM:2464996.2465013} used techniques to 
improve SIMD efficiency and performance on many-core and GPU architectures. 
Recent work can be found, \eg
in~\cite{li2017hbm,Liu:2015:CES:2751205.2751209,liu2015spmv}.
 Previous work on  \acrshort{SpMV}  has also focused on reducing
 communication volume for distributed-memory parallelization, often by using
 variants of graph or hypergraph partitioning
 techniques~\cite{Catalyurek:1999}. Yzelman and
 Bisseling~\cite{doi:10.1137/080733243,Yzelman-thesis-2011} extended
 hypergraph partitioning techniques in a cache-oblivious method,
 permuting rows and columns of the input matrix using a recursive
 hypergraph-based sparse matrix partitioning scheme so that the
 resulting matrix exhibits cache-friendly behavior during the
 \acrshort{SpMV}.



Despite \acrshort{SpMV} being a bandwidth-limited operation, not much work has 
been done to exploit the symmetry property of symmetric matrices to reduce
storage requirements and data transfers by using only the upper/lower triangular part of the matrix.
The major challenge here is to resolve the potential write conflicts of explicit 
\acrfull{SymmSpMV} kernels in parallel processing.
There are general solutions for such problems like 
lock based methods and thread private target
arrays~\cite{sparseX,thread_private_symm_spmv,Krotkiewski:2010:PSS:1752612.1752682,Mironowicz:2015}. However they have in common that their overhead may increase with the degree of parallelism.
Another recent research direction is the use of specialized storage formats 
like CSB \cite{CSB}, RSB \cite{RSB}, CSX \cite{sparseX} combined with the use of bitmasked 
register blocking techniques as in \cite{Buluc:2011:RMA:2058524.2059503}. As pointed out 
by \cite{liu2015spmv} these approaches have drawbacks like missing backward compatibility and matrix conversion costs. 
Due to these problems there are only a very few standard libraries, like 
\acrshort{MKL}~\cite{MKL}, that support primitives for efficient \acrshort{SymmSpMV} operation.
Another potential way of tackling this inherent data dependency problem is using a \DTWO coloring 
 of the underlying undirected graph, which has not been investigated so far to the best of our knowledge.


\Acrfull{MC} reordering to tackle data dependencies is a very well established strategy in parallelization of iterative solvers. As it is applied to the underlying graph it is not bound to a specific data format and may use existing highly optimized (serial) kernels, \ie it is orthogonal to general code optimization strategies.  
Prominent examples for  \acrshort{MC} in iterative solvers are Gauss-Seidel, incomplete 
Cholesky factorization or Kaczmarz method~\cite{RBGS,MC,feast_mc}, where typically a \DONE or \DTWO coloring is applied subject to the underlying dependencies of the iterative scheme. However, coloring changes the evaluation order of the original solver and may lead to worse convergence rates. This is different when using  \acrshort{MC} methods for parallelization of \acrshort{SymmSpMV} where we only need to ensure that entries of the target vector is not written in parallel. Here we do not  require strict serial ordering to get to the same result as in serial processing.
In terms of hardware utilization long-standing \acrshort{MC} methods often generate colorings which lack efficiency on modern cache-based
processors. Studies have been made to increase their
performance and improve inherent heuristics; an overview of the methods can be found in~\cite{gebremedhin2000scalable,dist_k_def,COLPACK,equitable_color}. However, 
for irregular and/or large sparse 
matrices \acrshort{MC} may lead to load imbalance, frequent global synchronization, 
and loss of data locality, severely reducing (single-node) performance. 
These problems typically become more stringent for higher order distance
colorings and larger matrices.
The \acrfull{ABMC}~\cite{ABMC} proposed by Iwashita \etal in 2012 addresses some of these issues as it tries to increase data locality by applying graph partitioning (blocking) before coloring. 
Beyond the quality of the actual coloring, the time to generate it is also critical, especially for very large problems. Here, widely used and publicly available coloring packages such as COLPACK\cite{COLPACK}, Kokkos\cite{kokkos} and ZOLTAN\cite{BOZDAG2008515,doi:10.1137/080732158} speed-up the coloring process itself by parallelization and other heuristics.

Design and implementation of hardware efficient computational kernels can be
supported by a structured performance engineering process based on white-box
models. On the processor/node level, the most prominent model is the roof{}line
model~\cite{Williams_roofline}. Its basic applicability as a reasonable
light-speed estimate for \acrshort{SpMV} was  demonstrated already 
in~\cite{Gropp:1999}, including an extension to sparse matrix multiple vector
multiplication. The \acrshort{SpMV} performance model was refined
in~\cite{Moritz_sell} with a focus on modeling the performance impact
of irregular accesses to the right-hand side (RHS) vector. It has been
successfully used to model performance on CPUs and GPGPUs for \acrshort{SpMV}
kernels~\cite{Moritz_sell} and for augmented sparse matrix multiple vector kernels
for Chebyshev filter diagonalization~\cite{ISC2018:ChebFD}. However, there is no
extension towards explicit \acrshort{SymmSpMV}, which shows increased
computational intensity and irregular accesses to both involved vectors. Typically the
expectation is that \acrshort{SymmSpMV} should be approximately twice as fast
as \acrshort{SpMV} as only half of the matrix information needs to be stored and
accessed.

Finally, there is a clear hardware trend towards processors with advanced
vector-style processing, higher core counts and more complex cache
hierarchies. Also, attainable bandwidth may increase even for ``standard'' CPU
based systems through the use of high bandwidth memory solutions at the cost of
very restricted memory sizes. A first step into this direction was the Intel
Xeon Knights Landing processor. The specification of the ARM-based Fujitsu
A64FX processor (to be used in the Post-K computer) may provide another
blueprint for future processor configurations~\cite{Post-K:Processor}: A 48-core
processor supporting 512-bit SIMD execution units on top of 32 \GiB HBM2 main
memory, which provides a bandwidth of 1 \TB/s.  It is obvious that such hardware
trends call for revisiting existing, time-critical components in simulation
codes both in terms of scalability and hardware efficiency. Moreover, the
potential of \acrshort{SymmSpMV} to substantially reduce the memory footprint of
sparse solvers needs to be exploited to meet the constraint of very limited
memory space.

%% file: contribution.tex

This paper addresses the general problem of generating hardware efficient \DK coloring
 of undirected graphs for modern multicore processors. As an application we 
 choose parallelization of the \acrshort{SymmSpMV} operation. We cover thread-level 
 parallelization and focus on a single multicore processor. The main contributions
  can be summarized as follows: 
\begin{itemize}
\item A new recursive algebraic coloring scheme (RACE) is proposed, 
which generates hardware efficient \DK colorings of undirected graphs. 
Special emphasis in the design of RACE is put on achieving data locality, 
generating levels of parallelism matching the core count of the underlying 
multicore processor and load balancing for shared memory parallelization.
\item We propose shared memory parallelization of \acrshort{SymmSpMV}  
using a \DTWO coloring of the underlying undirected graph to avoid
 write conflicts and apply RACE for generating the colorings.
\item A comprehensive performance study of shared memory parallel \acrshort{SymmSpMV} 
using RACE demonstrates the benefit of our approach. Performance modeling
 is deployed to substantiate our performance measurements, and a comparison to
  existing coloring methods as well a vendor optimized library (Intel MKL) 
  are presented. The broad applicability and the sustainability is validated 
  by using a wide set of 31 test matrices and two very different generations 
  of Intel Xeon processors.
\item We extend the existing proven \acrshort{SpMV} performance modeling approach
 to the \acrshort{SymmSpMV} kernel. 
In the course of the 
 performance analysis we further demonstrate why in some cases the ideal speedup
 may not be achievable.
\end{itemize}
We have implemented our graph coloring algorithms in the open source library \acrfull{RACE}.\footnote{\href{http://tiny.cc/RACElib}{http://tiny.cc/RACElib}}
Information required to reproduce the performance numbers provided in this 
paper is also available.\footnote{\href{http://tiny.cc/RACElib-AD}{http://tiny.cc/RACElib-AD}}


This paper is organized as follows: Our software and hardware environment as well as
 the benchmark matrices are introduced in \Cref{Sec:test_bed}. 
In \Cref{Sec:test_kernels} we describe the properties
of the \acrshort{SpMV} and \acrshort{SymmSpMV} kernels, including
\roofline performance limits, and motivate the need for an advanced coloring scheme.
In \Cref{Sec:race} we detail the steps of the \acrshort{RACE} algorithm
via an artificial stencil matrix and show how recursive level group construction
and coloring can be leveraged to exploit a desired level of parallelism
for \DK dependencies. The interaction between the parameters of the method
and their impact on the parallel efficiency is studied in \Cref{Sec:param_study}.
\Cref{Sec:expt} presents performance data for \acrshort{SymmSpMV}
for a wide range of matrices on two different multicore systems,
comparing \acrshort{RACE} with \acrshort{ABMC} and \acrshort{MC} as well as
Intel MKL, and also shows the efficiency of \acrshort{RACE} as
defined by the \roofline model yardstick.
\Cref{Sec:conclusion} concludes the paper and gives an outlook to
future work.

%% file: test_bed_and_others.tex
\subsection{Hardware test bed}
We conducted all benchmarks on a single CPU socket from Intel's \IVB and  \SKX families, respectively, since these  represent the oldest and the latest Intel architectures in active use within the scientific community at the time of writing:
\begin{itemize}
\item The \Intel \IVB architecture belongs to the class of ``classic'' designs with three inclusive cache levels. While the L1 and L2 caches are private to each core, the L3 cache is shared but scalable in terms of bandwidth. The processor supports the AVX instruction set extension, which is capable of 256-bit wide \SIMD execution.
\item Contrary to its predecessors, the \Intel \SKX architecture has a shared but noninclusive victim L3 cache and much larger private L2 caches. The model we use in this work supports AVX-512, which features 512-bit wide \SIMD execution.
 
\end{itemize}
Architectural details along with the attainable memory bandwidths are given in \Cref{tab:test_bed}. 
All the measurements were made with \CPU clock speeds fixed at the indicated base frequencies. Note that for the \SKX architecture the clock frequency is scaled down internally to 2.2 \GHZ when using multicore support and the AVX-512 instruction set; however, this is of minor importance for the algorithms discussed here.

\begin{figure}[t]
	\centering
	\subfloat[\emph{\IVB}]{\label{fig:ivy_size_vs_bw}\includegraphics[width=0.5\textwidth , height=0.16\textheight]{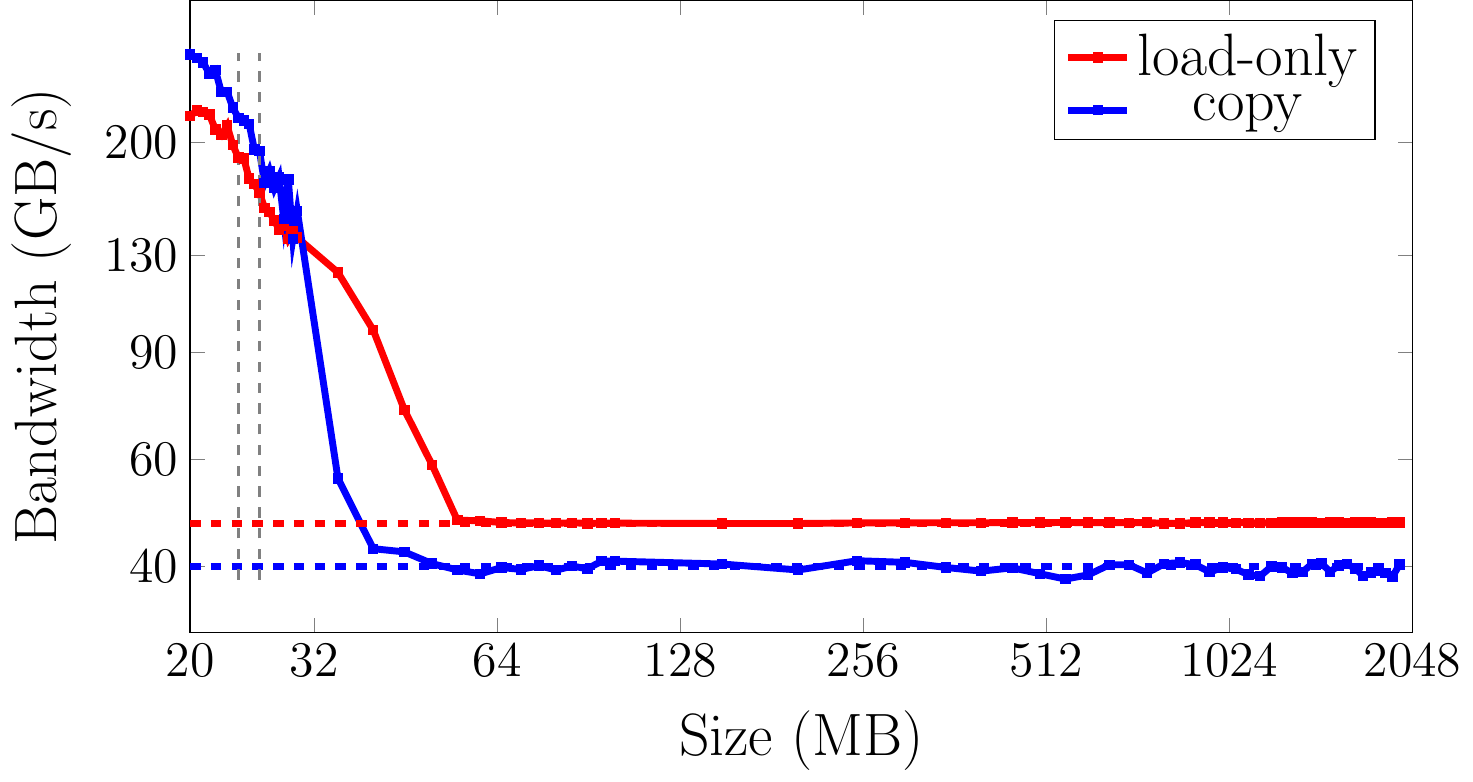}}
	\subfloat[\emph{\SKX}]{\label{fig:skx_size_vs_bw}\includegraphics[width=0.5\textwidth , height=0.16\textheight]{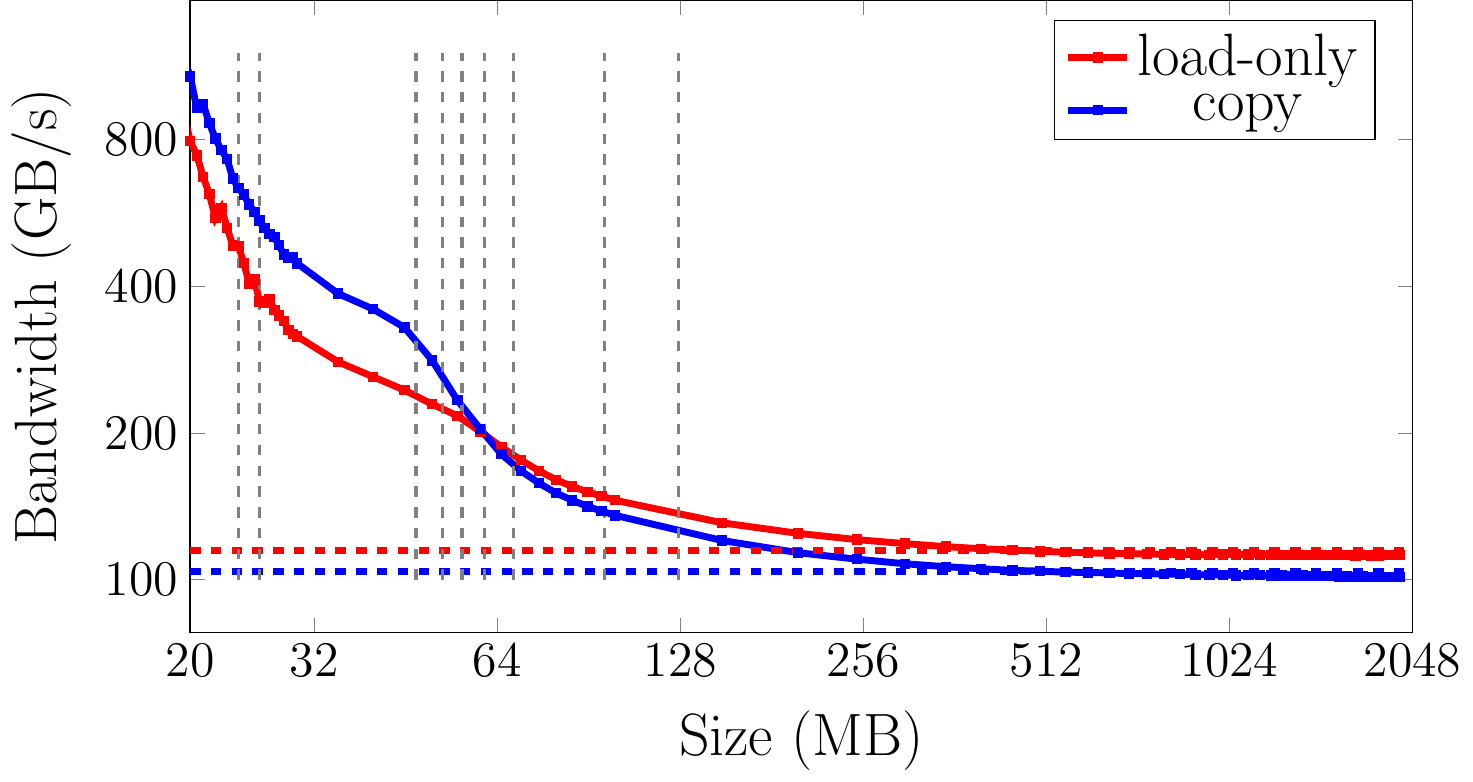}}
	\caption{Attained bandwidth versus total data size for a range from 20 \MB to 2 \GB. The dotted lines show the asymptotic bandwidth given in \Cref{tab:test_bed} for the load-only and copy benchmark. The benchmarks were performed on the full socket using the \likwidBench tool. The gray vertical lines correspond to the positions of matrices that might show caching effects; see \Cref{subsec:bench_mat}. Note the logarithmic scales.}
	\label{fig:size_vs_bw}
\end{figure}

\begin{table}[t]
	\centering
	\caption{Technical details (per socket) of the Intel CPUs used for the benchmarks.\label{tab:test_bed}}
	\begin{center}
		\begin{tabular}{l|cc}
			{Model name} & {Xeon\textsuperscript{\textregistered} E5-2660} & {Xeon\textsuperscript{\textregistered} Gold 6148} \\\midrule
			{Microarchitecture} & {\IVB} & {\SKX} \\\midrule
			{Base clock frequency} & {2.2 GHz} & {2.4 GHz}\\
			{Uncore clock frequency} & {2.2 GHz} & {2.4 GHz}\\
			{Physical cores per socket} & {10} & {20} \\
			{L1D cache} & {10 $\times$ 32 \KiB} & {20 $\times$ 32 \KiB}\\
			{L2 cache} & {10 $\times$ 256 \KiB} & {20 $\times$ 1 \MiB} \\
			{L3 cache} & {25 \MiB} & {27.5 \MiB}\\
			{L3 type} & {inclusive} & {noninclusive, victim}\\
			{Main memory} & {32 \GiB} & {48 \GiB}\\
			{Bandwidth per socket, load-only} & {47 \GBS} & {115 \GBS}\\ 
			{Bandwidth per socket, copy} & {40 \GBS} & {104 \GBS}\\
		\end{tabular}
	\end{center}
\end{table} 
As the attainable main memory bandwidth is the input parameter to the \roofline model used later, we have carefully measured this value depending on the data set size for two access patterns (copy and load-only). The data presented in \Cref{fig:size_vs_bw} basically show the characteristic performance drop if the data set size is too large to fit into the \acrfull{LLC}, which is an L3 cache on both architectures (cf. \Cref{tab:test_bed} for the actual sizes). 
Interestingly there is no sharp drop at the exact size of the \acrshort{LLC} but a rather steady performance decrease with enhanced data access rates also for data set sizes up to twice the \acrshort{LLC} size on \IVB. For \SKX this effect is even more pronounced as the noninclusive victim L3 cache architecture only stores data which are not in the L2 cache; thus the available cache size for an application may be the aggregate sizes of the L2 and L3 caches on this architecture.  The final bandwidth for the  \roofline model is chosen as the asymptotic value depicted in \Cref{fig:size_vs_bw}. Of course caching effects are extremely sensitive to the data access pattern and thus the values presented here only provide simple upper bounds for the \acrshort{SymmSpMV} kernel with its potentially strong irregular data access. 

\subsection{External Tools and Software}
The \LIKWID \cite{LIKWID} tool suite in version 4.3.2 was used, specifically
\likwidBench for bandwidth benchmarks (see \Cref{tab:test_bed}), and
\likwidPerfctr for counting hardware events 
 and measuring derived metrics.
 \LIKWID validates the quality of it's performance metrics
  and validation data is publicly available.\footnote{\href{https://github.com/RRZE-HPC/likwid/wiki/TestAccuracy}{https://github.com/RRZE-HPC/likwid/wiki/TestAccuracy}} Overall the \LIKWID data traffic measurements can be considered as highly accurate. Only the L3 data traffic measurement on \SKX fails the quantitative validation but it still provides good qualitative results.\footnote{\href{https://github.com/RRZE-HPC/likwid/wiki/L2-L3-MEM-traffic-on-Intel-Skylake-SP-CascadeLake-SP}{https://github.com/RRZE-HPC/likwid/wiki/L2-L3-MEM-traffic-on-Intel-Skylake-SP-CascadeLake-SP}}
 
 For coloring we used the \COLPACK \cite{COLPACK} library
 and \METIS \cite{METIS} version 5.1.0 for graph partitioning
 with the \acrshort{ABMC} method. The \SPMP \cite{SpMP} library was employed for
  \acrshort{RCM} bandwidth reduction, and the \acrshort{MKL} version 19.0.2
 for some reference computations and comparisons.

All code was compiled with the Intel compiler in version 19.0.2 and the following compiler flags: {\tt -fno-alias -xHost -O3} for \IVB and {\tt -fno-alias -xCORE-AVX512 -O3} for \SKX.

\subsection{Benchmark Matrices}
\label{subsec:bench_mat}
Most test matrices were taken from the Suite\-Sparse Matrix Collection (formerly University of Florida Sparse Matrix Collection)~\cite{UOF} combining sets from two related papers \cite{RSB,park_ls}, which allows the reader to make a straightforward comparison of results.  We also added some matrices from the Scalable Matrix Collection (ScaMaC) library\cite{ScaMaC}, which allows for scalable generation of large matrices related to quantum physics applications. A brief description of the background of these matrices can be found in ScaMaC documentation.\footnote{\href{https://alvbit.bitbucket.io/scamac_docs/_matrices_page.html}{https://alvbit.bitbucket.io/scamac\_docs/\_matrices\_page.html}} All the matrices considered are real, although our underlying software would also support complex matrices.
As mentioned before, we restrict ourselves to matrices representing fully connected undirected graphs.
\Cref{table:bench_matrices} gives an overview of the most important matrix properties like \acrfull{nrows}, \acrfull{nnz}, \acrfull{NNZR}, along with the bandwidth of the matrix without ($bw$) and with ($bw_{RCM}$) \acrshort{RCM} preprocessing. 

Due to the extended cache size as seen in \Cref{fig:size_vs_bw} it
might happen that some of the matrices attain higher effective
bandwidths due to partial/full caching, especially on \SKX. The ten potential candidates
for the \SKX chip in terms of symmetric and full storage ($< 128$ \MB)
are marked with an asterisk in \Cref{table:bench_matrices}, while only
two among these (\texttt{offshore} and \texttt{parabolic\_fem}) 
satisfy the criteria for \IVB ($< 40$ \MB). The corresponding data set
size for storing the upper triangular part of these matrices have been
labeled in \Cref{fig:size_vs_bw}.

\begin{table}[t]
	\centering
	\caption{Details of the benchmark matrices. \acrshort{nrows} is the number of matrix rows and \acrshort{nnz} is the number of nonzeros. $\acrshort{NNZR}=\acrshort{nnz}/\acrshort{nrows}$ is the average number of nonzeros per row. $bw$ and $bw_{RCM}$ refer to the matrix bandwidth without and with \acrshort{RCM} preprocessing. The letter ``C'' in the parentheses of the matrix name indicates a corner case matrix that will be discussed in detail, while the letter ``Q'' marks a matrix from quantum physics that is not part of the SuiteSparse Matrix Collection. With an asterisk (*) we have labeled all the matrices which are less than 128 \MB, which could potentially lead to some caching effects especially on the \SKX architecture. \label{tab:test_mtx}	\label{table:bench_matrices}}
	\begin{center}
		\input{pics/matrices/table.tex}

	\end{center}
\end{table}



%% file: pics/matrices/table.tex
\begin{tabular}{|l|l|S[table-format=7.0, table-space-text-pre=(, table-space-text-post=)]|S[table-format=8.0, table-space-text-pre=(, table-space-text-post=)]|S[round-mode=places,round-precision=2]|S[table-format=7.0, table-space-text-pre=(, table-space-text-post=)]|S[table-format=5.0, table-space-text-pre=(, table-space-text-post=)]|}
\toprule
{Index} & {Matrix name} &  {\acrshort{nrows}} & {\acrshort{nnz}} & {\acrshort{NNZR}}  & {$bw$} & {$bw_{RCM}$} \\
\midrule
{1}& {crankseg\_1* (C)}	& 52804	& 10614210	& 201.011	& 50388	& 5126	\\
{2}& {ship\_003*}	& 121728	& 8086034	& 66.427	& 3659	& 3833	\\
{3}& {pwtk*}	& 217918	& 11634424	& 53.389	& 189331	& 2029	\\
{4}& {offshore*}	& 259789	& 4242673	& 16.331	& 237738	& 19534	\\
{5}& {F1}	& 343791	& 26837113	& 78.062	& 343754	& 10052	\\
{6}& {inline\_1 (C)}	& 503712	& 36816342	& 73.09	& 502403	& 6002	\\
{7}& {parabolic\_fem* (C)}	& 525825	& 3674625	& 6.988	& 525820	& 514	\\
{8}& {gsm\_106857*}	& 589446	& 21758924	& 36.914	& 588744	& 17865	\\
{9}& {Fault\_639}	& 638802	& 28614564	& 44.794	& 19988	& 19487	\\
{10}& {Hubbard-12* (Q)}	& 853776	& 11098164	& 12.999	& 232848	& 38780	\\
{11}& {Emilia\_923}	& 923136	& 41005206	& 44.419	& 17279	& 14672	\\
{12}& {audikw\_1}	& 943695	& 77651847	& 82.285	& 925946	& 35084	\\
{13}& {bone010}	& 986703	& 71666325	& 72.632	& 13016	& 14540	\\
{14}& {dielFilterV3real}	& 1102824	& 89306020	& 80.979	& 1036475	& 25637	\\
{15}& {thermal2*}	& 1228045	& 8580313	& 6.987	& 1226000	& 797	\\
{16}& {Serena}	& 1391349	& 64531701	& 46.381	& 81578	& 84947	\\
{17}& {Geo\_1438}	& 1437960	& 63156690	& 43.921	& 26018	& 30623	\\
{18}& {Hook\_1498}	& 1498023	& 60917445	& 40.665	& 29036	& 28994	\\
{19}& {Flan\_1565}	& 1564794	& 117406044	& 75.03	& 20702	& 20849	\\
{20}& {G3\_circuit*}	& 1585478	& 7660826	& 4.832	& 947128	& 5068	\\
{21}& {Anderson-16.5* (Q)}	& 2097152	& 14680064	& 7.0	& 1198372	& 24620	\\
{22}& {FreeBosonChain-18 (Q)}	& 3124550	& 38936700	& 12.462	& 2042975	& 131749	\\
{23}& {nlpkkt120}	& 3542400	& 96845792	& 27.339	& 1814521	& 86876	\\
{24}& {channel-500x100x100-b050}	& 4802000	& 90164744	& 18.776	& 600299	& 23766	\\
{25}& {HPCG-192}	& 7077888	& 189119224	& 26.72	& 37057	& 110017	\\
{26}& {FreeFermionChain-26 (Q)}	& 10400600	& 140616112	& 13.52	& 5490811	& 434345	\\
{27}& {Spin-26 (Q)}	& 10400600	& 145608400	& 14.0	& 709995	& 211828	\\
{28}& {Hubbard-14 (Q)}	& 11778624	& 176675928	& 15.0	& 3171168	& 425415	\\
{29}& {nlpkkt200}	& 16240000	& 448225632	& 27.6	& 8240201	& 240796	\\
{30}& {delaunay\_n24}	& 16777216	& 100663202	& 6.0	& 16769102	& 32837	\\
{31}& {Graphene-4096 (C,Q)}	& 16777216	& 218013704	& 12.995	& 4098	& 6145	\\
\bottomrule
\end{tabular}

%% file: kernels.tex
We evaluate our methods by parallelization of the \acrshort{SymmSpMV}
kernel using \DTWO coloring, which avoids concurrent updates of the
same vector entries by different threads.


Since the kernel is closely related to the \acrfull{SpMV} kernel by
structure and computational intensity, we start with a discussion
of \acrshort{SpMV} and extend it towards \acrshort{SymmSpMV} later. In
all cases the aim is to derive realistic upper performance bounds,
which can be estimated once the computational intensity and main
memory bandwidth (\acrshort{b_s}; see \Cref{tab:test_bed}) are
known \cite{Williams_roofline}, \ie
\begin{align}
   	\label{eq:upper_performance}
   	P_\mathrm{kernel}  &= I_\mathrm{kernel}  \times b_S.
  \end{align}
Since $b_S$ depends on the ratio of load to store streams we present
the model for both upper (load-only) and lower bound (copy) bandwidth
cases.
In the following we choose the \acrfull{CRS} format for the
implementation of \acrshort{SpMV} as well as \acrshort{SymmSpMV} and
assume symmetric matrices.

\subsection{\acrshort{SpMV}}
A baseline \acrshort{SpMV} kernel is presented in \Cref{alg:SpMV}. It
has no loop-carried dependencies, so parallelization of the outer loop
using, \eg OpenMP, is straightforward.
\begin{algorithm}[t]
	\caption{\acrshort{SpMV} using the \acrshort{CRS} format: $b=A x$} 
	\label{alg:SpMV}
	\begin{algorithmic}[1]
	    \STATE{$double:: A[nnz], b[nrows], x[nrows]$}
	    \STATE{$integer:: col[nnz], rowPtr[nrows+1], tmp$}
		\FOR{$row=1:nrows$}
			\STATE{$tmp=0$}
			\FOR{$idx=rowPtr[row]:(rowPtr[row+1]-1)$}
				\STATE{$tmp += A[idx]*x[col[idx]]$} 
			\ENDFOR
			\STATE{$b[row] = tmp$}
		\ENDFOR
	\end{algorithmic}
\end{algorithm}
Following the discussion in~\cite{Moritz_sell}, its computational intensity is
\begin{equation}
\label{eq:SpMV_intensity}
I_\mathrm{\acrshort{SpMV}} (\alpha)= \frac{2}{8+4+8\alpha+20/\acrshort{NNZR}} \frac{\FLOP}{\BYTE}\eos
\end{equation}
Here we assume that the matrix data ($A[], col[]$), the left-hand side
(LHS) vector ($b[]$), and the row pointer information ($rowPtr[]$) are
loaded only once from main memory, since these data structures are
consecutively accessed. The intensity is calculated from the average
cost of performing all computations required for one nonzero element
of the matrix. Thus, contributions which are independent of the inner
(short) loop are rescaled by $\acrshort{NNZR}$, which is the average
number of nonzeros per row (i.e., the average length of the inner
loop).

The $8\alpha$ term quantifies the data traffic caused by accessing the
RHS vector ($x[]$). The value of $\alpha$ depends on
the matrix structure as well as on the RHS vector data set size and
the available cache size. The minimum value of
$\alpha=\acrshort{NNZR}^{-1}$ is attained if the RHS vector is only
loaded once from main memory to the cache and all subsequent accesses
in the same \acrshort{SpMV} are cache hits. This limit is typically
observed for matrices with low bandwidth (high access locality) or if
the cache is large enough to hold the full RHS data during
one \acrshort{SpMV}. The actual value of $\alpha$ can be determined
experimentally by measuring the data traffic when executing
the \acrshort{SpMV}; see~\cite{Moritz_sell} for more
details.\footnote{In~\cite{Moritz_sell} the traffic for the row
pointer was not accounted for, \ie the denominator in
(\ref{eq:SpMV_intensity}) is larger by
$\frac{4}{\acrshort{NNZR}}\,\BYTE$. This error is only significant
when $\acrshort{NNZR}$ is small.}  The optimal value of
$\alpha=\acrshort{NNZR}^{-1}$ together with the corresponding
computational intensities for all matrices is shown in 
\Cref{table:alpha_values}. The measured $\alpha_{\acrshort{SpMV}}$
 is used as a sensible lower bound for $\alpha_{\acrshort{SymmSpMV}}$ 
 values (see \Cref{sect:SymmSpmv}) in cases where advanced cache
 replacement strategies  do not apply; therefore the table also presents the 
 corresponding measured $\alpha_{\acrshort{SpMV}}$ (= assumed $\alpha_{\acrshort{SymmSpMV}}$)
values for different matrices.
\begin{table}[t]
	\centering
	\caption{The optimal value of $\alpha_{\acrshort{SpMV}}$
	is shown in column three. Following \Cref{eq:upper_performance} the
	maximum \acrshort{SpMV} performance can be calculated for each
	architecture using the best intensity values
	($I_{\acrshort{SpMV}}(\alpha_{SpMV})$ in
	$\frac{\FLOP}{\BYTE}$) shown in the fourth
	column. The assumed $\alpha_{\acrshort{SymmSpMV}}$ on \SKX and 
	\IVB architectures are presented in columns five and six, respectively.
	The assumed $\alpha_{\acrshort{SymmSpMV}}$ is equal to 
	the measured $\alpha_{\acrshort{SpMV}}$ for all matrices
	except the ones marked with asterisk, where  $\alpha_{\acrshort{SymmSpMV}}$
	is set to optimal $\alpha_{\acrshort{SymmSpMV}}$ (= 1/\acrshort{SymmNNZR}).
	 \label{table:alpha_values}}
    \begin{center} 
      \input{pics/results/alpha_table_generator/table.tex}

    \end{center}
\end{table}
Choosing the matrices 10, 22, and 31, which have approximately the
same optimal $\alpha_{\acrshort{SpMV}}$, one can study the delicate
influence of matrix structure (\ie matrix bandwidth and number of
rows; see \Cref{tab:test_mtx}) and the cache size on the actual data
traffic, \ie the measured values of $\alpha_{\acrshort{SpMV}}$.
 
For most of the ten candidate matrices on the \SKX architecture that could
potentially show a caching effect (see \Cref{table:bench_matrices}) we
observe the measured $\alpha_{\acrshort{SpMV}}$ to be lower than
optimal. In this case we set their $\alpha_{\acrshort{SymmSpMV}}$ values to the optimal alpha value
of \acrshort{SymmSpMV} ($\alpha_{\acrshort{SymmSpMV}}$) which will be
defined in the following \Cref{sect:SymmSpmv}. These cases are marked
with an asterisk in \Cref{table:alpha_values}.

\subsection{\acrshort{SymmSpMV}}
\label{sect:SymmSpmv}

\Acrshort{SymmSpMV} exploits the symmetry of the matrix ($A_{ij}=A_{ji}$)
to reduce storage size for matrix data and reduce the overall memory
traffic by operating on the upper (or lower) half of the matrix
only. Thus for every nonzero matrix entry we need to update two
entries in the LHS vector ($b[]$) as shown in~\Cref{alg:SymmSpMV}.

\begin{algorithm}[t]
	\caption{SymmSpMV $b=Ax$, where $A$ is an upper triangular matrix} 
	\label{alg:SymmSpMV}
	\begin{algorithmic}[1]
		\FOR{$row=1:nrows$}
		\STATE{$diag\_idx=rowPtr[row]$}
		\STATE{$b[row] += A[diag\_idx]*x[row]$}
			\STATE{$tmp = 0$}
			\FOR{$idx=(rowPtr[row]+1):(rowPtr[row+1]-1)$}
				\STATE{$tmp += A[idx]*x[col[idx]]$}
				\STATE{$b[col[idx]] += A[idx]*x[row]$} 
			\ENDFOR
			\STATE{$b[row] += tmp$}
		\ENDFOR
	\end{algorithmic}
\end{algorithm}
In line with the discussion above, the computational intensity of \acrshort{SymmSpMV} is
\begin{align}
\label{eq:SymmSpMV_intensity}
I_\mathrm{\acrshort{SymmSpMV}} (\alpha) &= \frac{4}{8+4+24\alpha+4/\acrshort{SymmNNZR}} \frac{\FLOP}{\BYTE}\cma\\
\label{eq:NNZR_symm}
\text{ where  } \acrshort{SymmNNZR} &= (\acrshort{NNZR}-1)/2 + 1\eos
\end{align}

For a given nonzero matrix element $4~\FLOP$ are performed, which is
twice the amount of work than in \acrshort{SpMV}. In addition we have
indirect access to the LHS vector (read and write) which triples the
traffic contribution quantified by $\alpha$\@. The only term scaled
with \acrshort{SymmNNZR} (number of nonzeros per row in the upper
triangular part of the matrix) is the row pointer. The most optimistic
value of $\alpha$ ($\alpha_{\acrshort{SymmSpMV}}$) in this case is
$1/{\acrshort{SymmNNZR}}$, which corresponds to a one time transfer of
the LHS and RHS vectors.  Note that the $\alpha$ for \acrshort{SpMV}
and \acrshort{SymmSpMV} may be different even for the same matrix and
the same compute device, as in the latter case the two vectors are
accessed irregularly and compete for cache. Thus we
can assume that the $\alpha$ value measured for \acrshort{SpMV}
($\alpha_\mathrm{\acrshort{SpMV}}$) is a lower bound
for \acrshort{SymmSpMV}. \Cref{table:alpha_values} show the 
 assumed $\alpha_{\acrshort{SymmSpMV}}$ values taken for performance
modeling. Since an upper bound for the performance is the
product of computational intensity and main memory bandwidth
(see \Cref{eq:upper_performance}), this approach provides an upper
performance bound for \acrshort{SymmSpMV}.  However, note that the
performance models derived for matrices having caching effects
(see \Cref{table:alpha_values}) need not be strictly upper bound, as
they heavily depends on the caching strategy of the underlying
architecture.

Comparing~\Cref{eq:SymmSpMV_intensity} and~\Cref{eq:SpMV_intensity} it is obvious that the perfect speedup of 2$\times$ when using \acrshort{SymmSpMV} instead of \acrshort{SpMV} is only attainable in the limit of small $\alpha$\@. 
Considering the large prefactor of the $\alpha$ contribution, any implementation of \acrshort{SymmSpMV} must aim at ensuring high data locality. The indirect update of the LHS also has a large impact on parallelization strategy as two rows which have a nonzero in the same column cannot be computed in parallel. In a graph-based approach to this problem, this is equivalent to the constraint that only vertices which have at least distance two can be computed in parallel.

%% file: pics/results/alpha_table_generator/table.tex
\begin{tabular}{|l|l|S[round-mode=places,round-precision=4]|S[round-mode=places,round-precision=4]|S[round-mode=places,round-precision=4]|S[round-mode=places,round-precision=4]|}
\toprule
\multirow{2}{*}{Index} & \multirow{2}{*}{Matrix name} & \multicolumn{1}{c|}{$\alpha_{SpMV}$} & {$I_{\acrshort{SpMV}}(\alpha_{SpMV})$} & \multicolumn{2}{c|}{Assumed $\alpha_{SymmSpMV}$} \\
\cline{3-6}
& &  {Optimal} & {Optimal} & {SKX} & {IVB}  \\
\midrule
{1}& {	crankseg\_1                }	& 0.004974840341951422	& 0.16475420629866486	& 0.009900427637091272*	& 0.017876	\\
{2}& {	ship\_003                  }	& 0.015054104375856307	& 0.16101095659221026	& 0.029661678743938248*	& 0.039038	\\
{3}& {	pwtk                      }	& 0.018730450092715727	& 0.1596876177714501	& 0.03677214142565592*	& 0.038276	\\
{4}& {	offshore                  }	& 0.061232390023872055	& 0.14583098113326293	& 0.11539864519682566*	& 0.105831	\\
{5}& {	F1                        }	& 0.012810282496064584	& 0.16182947693011468	& 0.025296509558520128*	& 0.043622	\\
{6}& {	inline\_1                  }	& 0.013681750417974054	& 0.16151058900649082	& 0.013709	& 0.034046	\\
{7}& {	parabolic\_fem             }	& 0.14309623622555628	& 0.12494772020022805	& 0.25036603514337963*	& 0.224973	\\
{8}& {	gsm\_106857                }	& 0.02708985058701268	& 0.15675804527541276	& 0.052750692788036055*	& 0.094584	\\
{9}& {	Fault\_639                 }	& 0.022324366119157866	& 0.15841480951843234	& 0.045281	& 0.086085	\\
{10}& {	Hubbard-12                }	& 0.07692947982285911	& 0.14130255800224512	& 0.14286818452683273*	& 0.231786	\\
{11}& {	Emilia\_923                }	& 0.022512653462004855	& 0.15834868547473438	& 0.08265	& 0.085462	\\
{12}& {	audikw\_1                  }	& 0.012152898336217176	& 0.16207086168751325	& 0.062422	& 0.063762	\\
{13}& {	bone010                   }	& 0.013768014517655372	& 0.16147909155409917	& 0.049208	& 0.052338	\\
{14}& {	dielFilterV3real          }	& 0.01234882033880347	& 0.16199884583462107	& 0.072827	& 0.067509	\\
{15}& {	thermal2                  }	& 0.14312355713563962	& 0.12494174903463444	& 0.2504078517886007*	& 0.227709	\\
{16}& {	Serena                    }	& 0.02156070528689192	& 0.15868356434880437	& 0.100582	& 0.115621	\\
{17}& {	Geo\_1438                  }	& 0.022768134283905977	& 0.15825905217944752	& 0.089589	& 0.091725	\\
{18}& {	Hook\_1498                 }	& 0.024591034497360605	& 0.1576224362116434	& 0.103075	& 0.094818	\\
{19}& {	Flan\_1565                 }	& 0.013328053114104274	& 0.161639862432339	& 0.054135	& 0.052516	\\
{20}& {	G3\_circuit                }	& 0.20695912474502637	& 0.11239203379889182	& 0.34294305499160477*	& 0.335974	\\
{21}& {	Anderson-16.5             }	& 0.14285714285714285	& 0.125	& 0.363368	& 0.318715	\\
{22}& {	FreeBosonChain-18         }	& 0.08024691655235494	& 0.14038128167567254	& 0.27076	& 0.262774	\\
{23}& {	nlpkkt120                 }	& 0.03657773850304069	& 0.15356057042478993	& 0.160002	& 0.165642	\\
{24}& {	channel-500x100x100-b050  }	& 0.05325806761196896	& 0.14824449726378677	& 0.173504	& 0.133898	\\
{25}& {	HPCG-192                  }	& 0.03742553488106633	& 0.15328119500901655	& 0.135801	& 0.139089	\\
{26}& {	FreeFermionChain-26       }	& 0.07396449704142012	& 0.1421362489486964	& 0.387859	& 0.397282	\\
{27}& {	Spin-26                   }	& 0.07142857142857142	& 0.14285714285714285	& 0.367034	& 0.351781	\\
{28}& {	Hubbard-14                }	& 0.06666796002509115	& 0.14423039256024434	& 0.357508	& 0.359807	\\
{29}& {	nlpkkt200                 }	& 0.036231752783504406	& 0.15367487636455557	& 0.16692	& 0.172028	\\
{30}& {	delaunay\_n24              }	& 0.1666668333335	& 0.1199999663999758	& 0.406459	& 0.319197	\\
{31}& {	Graphene-4096             }	& 0.0769548711240621	& 0.14129546073388705	& 0.160392	& 0.127774	\\
\bottomrule
\end{tabular}

%% file: motivation.tex

\begin{figure}[t]
  	\centering
    \subfloat[SymmSpMV]{\label{fig:motivation_symm_spmv}\includegraphics[width=0.248\textwidth, height=0.20\textheight]{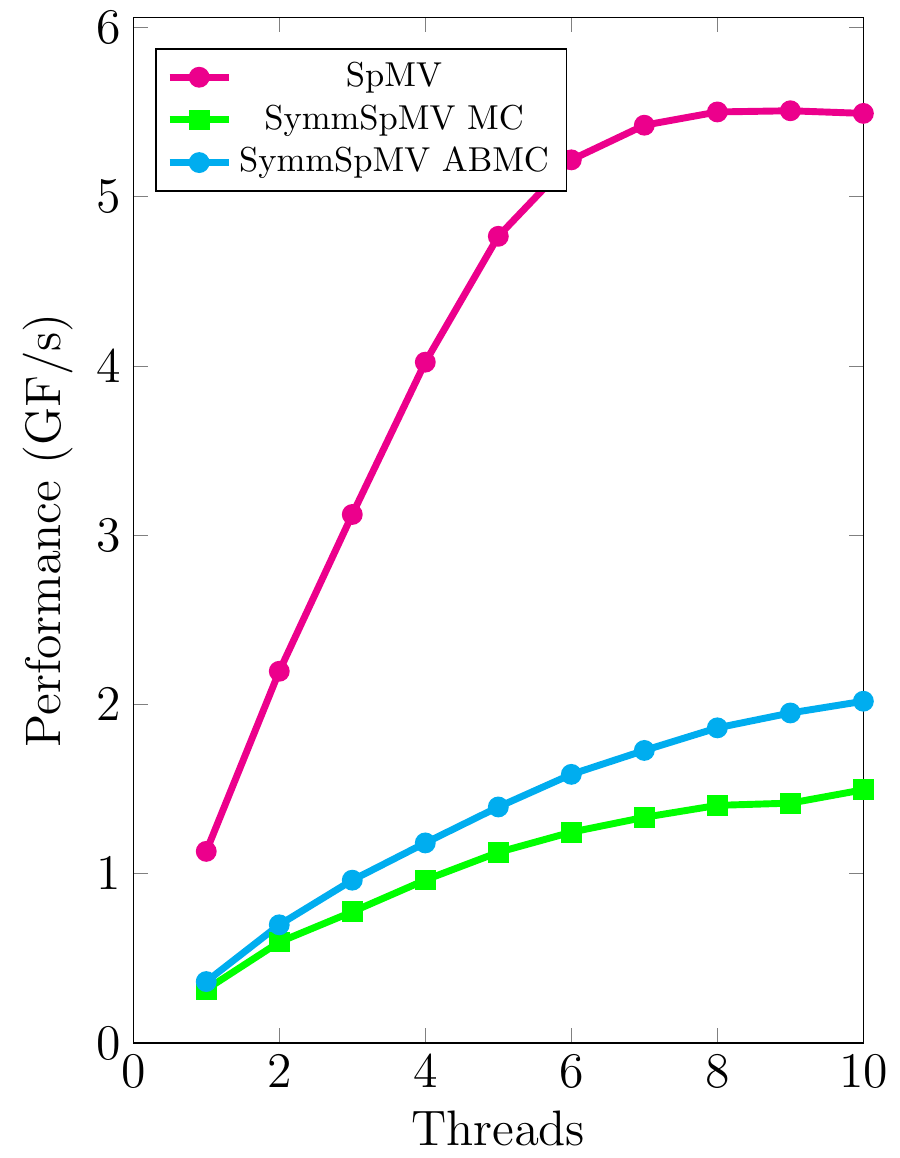}}
  	\subfloat[Data traffic]{\label{fig:motivation_data}\includegraphics[width=0.248\textwidth, height=0.20\textheight]{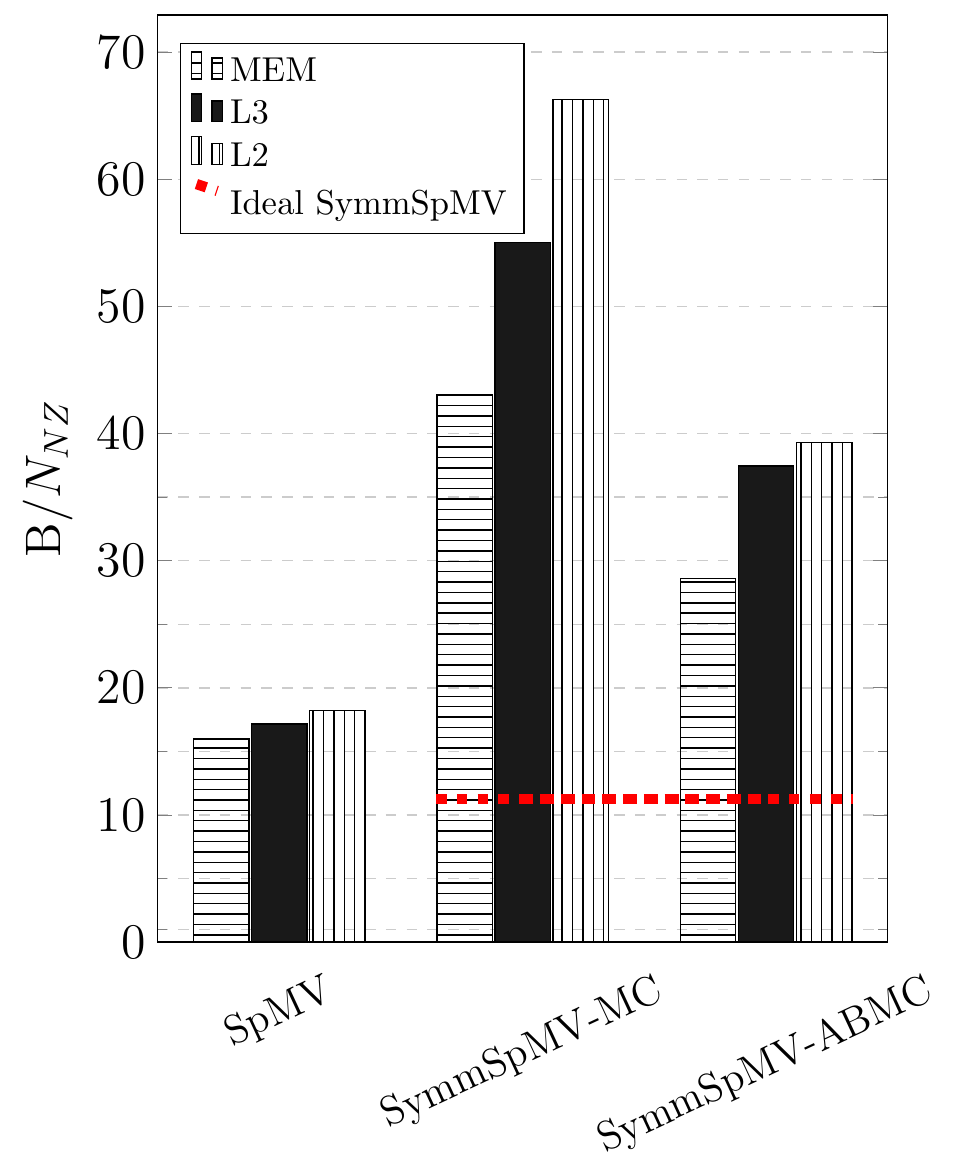}} 
    \subfloat[SymmSpMV]{\label{fig:motivation_symm_spmv_skx}\includegraphics[width=0.248\textwidth, height=0.20\textheight]{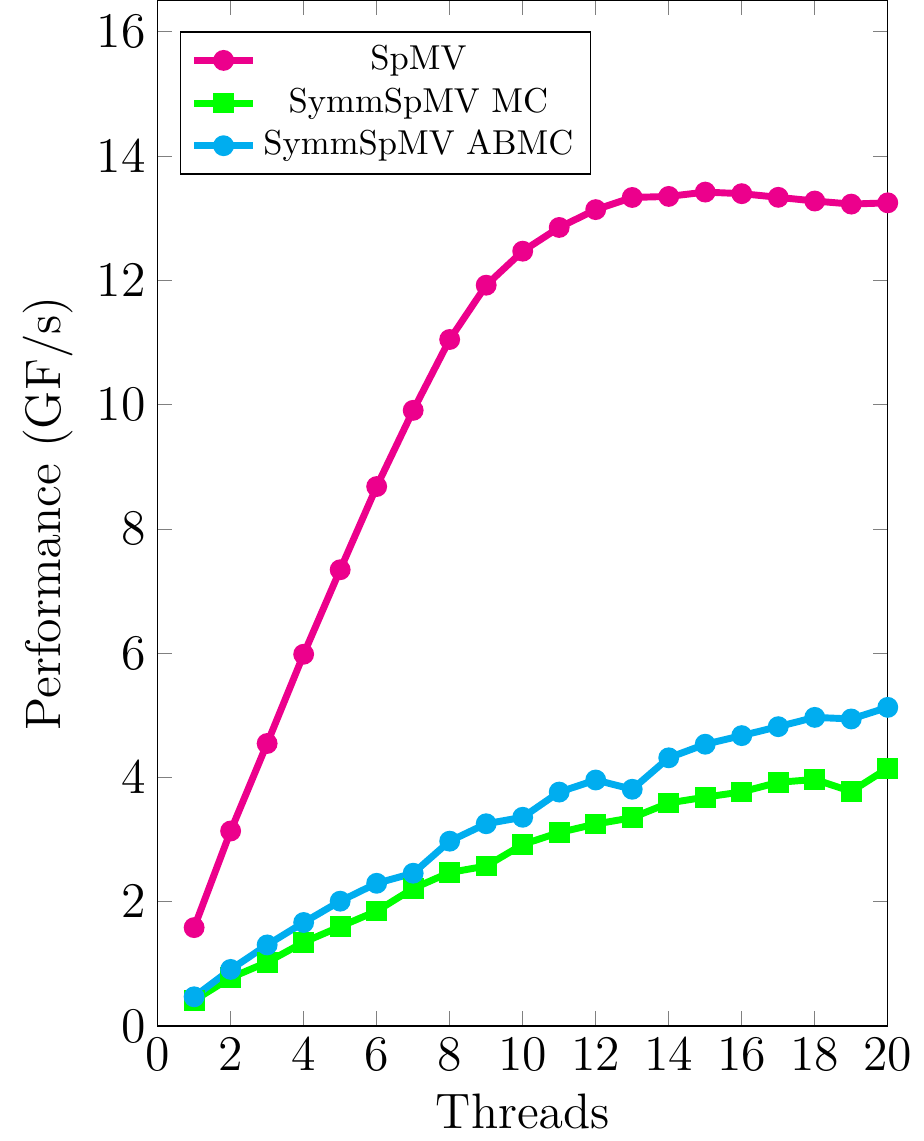}}
    \subfloat[Data traffic]{\label{fig:motivation_data_skx}\includegraphics[width=0.248\textwidth, height=0.20\textheight]{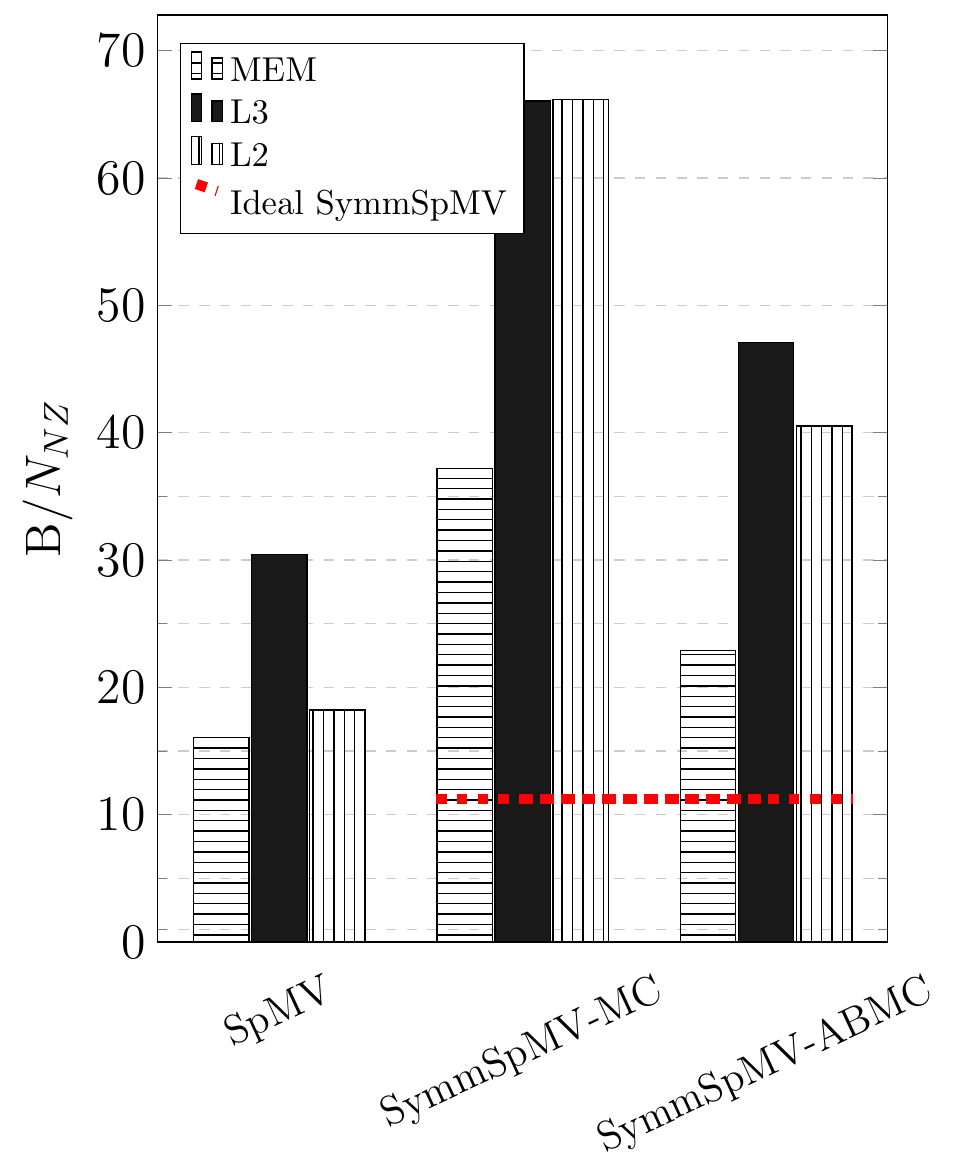}}
  	\caption{Scaling performance of \acrshort{SymmSpMV} with \acrshort{MC} and \acrshort{ABMC} compared to \acrshort{SpMV} on one socket of \IVB and \SKX is shown in \Cref{fig:motivation_symm_spmv,fig:motivation_symm_spmv_skx} respectively. \Cref{fig:motivation_data,fig:motivation_data_skx} show average data traffic per nonzero entry ($\acrshort{nnz}$) of the full matrix as measured with \LIKWID for all cache levels and main memory on full socket of \IVB and \SKX respectively. The Spin-26 matrix was prepermuted with \acrshort{RCM}.}
  	\label{fig:motivation}
  \end{figure}

As pointed out previously, the parallelization of
the \acrshort{SymmSpMV} kernel can be done
via \DTWO coloring of the corresponding graph. The computational
intensity, and hence the performance, depends on the data access
patterns to the LHS and RHS vectors. As coloring schemes change those
patterns, they may change the computational intensity, and we have to
investigate this effect in more detail. We apply the basic
\acrshort{MC} scheme generated by COLPACK~\cite{COLPACK} to
parallelize \acrshort{SymmSpMV} and compare it with \acrshort{SpMV},
which serves as our performance yardstick. Note that any required
preprocessing is excluded from the timings. In~\Cref{fig:motivation}
we present performance and data transfer volumes for
the \emph{Spin-26} matrix on a single socket of the \IVB
and \SKX systems. For \acrshort{SpMV} we recover the well-known memory bandwidth
saturation pattern as we fill the chip (\Cref{fig:motivation_symm_spmv,fig:motivation_symm_spmv_skx}).
 Measuring the actual data volume from main memory using \LIKWID
we find $16.24$ and $16.36$ \BYTE per nonzero matrix entry (\Cref{fig:motivation_data,fig:motivation_data_skx})
on \IVB and \SKX architectures.
 This corresponds to the denominator of
$I_\mathrm{\acrshort{SpMV}}$ in~\Cref{eq:SpMV_intensity}, so we can
determine $\alpha_\mathrm{\acrshort{SpMV}}=0.351$ for \IVB and $0.367$ for \SKX, thus we can calculate an 
optimistic bound for the intensity of \acrshort{SymmSpMV} according to ~\Cref{eq:SymmSpMV_intensity}.
 Using the copy and the load-only bandwidth of \IVB (see~\Cref{tab:test_bed}) 
 in \Cref{eq:upper_performance} we find a maximum
attainable {\acrshort{SymmSpMV}} performance range for this matrix of
$P_\mathrm{\acrshort{SymmSpMV}}=7.63,\ldots,8.96\,\GF$, while for \SKX we expect  $P_\mathrm{\acrshort{SymmSpMV}}=19.49,\ldots,21.55\,\GF$ . This indicates a possible
speedup of approximately 1.4$\times$ -- 1.6$\times$ compared to the \acrshort{SpMV}
baseline ($5.5\,\GF$ and $13.41\,\GF$ on \IVB and \SKX), the \acrshort{SymmSpMV} implementation
using \acrshort{MC} falls short of this expectation and is more than
three times slower than \acrshort{SpMV}.

  \begin{figure}[t]
  	\centering
  	\includegraphics[scale=0.45]{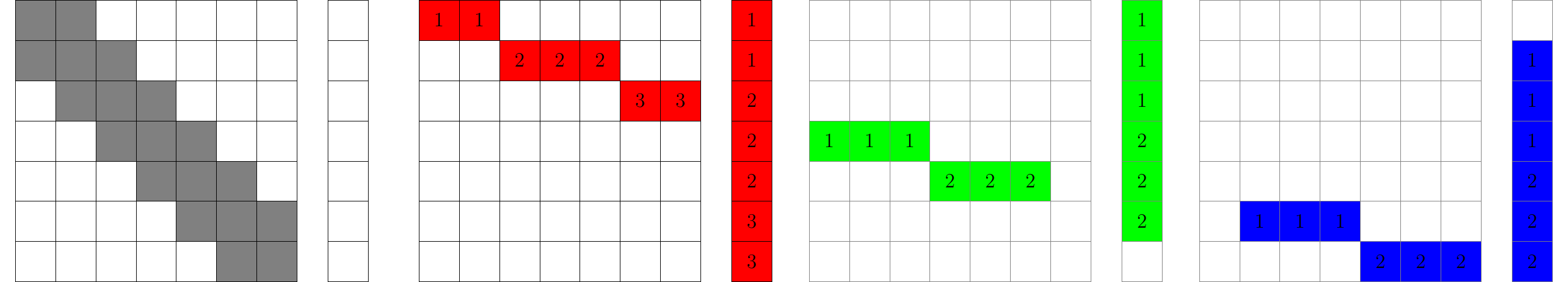}
  	\caption{Illustration of the increase of $\alpha$ by \acrshort{MC}. Numbers represent thread ids. Note that this figure shows only rows of the matrix permuted according to \acrshort{MC}, but in practice one would permute both rows and columns.}
  	\label{fig:mc_alpha}
  \end{figure}
  
The reason for this decrease is the nature of the \acrshort{MC}
permutation. For \DTWO coloring, sets of structurally orthogonal
rows have to be determined \cite{dist_k_def}, \ie rows that do not
overlap in any column entry. These sets are referred to as colors. \Cref{fig:mc_alpha} shows the corresponding permutation
and the obtained sets of colors applied to a toy problem with high
data locality. Different rows of the same color can be executed in
parallel, but colors are operated one after the
other. After \acrshort{MC} a color may contain rows from very distant
parts of the matrix, potentially destroying data locality.  Assuming
that the \acrshort{LLC} can hold a maximum of six elements, we find
that the RHS vector must be loaded every time for each
new color. This is the reason why we observe 3$\times$ more bytes
per nonzero for \acrshort{SymmSpMV} with \acrshort{MC} compared
to \acrshort{SpMV}, as seen in~\Cref{fig:motivation_data,fig:motivation_data_skx}.  However,
our performance model indicates that \acrshort{SymmSpMV} should
exhibit only 0.7$\times$ the data traffic of \acrshort{SpMV} (see
red dotted line in \Cref{fig:motivation_data,fig:motivation_data_skx}). Of course this
effect strongly depends on the matrix structure, the matrix size, and
the cache size.
        
\Acrshort{ABMC}\cite{ABMC} tries to preserve data locality by first partitioning
the entire matrix into blocks of specified size and then
applying \acrshort{MC} to these blocks. Threads then work in parallel
between blocks of the same color. Along the lines of \cite{Park_HPCG} 
we use \METIS \cite{METIS} to partition the matrix into blocks, and COLPACK 
for \acrshort{MC}. The size of blocks for \acrshort{ABMC} is
determined by a parameter scan (range 4 \ldots 128;
see~\cite{ABMC})\@. As stated above, the timing for the performance measurements
excludes preprocessing and the parameter search. This method reduces the 
data traffic (see \Cref{fig:motivation_data,fig:motivation_data_skx}) as there is better data
locality within a block. Consequently, the performance improves
over plain \acrshort{MC} (see \Cref{fig:motivation_symm_spmv,fig:motivation_symm_spmv_skx}). However,
we are far off the performance model prediction. In
addition to data locality, other factors like global synchronizations
and false sharing also contribute to this failure. These
effects strongly depend on the number of colors and in general
increase with chromatic number. In the case of the Spin-26 matrix the overhead of
synchronization is roughly 10\% for the \acrshort{MC} method.  For most of
the matrices considered in this work one can also observe a strong
positive correlation between false sharing and the number of threads
for \acrshort{SymmSpMV} kernels due to the indirect writes
in \acrshort{SymmSpMV}.

%% file: race_method.tex

Our advanced coloring algorithm is based on three steps:
\begin{enumerate}
	\item level construction,
	\item \DK coloring,
	\item load balancing.
\end{enumerate}
In the first step we apply a bandwidth reduction algorithm
including level construction and matrix reordering. We then use the
information from the level construction step to form subsets of levels
which allow for hardware efficient \DK coloring of the graph. Finally
we present a concept to ensure load balancing between threads. These steps
are applied recursively if required.

\begin{figure}[t]
	\centering
	\subfloat[\Stex]{\label{fig:2d-7pt-a}\includegraphics[width=0.17\textwidth , height=0.13\textheight]{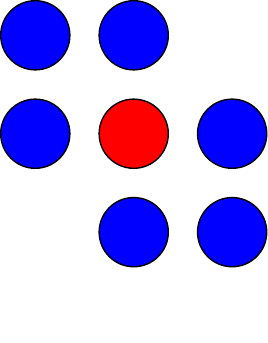}}
	\hspace{0.8em}
	\subfloat[Sparsity
	 pattern]{\label{fig:2d-7pt-b}\includegraphics[width=0.38\textwidth , height=0.19\textheight]{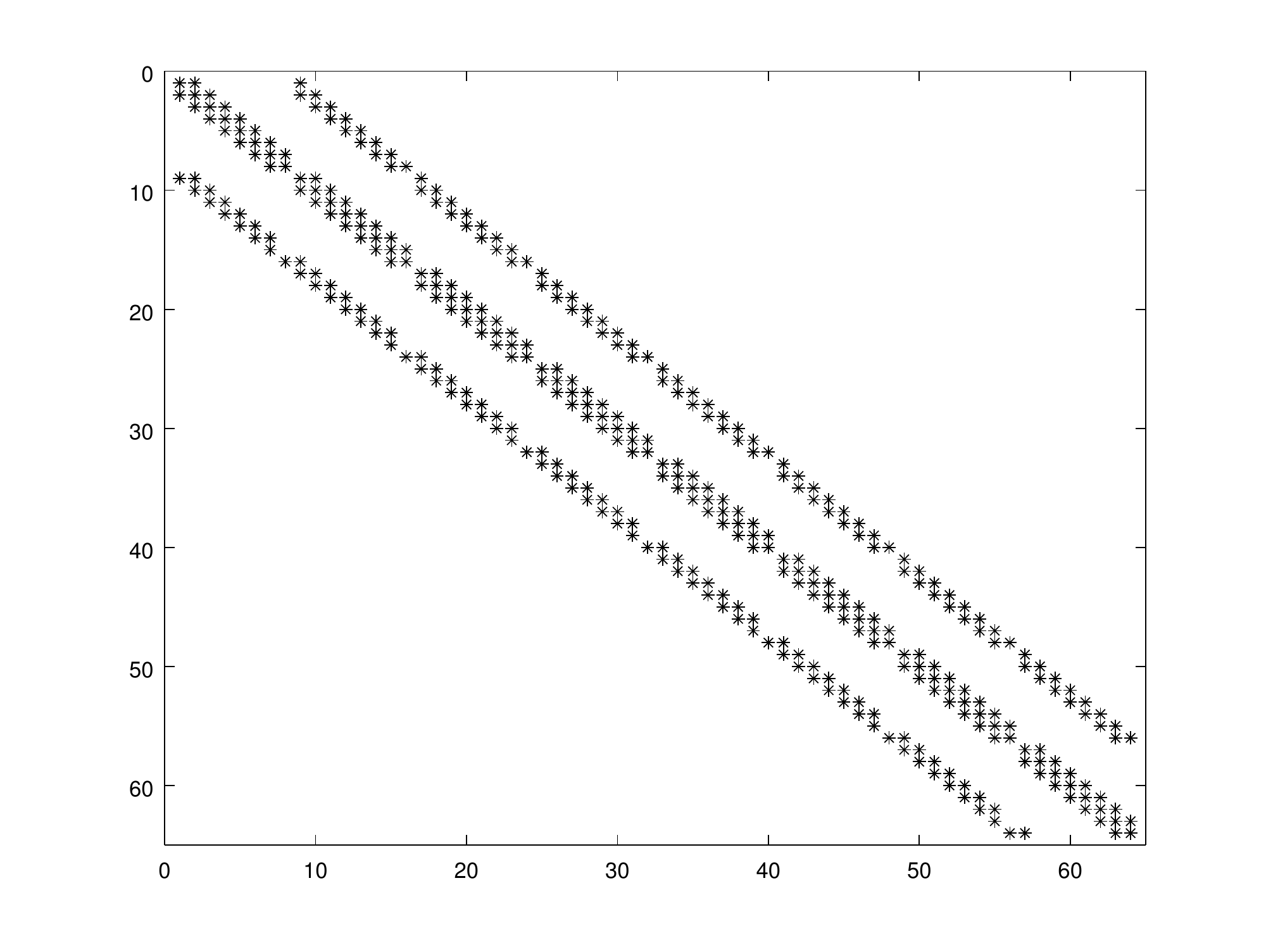}}
	\hspace{1em}
	\subfloat[Graph]{\label{fig:2d-7pt-c}\includegraphics[width=0.3\textwidth , height=0.18\textheight]{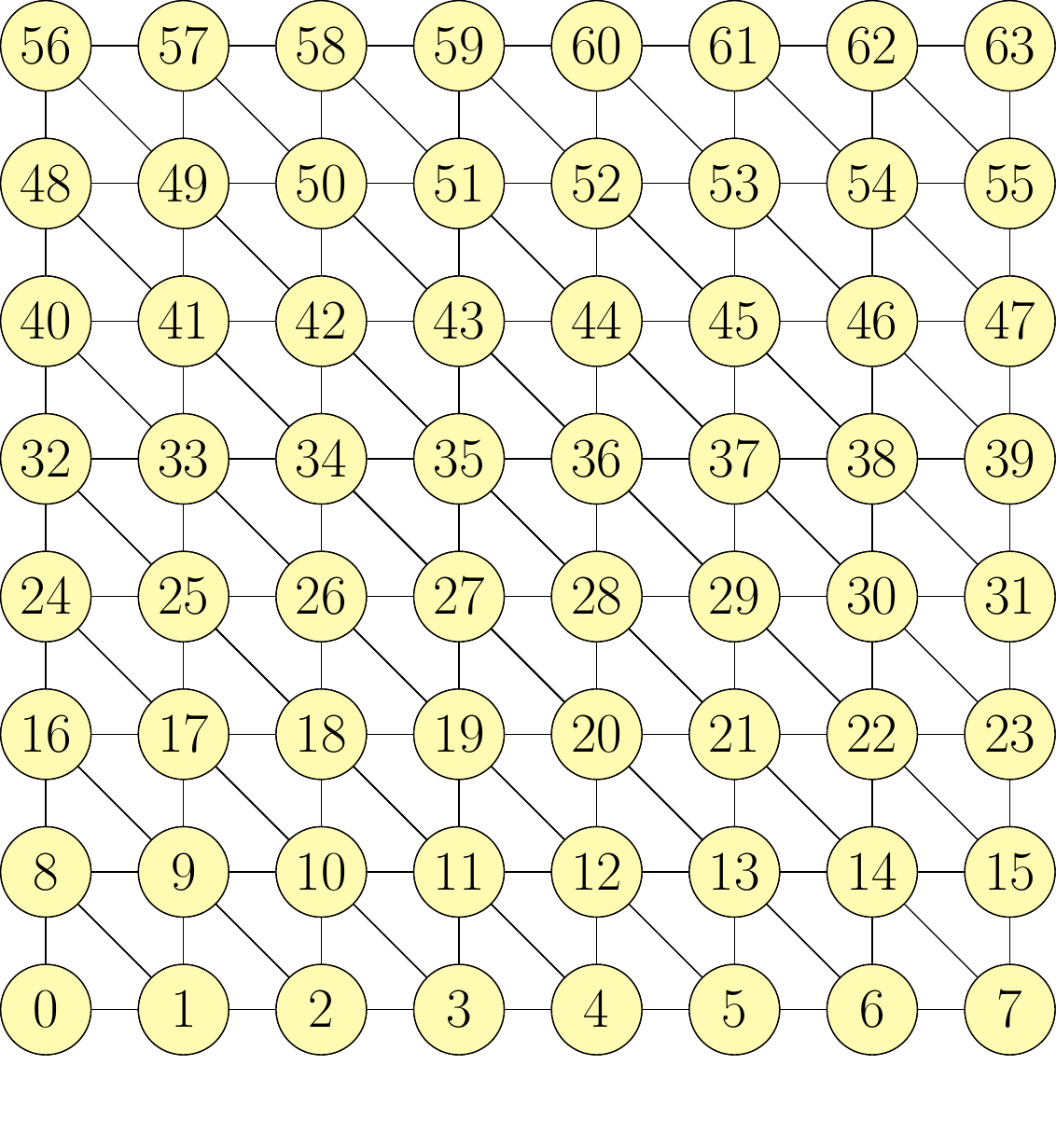}}
	\caption{\sref{fig:2d-7pt-a} Structure of an artificially designed stencil,
	\sref{fig:2d-7pt-b} corresponding sparsity pattern of its matrix
	representation on an $8\times 8$ lattice with Dirichlet
	boundary conditions, and \sref{fig:2d-7pt-c} the graph representation of the
	matrix. The stencil
	structure was chosen for illustration purposes and does not
	represent any specific application scenario.}
	\label{fig:2d-7pt}
\end{figure}

To illustrate the method we choose a
simple matrix which is associated with an artificially constructed
two-dimensional stencil as shown in \Cref{fig:2d-7pt-a}. The
corresponding sparsity pattern and the graph of the matrix are shown
in \Cref{fig:2d-7pt-b,fig:2d-7pt-c} respectively.

\subsection*{Definitions}
We need the following definitions from graph theory:
\begin{itemize}
	\item \textbf{Graph: } $G = (V,E)$ represents a graph, with $V(G)$
              denoting its set of vertices and $E(G)$ denoting its edges. Note that
              we restrict ourselves to irreducible undirected graphs.
	\item \textbf{Neighborhood:} $N(u)$ is the neighborhood of a vertex $u$ and is defined as
	\begin{equation*}
	  N(u) = \set{ v \in V(G) : (u,v) \in E(G)}\eos
	\end{equation*}
	\item \textbf{$k$th Neighborhood:} $N^{k}(u)$ of a vertex $u$ is defined as
	 \begin{align*}
	 	N^2(u) &= N(N(u))  \\
	 	N^3(u) &= N^2(N(u)) \\
	 	\vdots\\
	 	N^k(u) &= N^{k-1}(N(u)) \eos
	 \end{align*}
	\item \textbf{Subgraph:} In this paper a subgraph $H$ of $G$ specifically
              refers to the subgraph induced by vertices $V' \subseteq V(G)$ and is defined as
	\begin{equation*}
		H = (V', \set{ (u,v) : (u,v) \in E(G) \text{ and } u,v \in V'})\eos
	\end{equation*}
\end{itemize}

\subsection{Level Construction}\label{subsec:LEVEL_CONST}

The first step of \acrshort{RACE} is to determine
different \textit{\levels} in the graph and permute the
graph data structure. This we achieve using
well-known bandwidth reduction algorithms such as \acrfull{RCM} \cite{RCM}
or \acrfull{BFS} \cite{BFS}\@. Although the RCM method is
also implemented in \acrshort{RACE}, we use the \acrshort{BFS}
reordering in the following for simpler illustration.

\begin{figure}[t]
	\centering
	\subfloat[Level construction]{\label{fig:2d_7pt_level_construction}\includegraphics[height=0.18\textheight,width=0.32\textwidth]{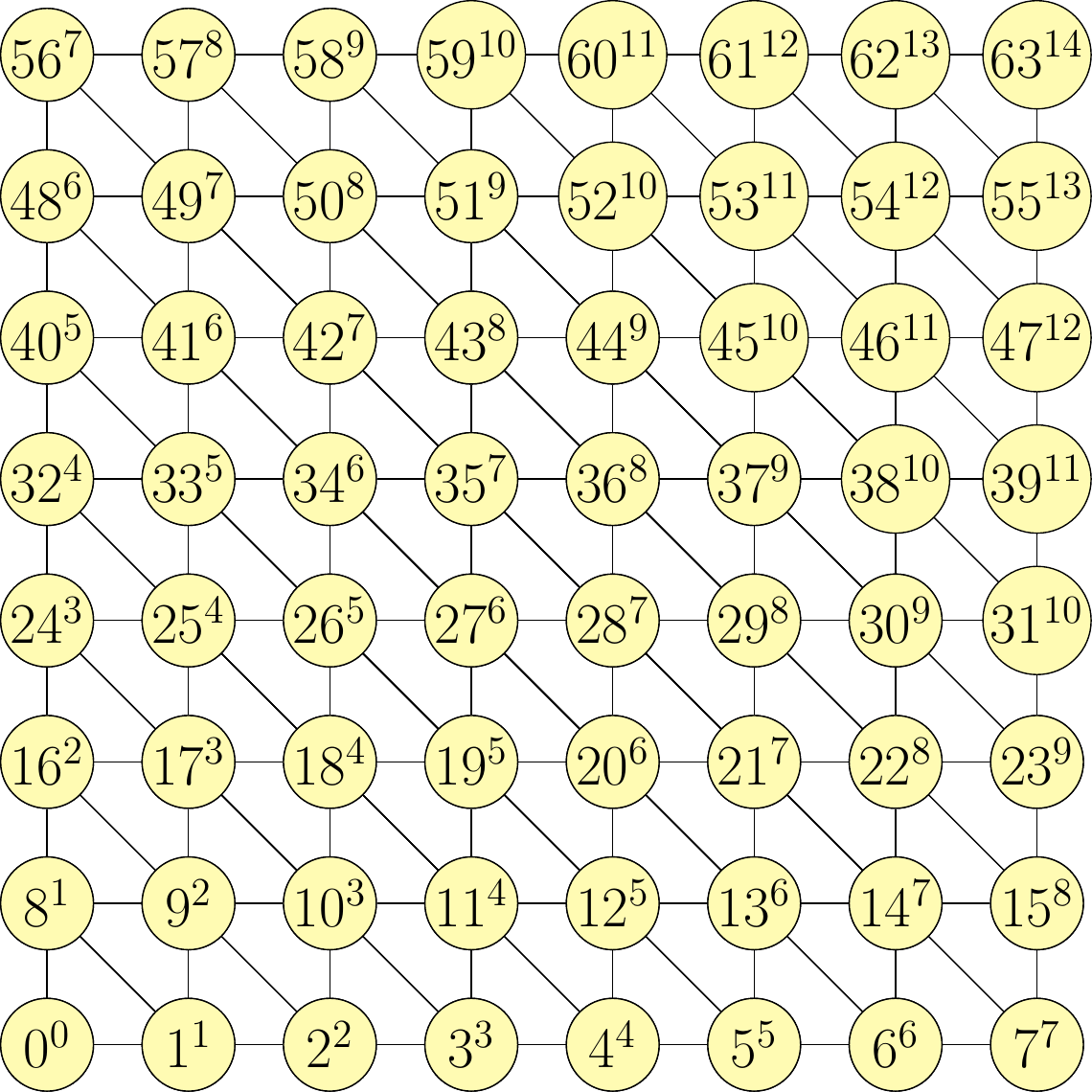}
			\begin{picture}(0,0)
			\put(-44,68){\fbox{\includegraphics[height=1.4cm]{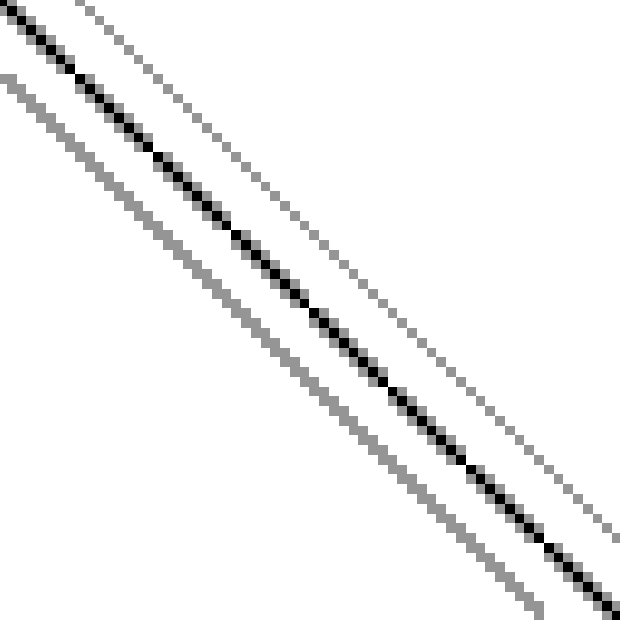}}}
			\end{picture}
		}
	\hspace{1em}
	\subfloat[Permuted graph ($G'$)]{\label{fig:2d_7pt_perm}\includegraphics[height=0.18\textheight,width=0.32\textwidth]{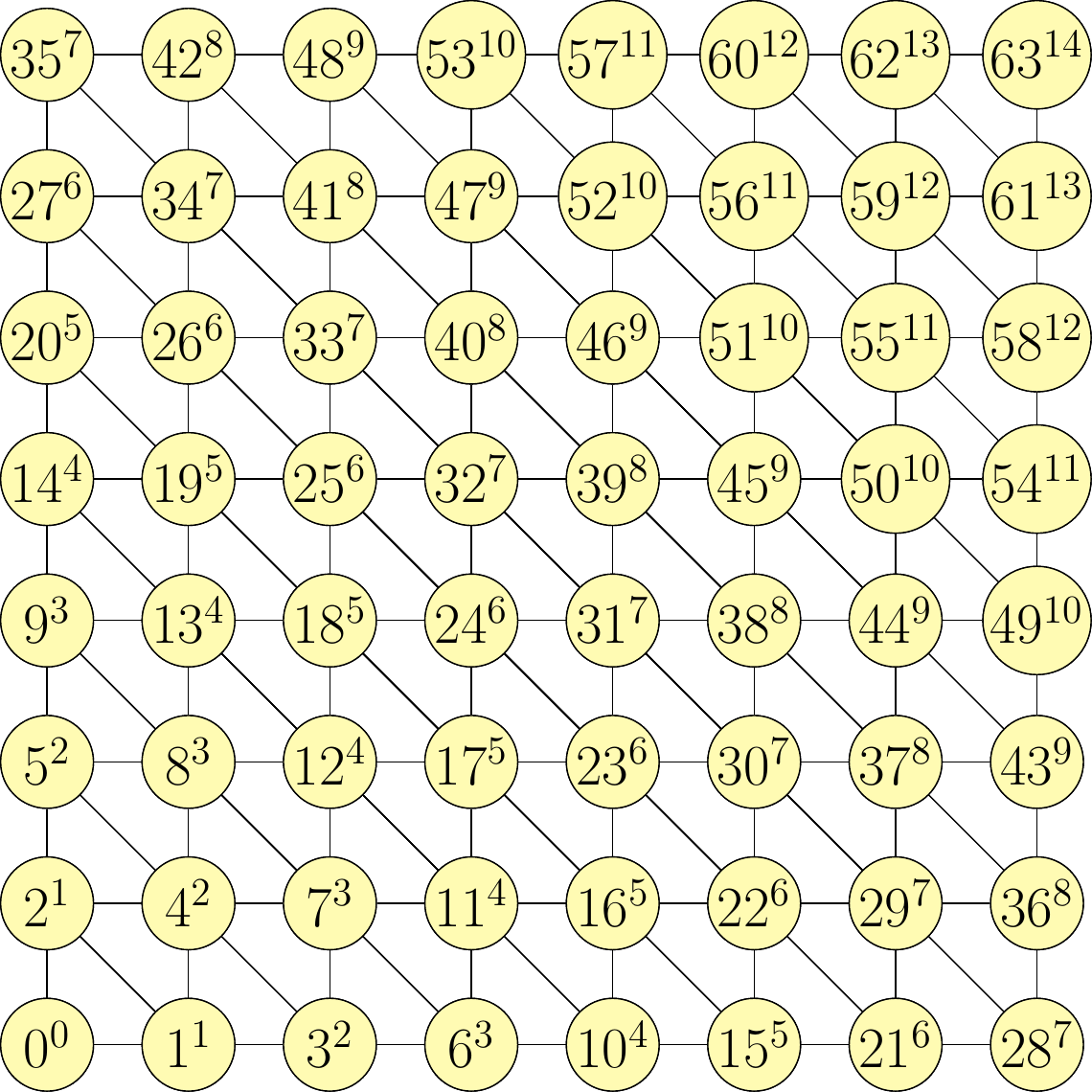}
			\begin{picture}(0,0)
			\put(-43.5,68){\fbox{\includegraphics[height=1.4cm]{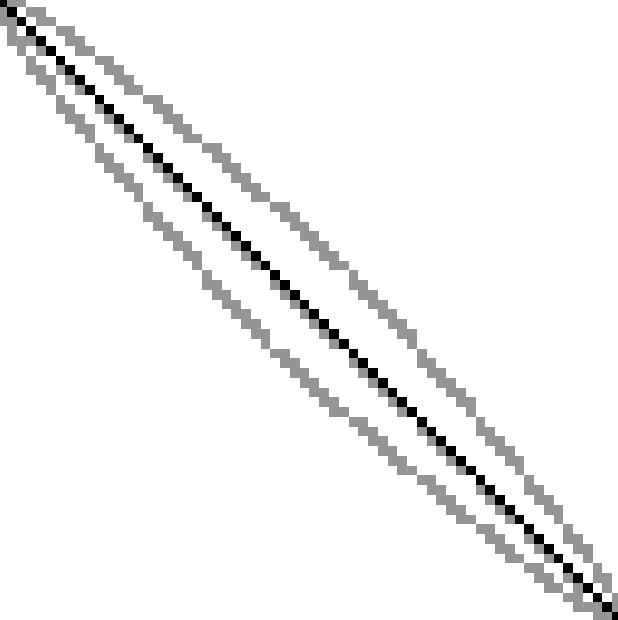}}}
			\end{picture}
		}
	\hspace{1em}
	\subfloat[]{\label{fig:2d_7pt_levelPtr}\includegraphics[height=0.18\textheight,width=0.07\textwidth]{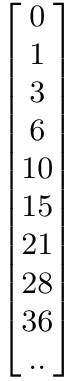}}
	\caption{\sref{fig:2d_7pt_level_construction} Levels of the original graph and \sref{fig:2d_7pt_perm} the permuted
          graph for the \stex. Insets show the corresponding sparsity patterns.
          \sref{fig:2d_7pt_levelPtr} Shows the entries of the \levelPtr array associated
          with $G'$.}
	\label{fig:2d-7pt_step_1_2}
\end{figure}

First we choose a \emph{root} vertex and assign it to the
first \level, $L(0)$\@. For $i>0$, \level \acrshort{L_i}
is defined to contain vertices that are in the neighborhood of vertices
in $L(i-1)$ but not in the neighborhood of vertices
in $L(i-2)$ \cite{BFS_level_def}, \ie
\begin{equation}\label{eq:level}
L(i) = 
\begin{cases}
	 root & \text{ if } i = 0, \\
	 u : u \in N(L(i-1))  & \text{ if } i = 1, \\
	 u : u \in N(L(i-1)) \cap \overline{N(L(i-2))}  & \text{otherwise}.
\end{cases}   
\end{equation}
From \Cref{eq:level} one finds that the $i$th \level consists of all
vertices that have a minimum distance $i$ from the root node.
\Cref{alg:BFS} shows how to determine this distance and thus set up the
\levels $L(i)$\@. We refer to the total number of \levels obtained for a particular graph
as \acrshort{totalLvl}. \Cref{fig:2d_7pt_level_construction} shows the
\acrshort{totalLvl}=14 \levels of our artificial stencil
operator, where the index of each vertex ($v$) is the
vertex number and the superscript represents the \level number, \ie
\begin{equation}\label{eq:node_notation}
	v^i \implies v \in L(i)\eos
\end{equation}
Note that the $L(i)$ are  substantially different from the \levels used in
the ``level-scheduling'' \cite{saad} approach, which applies ``depth first
search.''

\setlength{\fboxsep}{0pt}%

After the \levels have been determined, the matrix is permuted in the
order of its \levels, such that the vertices in $L(i)$ are stored
consecutively and appear before those of
$L(i+1)$. \Cref{fig:2d-7pt_step_1_2} shows the graph ($G' = P(G)$)
of the \stex after applying this permutation ($P$) and demonstrates
the enhanced spatial locality of the vertices within and between
\levels (see \Cref{fig:2d_7pt_perm}) as compared to the original
(lexicographic) numbering (see \Cref{fig:2d_7pt_level_construction}).
Until now the procedure is the same as \acrshort{BFS} (or
\acrshort{RCM}).

As \acrshort{RACE} uses information about the \levels for resolving
dependencies in the coloring step, we store the index of the entry point to each
\level in the permuted data structure (of $G'$) in an array
$\levelPtr[0:$ \acrshort{totalLvl}$]$, so that \levels on $G'$ can be
identified as
\begin{equation*}
  L(i) = \set{ u : u \in [\levelPtr[i]:(\levelPtr[i+1]-1)]
    \text{ and } u \in V(G')}\eos
\end{equation*}
The entries of \levelPtr for the \stex are shown in \Cref{fig:2d_7pt_levelPtr}. 
 
\subsection{Distance-k coloring} \label{subsec:DK}

The data structure generated above serves as the basis for our \DK coloring
procedure as it contains information about the neighborhood relation
between the vertices of any two \levels. Following the definition
in~\cite{dist_k_def}, two vertices are called \DK neighbors if the
shortest path connecting them consists of at most $k$ edges.
This implies that $u$ is a \DK neighbor of $v$ (referred to as
$u\xrightarrow{k}v$) if
\begin{equation}\label{eq:dk}
  u\xrightarrow{k}v  \iff  v \in \set{ u \cup N(u) \cup N^2(u) \cup \cdots N^k(u) }\eos
\end{equation}
For the undirected graphs as used here, $u\xrightarrow{k}v$
also implies $v\xrightarrow{k}u$. Based on this definition we consider
two vertices to be \DK independent if they are not \DK
neighbors, and two \levels are said to be \DK independent 
if their vertices are mutually \DK independent.
Thus, \levels $L(i)$ and $L(i\pm(k+j))$ of the permuted
graph $G'$ are \DK independent for all $j\geq1$, denoted as
\begin{equation}\label{corollary_dk}
	L(i) \not{\xrightarrow{k}} L(i\pm(k+j)) \forall j\geq1 \eos
\end{equation} 
\Cref{corollary_dk} implies that if there is a gap of at least
one \level between any two \levels (e.g., $L(i) \mbox{ and } L(i+2)$) all
pairs of vertices between these two levels are \DONE independent. Similarly
if the gap consists of at least two \levels between any two
\levels (e.g., $L(i) \mbox{ and } L(i+3)$) we have \DTWO independent
\levels, and so on.
 \begin{figure}[t]
 	\centering
 	\subfloat[Distance-1 independent \levelGroups]{\label{fig:2d_7pt_d1}\includegraphics[height=0.2\textheight,width=0.4\textwidth]{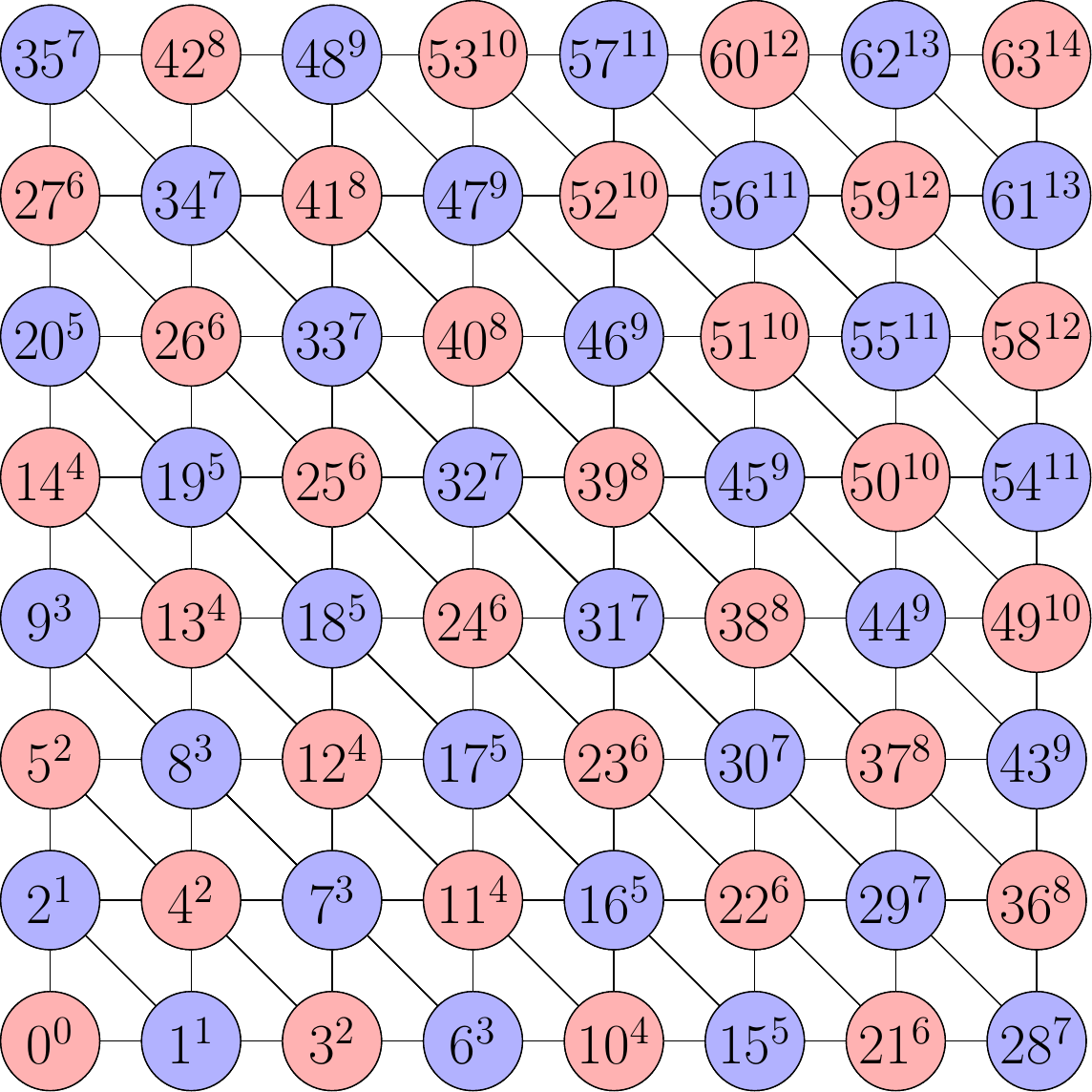}}
 	\hspace{2.5em}
 	\subfloat[Distance-2 independent \levelGroups]{\label{fig:2d_7pt_d2}\includegraphics[height=0.23\textheight,width=0.48\textwidth]{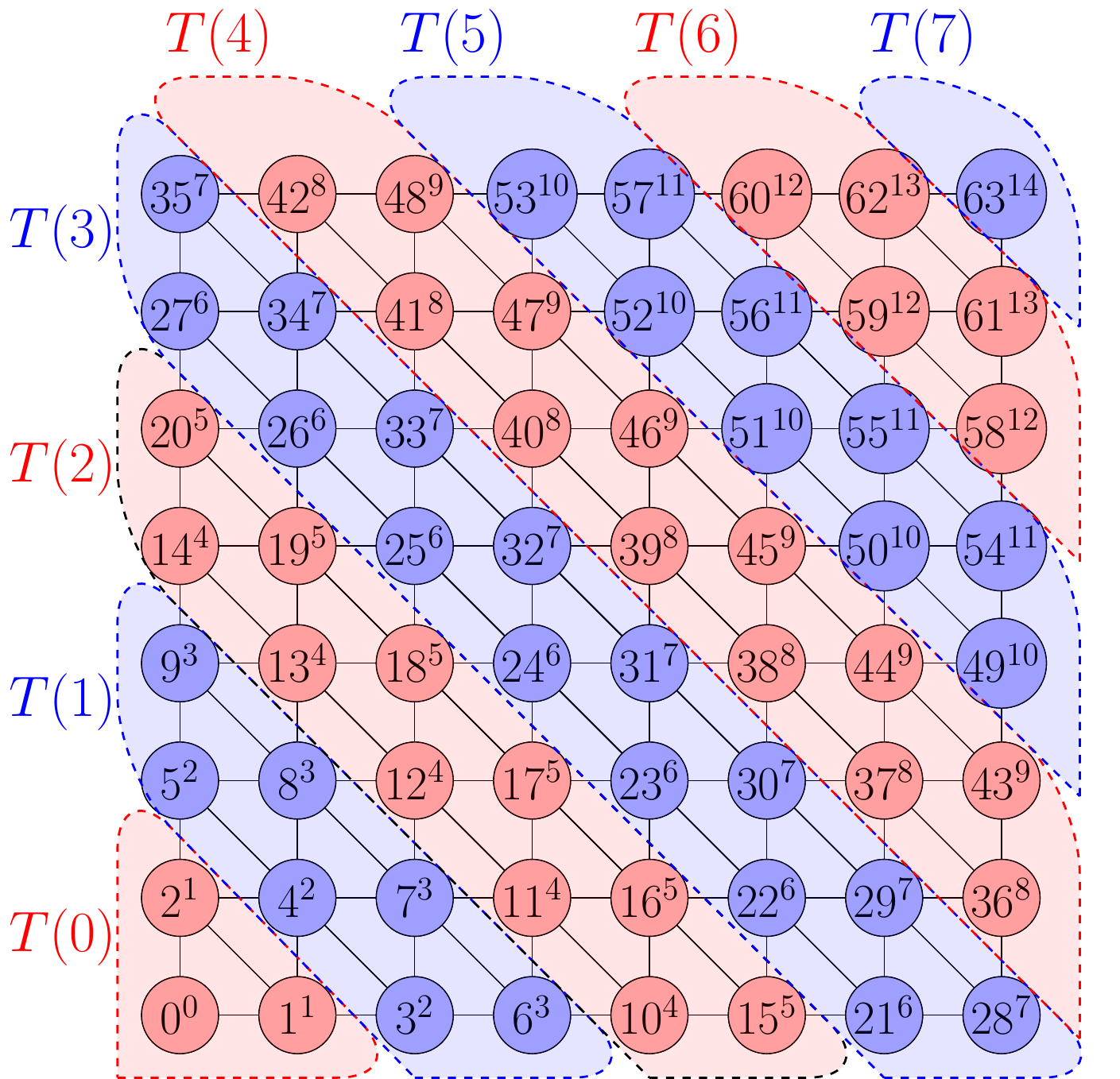}}
 	\caption{Forming \DONE and \DTWO independent \levelGroups for the \stex.}
 	\label{fig:2d-7pt_d1_d2}
 \end{figure}

The definition used in \Cref{corollary_dk} offers many choices for
forming \DK independent sets of vertices, which can then be executed
in parallel.  In \Cref{fig:2d-7pt_d1_d2} we present one example each
for \DONE (\Cref{fig:2d_7pt_d1}) and \DTWO (\Cref{fig:2d_7pt_d2})
colorings of our \stex. The \DONE coloring uses a straightforward
approach by assigning two colors to alternating \levels, i.e., \levels
of a color can be calculated concurrently. In case of \DTWO
independence we do not use three colors but rather aggregate two
adjacent \levels to form a \textit{\levelGroup} (denoted by
\acrshort{T_i}) and perform a \DONE coloring on top of those
groups. This guarantees that vertices of two \levelGroups of the same
color are \DTWO independent and can be executed in parallel. Here, the
vertices in $T(0)$, $T(2)$, $T(4)$, and $T(6)$ can be operated on by
four threads in parallel, i.e., one thread per \levelGroup.  After
synchronization the remaining four blue \levelGroups can also be
executed in parallel. This idea can be generalized such that for \DK
coloring, each \levelGroup contains $k$ adjacent \levels.
Thus formed \levelGroups are then \DONE
colored. Then, all \levelGroups within a color can be executed in
parallel. This simple approach allows one to generate workload for
a maximum of ${\acrshort{totalLvl}}/{2 k}$ threads if \DK coloring is
requested.\footnote{This implies that as the number of levels increases,
so does the parallelism. E.g., if the matrix contains at least one dense row,
 there is parallelism as $\acrshort{totalLvl}=2$ in this case.} 
Note that in all cases, vertices within a single \levelGroup
 are computed in their original order, which allows for good
spatial access locality.
 
Choosing the same number of \levels for each \levelGroup may,
however, cause severe load imbalance depending on the matrix structure. In
particular, the use of bandwidth reduction schemes such as BFS or RCM
will further worsen this problem due to the lenslike shape of the
reordered matrix (see inset of \Cref{fig:2d_7pt_perm}), leading to low
workload for \levelGroups containing the top and bottom rows of the
  matrix. Compare, \eg $T(0)$ and $T(7)$ with $T(3)$ and $T(4)$ in
\Cref{fig:2d_7pt_d2}. However, \Cref{corollary_dk} does not require
\emph{exactly} $k$ levels to be in a \levelGroup but  
only \emph{\atleast} $k$. In
the following we exploit this to alleviate the imbalance
problem.

\subsection{Load balancing}\label{subsec:LB} 
The RACE load balancing scheme tries to balance the workload across level
groups within each color for a given number of threads while maintaining data
locality and the \DK constraint between the two colors. To achieve this we use
an idea similar to incremental graph partitioning \cite{load_balancing}.  The
\levelGroups containing low workload ``grab'' adjacent levels from neighboring
level groups; overloaded \levelGroups shift levels to adjacent \levelGroups.
One can either balance the number of rows (\ie
vertices) $\acrshort{nrows}(T(i))$ or the number of
nonzeros (\ie edges)  $\acrshort{nnz}(T(i))$. Both variants
are supported by our implementation, and we choose balancing by number of
rows in the following to demonstrate the method (see
\Cref{alg:LB}).
 
\begin{figure}[t]
   	\centering
   	\includegraphics[width=\textwidth]{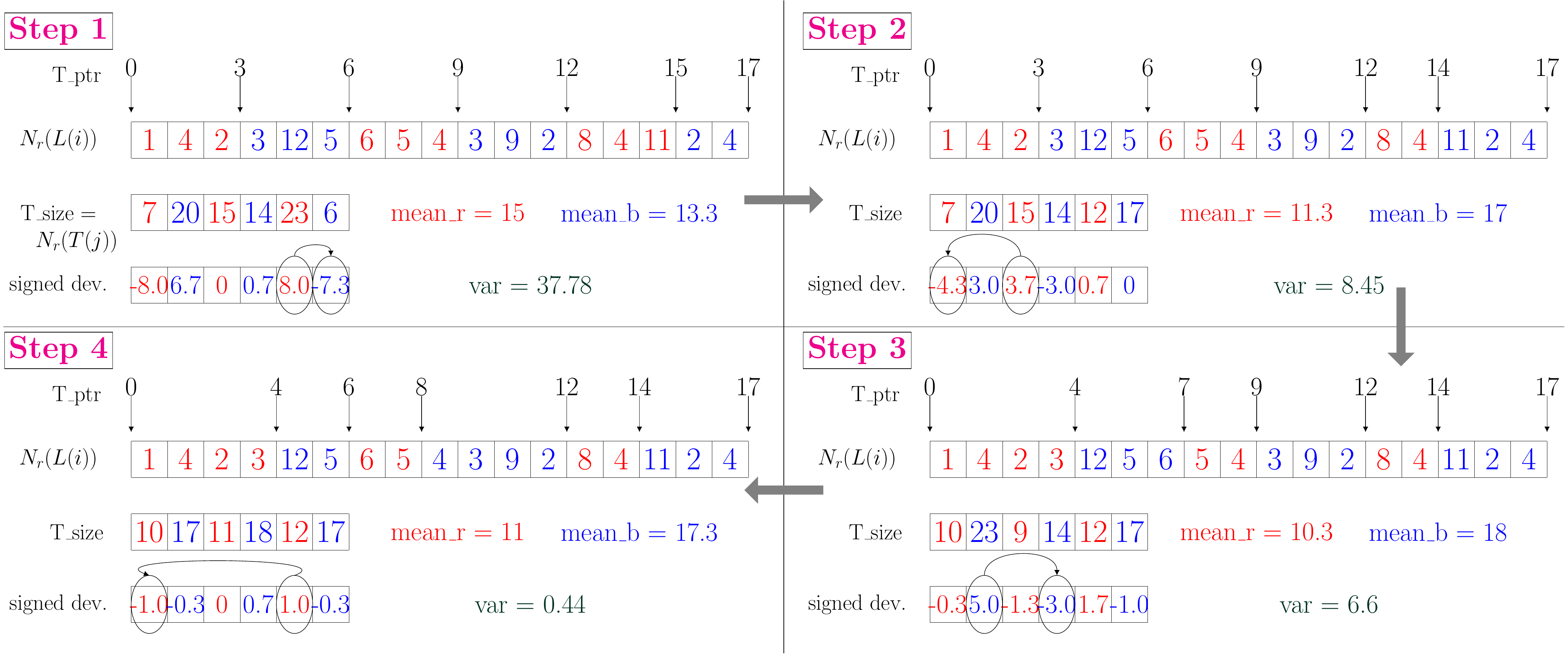}
   	\caption{All steps of the load balancing scheme, applied to an
          arbitrarily chosen initial distribution of 17 levels into six level
          groups for \DTWO coloring. Rebalancing steps are performed clockwise
          starting from top left. $mean\_r$ and $mean\_b$ denote the current
          average number of rows per \levelGroup and color. $var$ is the overall
          variance.}
   	\label{fig:lb_alg}
   \end{figure}

\begin{figure}[t]
	   	\centering
	   	\subfloat[Five threads]{\label{fig:2d_7pt_lb}\includegraphics[width=0.48\textwidth, height=0.28\textheight]{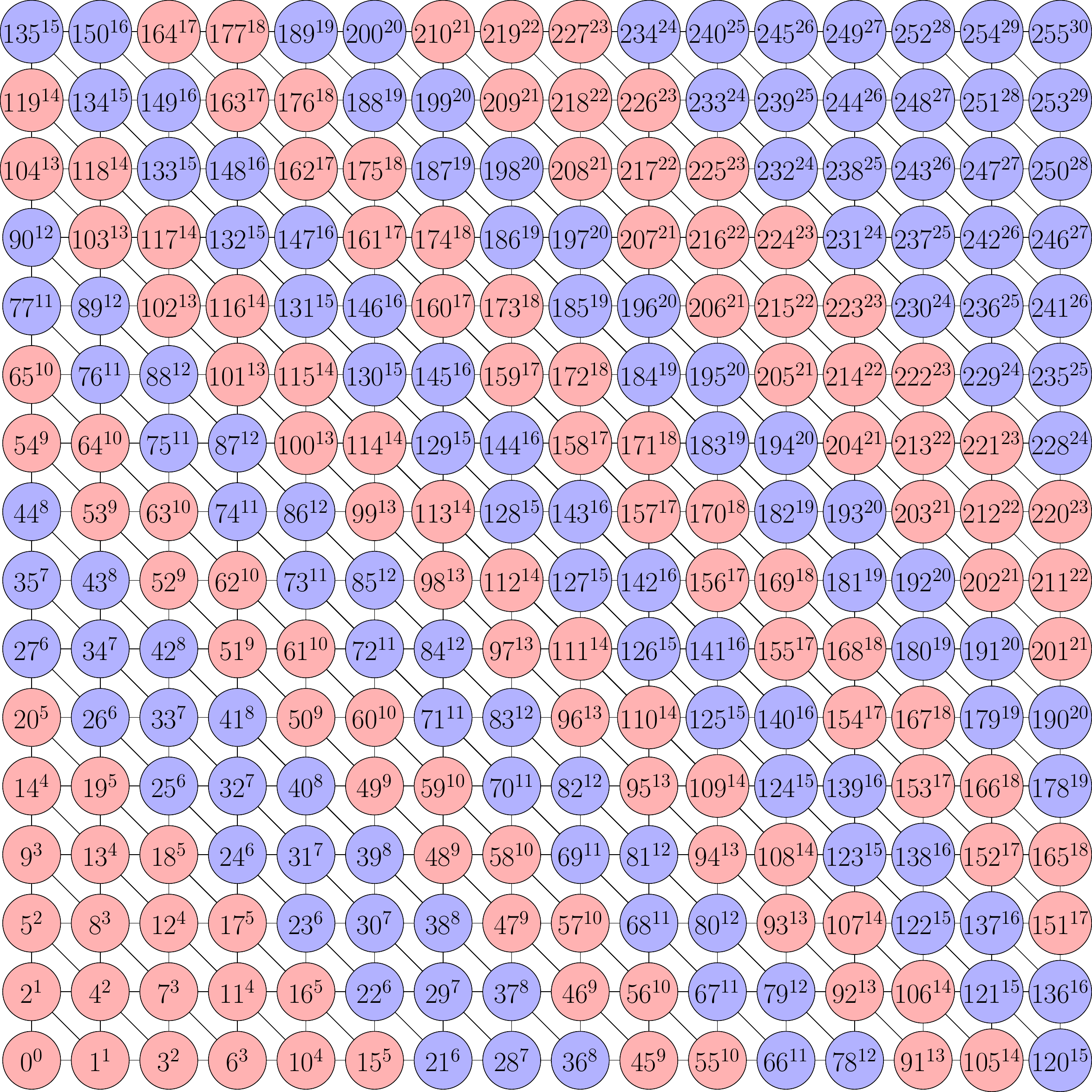}}
	   	\hspace{0.2em}
	   	\subfloat[Eight threads]{\label{fig:2d_7pt_lb_8_threads}\includegraphics[width=0.48\textwidth, height=0.28\textheight]{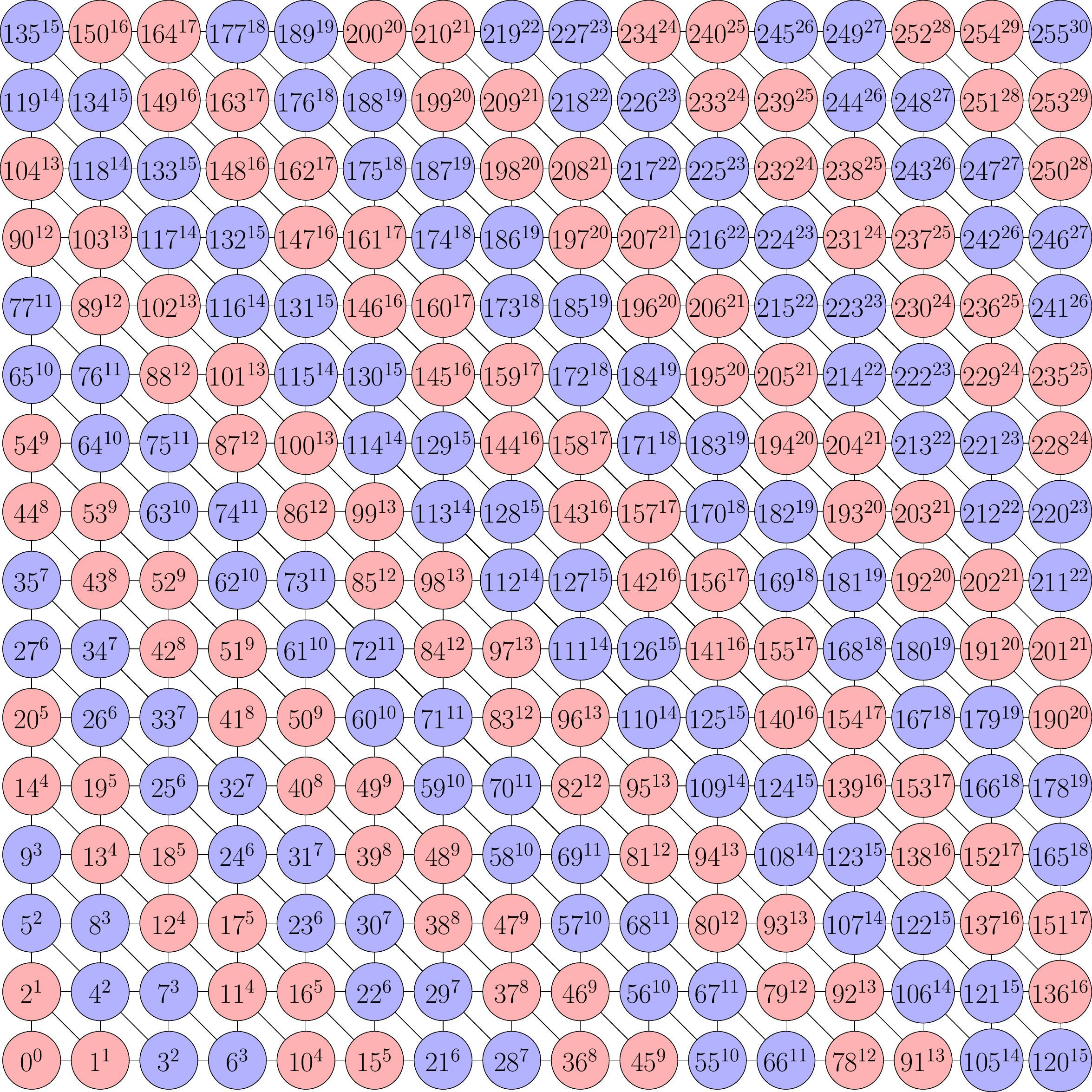}}
		 \caption{Domain size $16 \times 16$ for the \stex and \DTWO dependency, \sref{fig:2d_7pt_lb} after load balancing for five threads, 
		 \sref{fig:2d_7pt_lb_8_threads} after load balancing for eight threads.
		}
	\end{figure}

For a given set of \levelGroups we calculate the mean and variance of
$\acrshort{nrows}(T(i))$ within each color (red and blue).
The overall variance,
which is the target of our minimization procedure, is then found by summing up
the variances across colors. \Inorder to reduce this value we first select the
two \levelGroups with largest negative/positive deviation from the mean (which
is $T(5)$ and $T(4)$ in step 1 of \Cref{fig:lb_alg}) and try to add/remove
levels to/from them (see top row of \Cref{fig:lb_alg}). When
removing \levels from a \levelGroup, the \DK coloring is strictly maintained 
by keeping at least $k$ levels in it. The shift of
\levels is done with the help of an array  $T\_ptr[]$, which holds pointers to the
beginning of each \levelGroup (see \Cref{fig:lb_alg}), avoiding any copy
operation. If shifting levels between the two level groups with the largest
deviation does not lead to a lower overall variance, no levels are exchanged and
we choose the next pair of level groups according to a ranking which is based on
the absolute deviation from the mean (see \Cref{alg:LB} for implementation
details) and continue. Following this process in an iterative way we finally end
up in a state of lowest overall variance at which no further moves are
possible, either because they would violate the \DK dependency or lead to
an increase in overall
variance. \Cref{fig:lb_alg} shows the load balancing procedure under a \DTWO
constraint for some initial mapping of 17 levels to six \levelGroups. Applying
the procedure to our \stex of size $16 \times 16$, requesting \DTWO coloring and
ten level groups leads to the mapping shown in \Cref{fig:2d_7pt_lb}. Note that
\levelGroups at the extreme ends have more \levels due to fewer vertices
(\acrshort{nrows}) in each \level, while \levelGroups in the middle, having more
vertices, maintain two levels to preserve the \DTWO constraint.

  
%

	\subsection{Recursion}\label{subsec:REC}
As discussed in \Cref{subsec:DK}, the maximum degree of parallelism is limited
by the total number of levels (\acrshort{totalLvl}) and may be further reduced 
by level aggregation. In case of our
$16\times16$ stencil example the maximum possible parallelism 
is eight threads, which may cause load imbalance as seen in 
\Cref{fig:2d_7pt_lb_8_threads}.
Hence, 
further parallelism must be found within the \levelGroups.
Compared to methods like \acrshort{MC} 
we do not require all vertices in a \levelGroup to be \DONE (or \DK in general) 
independent. This is a consequence of our \level-based approach, which
requires \DK independence between vertices of different levels but not 
within a level (see \Cref{corollary_dk}). There may be more parallelism hidden within 
the \levelGroups,  which can be interpreted as \subgraphs.  Thus we apply the three 
steps of our method recursively on selected \subgraphs to exploit the parallelism 
within them.  

In the following section we first demonstrate the basic
idea in the context of \DONE dependencies, which can be resolved
within the given \levelGroup by design. However, for $k>1$, vertices in a
\levelGroup may have \DK dependencies via vertices in adjacent
\levelGroups. We generalize our procedure to \DK dependencies as a second step
in \Cref{subsec:Dk_dependency}. Finally, in \Cref{subsec:subgraph_selection} we
apply the recursive scheme to our \stex and introduce proper \subgraph selection
as well as global load balancing strategies.

In order to visualize the basic concepts easily and discuss important corner
cases of the recursive approach we start with the simple graph shown in
\Cref{fig:rec_d1_s1_a}, which is not related to our \stex. To distinguish
between \levelGroups at different stages \acrshort{s} of the recursive procedure
we add a subscript to the levels (\acrshort{L_si}) and \levelGroups
(\acrshort{T_si}) indicating the stage of recursion at which they are generated,
with $s=0$ being the original distribution before recursion is applied to any
\subgraph.

	
	\subsubsection{Distance-$1$ dependency} \label{subsec:D1_dependency}

\begin{figure}[t]
     	\centering
     	\subfloat[Example graph]{\label{fig:rec_d1_s1_a}\includegraphics[width=0.26\textwidth, height=0.14\textheight]{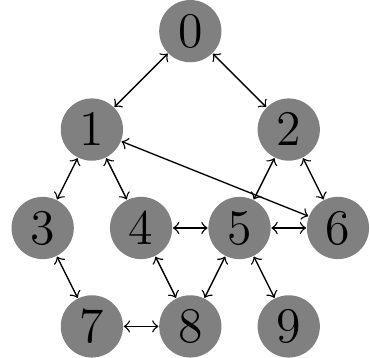}}
     	\hspace{1.5em}
     	\subfloat[Stage 0, levels in graph]{\label{fig:rec_d1_s1_b}\includegraphics[width=0.32\textwidth, height=0.14\textheight]{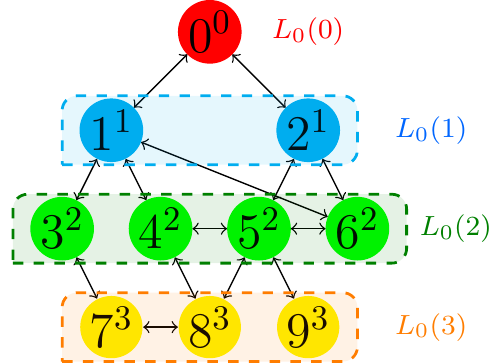}}
     	\hspace{1.5em}
     	\subfloat[\DONE coloring]{\label{fig:rec_d1_s1_c}\includegraphics[width=0.32\textwidth, height=0.14\textheight]{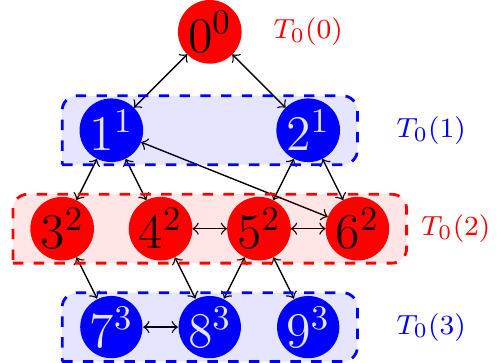}}
        \caption{Exposing potential for more parallelism in a graph with \DONE coloring.
          $T_0(1),T_0(2),$ and
          $T_0(3)$ have internal unexposed parallelism.
          Note that the graph shown here is not
          related to the previous \stex.}
     	\label{fig:rec_d1_s1}
     \end{figure}

For the \DONE coloring of the graph in \Cref{fig:rec_d1_s1} we find that
three of the four \levelGroups of the initial stage still contain \DONE-independent vertices; e.g., in $T_0(2)$ we have vertices $3 \not{\xrightarrow{1}} 4$ ($3$ \DONE
independent to $4$), $3 \not{\xrightarrow{1}} 5$, $3 \not{\xrightarrow{1}} 6$,
and $4 \not{\xrightarrow{1}} 6$, implying each of these pairs can be computed in
parallel without any \DONE conflicts. This parallelism is not exposed at
the first stage ($s=0$) as vertices in $L_0(i)$ are chosen such
that they are \DONE neighbors of $L_0(i-1)$, ignoring any vertex relations
\emph{within} $L_0(i)$.

Recursion starts with the selection of a \subgraph of the matrix, which is
discussed in more detail later (see \Cref{subsec:subgraph_selection}). Here we
choose the \subgraph induced by $T_0(2)$. It can be isolated from the
rest of the graph since the \DONE coloring step in stage 0 has already produced
independent \levelGroups. Now we just need to repeat the three
steps explained previously (\Cref{subsec:LEVEL_CONST}--\Cref{subsec:LB}) on this
\subgraph.
     \begin{figure}[t]
     	\centering
     	\subfloat[]{\label{fig:rec_d1_s2_a}\includegraphics[width=0.28\textwidth, height=0.14\textheight]{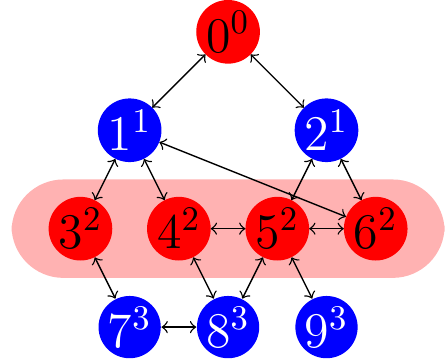}}
     	\hspace{2.25em}
     	\subfloat[]{\label{fig:rec_d1_s2_b}\includegraphics[width=0.07\textwidth, height=0.14\textheight]{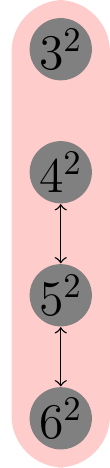}}
     	\hspace{1.75em}
     	\subfloat[]{\label{fig:rec_d1_s2_c}\includegraphics[width=0.065\textwidth, height=0.14\textheight]{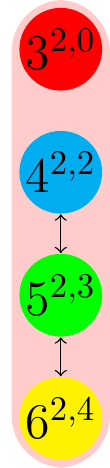}}
     	\hspace{1.75em}
     	\subfloat[]{\label{fig:rec_d1_s2_d}\includegraphics[width=0.065\textwidth, height=0.14\textheight]{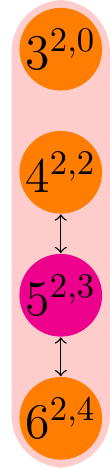}}
	     \hspace{1.75em}
	     \subfloat[]{\label{fig:rec_d1_s2_e}\includegraphics[width=0.065\textwidth, height=0.14\textheight]{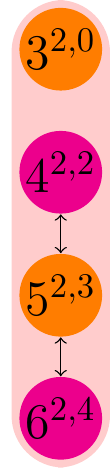}}
     	\caption{Applying recursion to the \subgraph induced by
          $T_0(2)$. \Cref{fig:rec_d1_s2_b} shows the isolated \subgraph,
          while \Cref{fig:rec_d1_s2_c} presents the level construction step on the
          \subgraph. Two potential \DONE colorings of this \subgraph are shown
          in \Cref{fig:rec_d1_s2_d,fig:rec_d1_s2_e}.}
     	
     	\label{fig:rec_d1_s2}
     \end{figure}
     
     \Cref{fig:rec_d1_s2} shows an illustration of applying the first recursive step ($s=1$) on $T_0(2)$, where we extend the definition of the vertex numbering in \Cref{eq:node_notation} to the following:
	 \begin{equation}
	    v^{i,j,k...} \implies v \in \set{L_0(i) \cap L_1(j) \cap L_2(k) \cap \cdots}.
	 \end{equation}
At the end of the recursion (\cf
\Cref{fig:rec_d1_s2_d,fig:rec_d1_s2_e}) on $T_0(2)$, we obtain parallelism for
two more threads in this case.
Note that the \subgraphs might have ``islands'' (groups of vertices
that are not connected to the rest of the graph); \eg vertex 3 and vertices
4,5,6 form two islands in \Cref{fig:rec_d1_s2_b}. Since an island is 
disconnected from the rest of the (sub)graph it can be executed independently
and in parallel to it. To take advantage of this, the starting node in the
next island is assigned a level number with an increment of two, as seen in
\Cref{fig:rec_d1_s2_c}. This allows for two different colorings of the island,
increasing the
number of valid \DONE configurations (\cf
\Cref{fig:rec_d1_s2_d,fig:rec_d1_s2_e}). The selection of the optimal
one will be done in the final load balancing step as described in
\Cref{subsec:LB}.
     
As this recursive process finds independent \levelGroups
($T_{s+1}$) within a \levelGroup of the previous stage ($T_s$), the
thread assigned to $T_s$ has to spawn threads to parallelize within
$T_{s+1}$.

\subsubsection{Distance-$k$ dependencies with $k>1$}  \label{subsec:Dk_dependency}

\begin{figure}[t]
     	\centering
     	\subfloat[]{\label{fig:rec_d2_wrong_a}\includegraphics[width=0.23\textwidth, height=0.13\textheight]{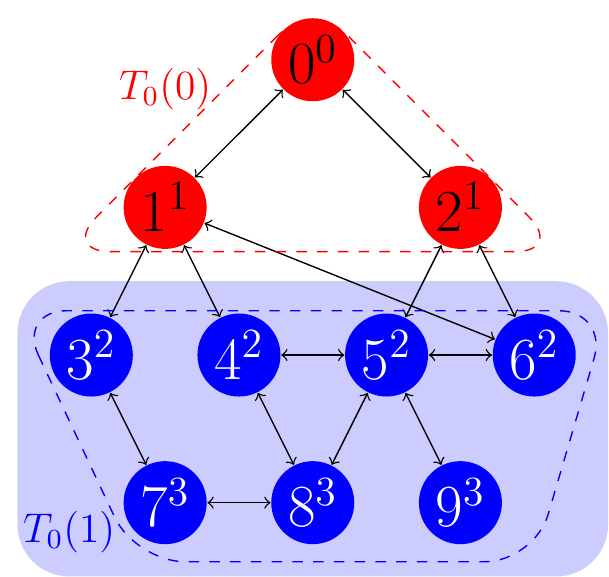}}
     	\hspace{0.6em}
     	\subfloat[]{\label{fig:rec_d2_wrong_b}\includegraphics[width=0.23\textwidth, height=0.07\textheight]{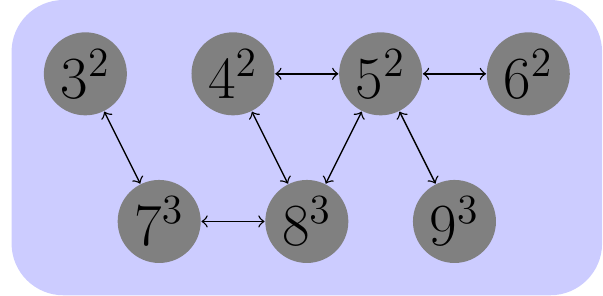}}
     	\hspace{0.6em}
     	\subfloat[]{\label{fig:rec_d2_wrong_c}\includegraphics[width=0.23\textwidth, height=0.07\textheight]{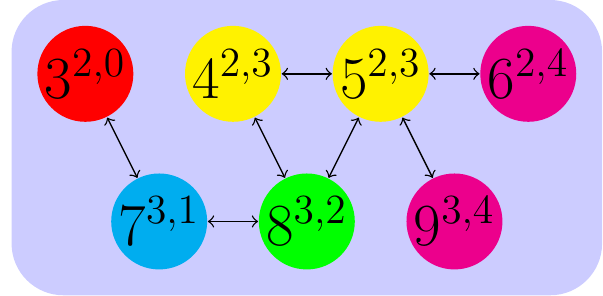}}
     	\hspace{0.6em}
     	\subfloat[]{\label{fig:rec_d2_wrong_d}\includegraphics[width=0.23\textwidth, height=0.13\textheight]{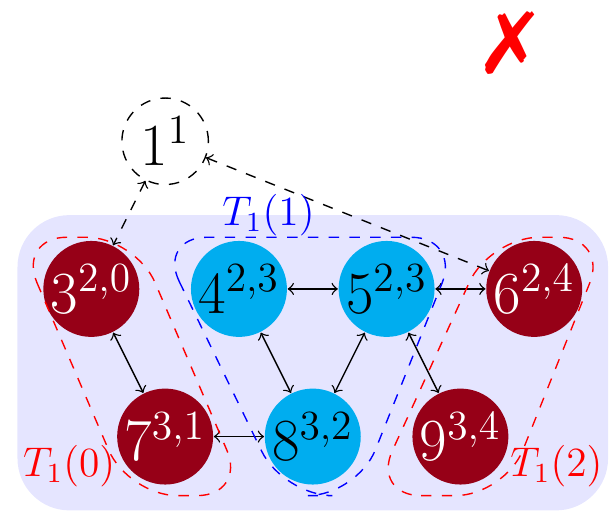}}
     	\caption{Two \levelGroups generated by a \DTWO coloring
          (\Cref{fig:rec_d2_wrong_a}). \Cref{fig:rec_d2_wrong_b} shows the
          \subgraph induced by \levelGroup $T_0(1)$. Level construction on the
          selected \subgraph is shown in \Cref{fig:rec_d2_wrong_c}. Forming
          \DTWO independent \levelGroups on these levels does not guarantee a
          \DTWO independence between the newly generated \levelGroups of the
          same sweep (color) as seen in \Cref{fig:rec_d2_wrong_d}.}
     	\label{fig:rec_d2_wrong}
     \end{figure}

In general, it is insufficient to consider only
the \subgraphs induced by \levelGroups in the recursion step, as can be seen in
\Cref{fig:rec_d2_wrong_a} for \DTWO coloring. Applying the three steps (see
\Cref{fig:rec_d2_wrong_b,fig:rec_d2_wrong_c,fig:rec_d2_wrong_d}) to the
\subgraph induced by $T_0(1)$ does not guarantee \DTWO independence between the
new \levelGroups $T_1(0)$ and $T_1(2)$. It is obvious that for general \DK
colorings two vertices $a,b$ within a \levelGroup might be connected by a shared
vertex $c$ outside the \levelGroup.  Thus, our three step procedure must be
applied to a \subgraph which contains the actual \levelGroup ($T_s(j)$) as well
as its all distance-$p$ neighbors, where $p=1,2,\ldots,(k-1)$.

\begin{figure}[t]
     	\centering
     	\subfloat[]{\label{fig:rec_d2_correct_a}\includegraphics[width=0.22\textwidth, height=0.13\textheight]{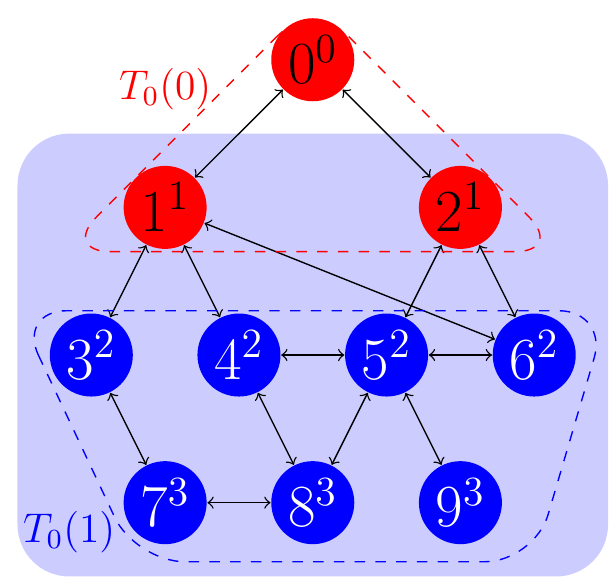}}
     	\hspace{0.6em}
     	\subfloat[]{\label{fig:rec_d2_correct_b}\includegraphics[width=0.22\textwidth, height=0.105\textheight]{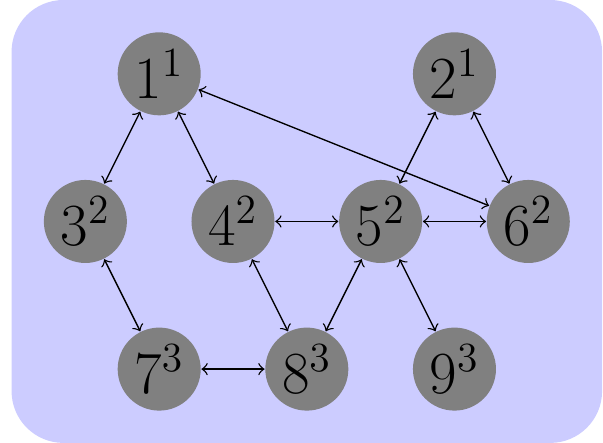}}
     	\hspace{0.6em}
     	\subfloat[]{\label{fig:rec_d2_correct_c}\includegraphics[width=0.22\textwidth, height=0.105\textheight]{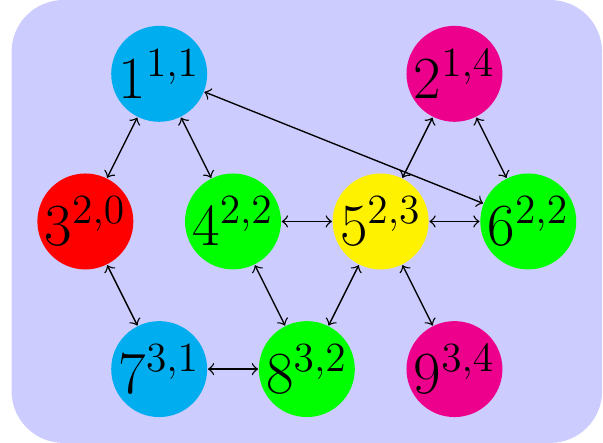}}
     	\hspace{0.6em}
     	\subfloat[]{\label{fig:rec_d2_correct_d}\includegraphics[width=0.22\textwidth, height=0.105\textheight]{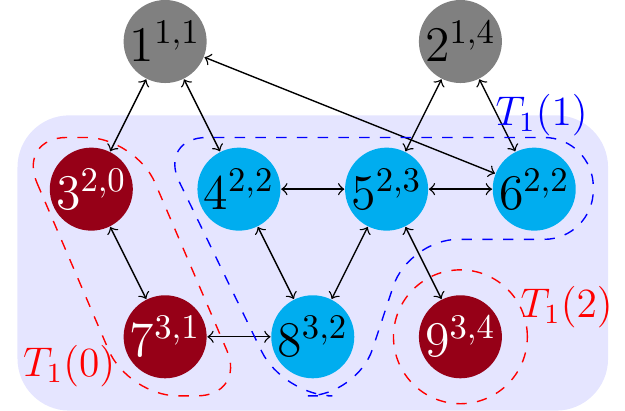}}
     	\hspace{0.6em}
     	\caption{Correct procedure for \DTWO coloring of \levelGroup
          $T_0(1)$. The \subgraph as shown in \Cref{fig:rec_d2_correct_b}
          contains \levelGroup $T_0(1)$ and its \DONE neighborhood. A level
          construction step is applied to this \subgraph in
          \Cref{fig:rec_d2_correct_c}. Distance-2 coloring by level aggregation
          leading to \levelGroups of stage 1 is shown in
          \Cref{fig:rec_d2_correct_d}; we get three \levelGroups at the end of
          the recursion on $T_0(1)$.}
     	\label{fig:rec_d2_correct}
     \end{figure}

This ensures that there is no vertex outside the \subgraph which can mediate a
\DK dependency between vertices in the embedded \levelGroup ($T_s(j)$). We can
now construct the new levels on this \subgraph considering the neighborhood, but
we only store the vertices in the new \levels $L_{s+1}(:)$ that are in the
embedded \levelGroup ($T_s(j)$). Next we apply \DK coloring by aggregation of
the new levels, leading to a set of \levelGroups $T_{s+1}(:)$ within
$T_s(j)$. \Cref{fig:rec_d2_correct} demonstrates this approach to resolve the conflict
shown in \Cref{fig:rec_d2_wrong_d}. \Cref{fig:rec_d2_correct_b} presents
the \subgraph containing the selected \levelGroup $T_0(1)$ and its \DONE
neighborhood.

Level construction is performed on the \subgraph (\Cref{fig:rec_d2_correct_c}),
but the new levels only contain vertices of $T_0(1)$, \ie $L_1(1) =
\{7^{3,1}\}$. Finally, \DTWO coloring by aggregation of two adjacent levels is
performed, leading to three \levelGroups of the second stage $s=1$
(\Cref{fig:rec_d2_correct_d}), \ie $T_1(0)=\{L_1(0) \cup L_1(1)\}$.  Now
vertices $3$ and $6$ are mapped to \levelGroups of different colors. Note that
the permutation step on the newly generated levels is not shown but is performed
as well to maintain data locality.
    
       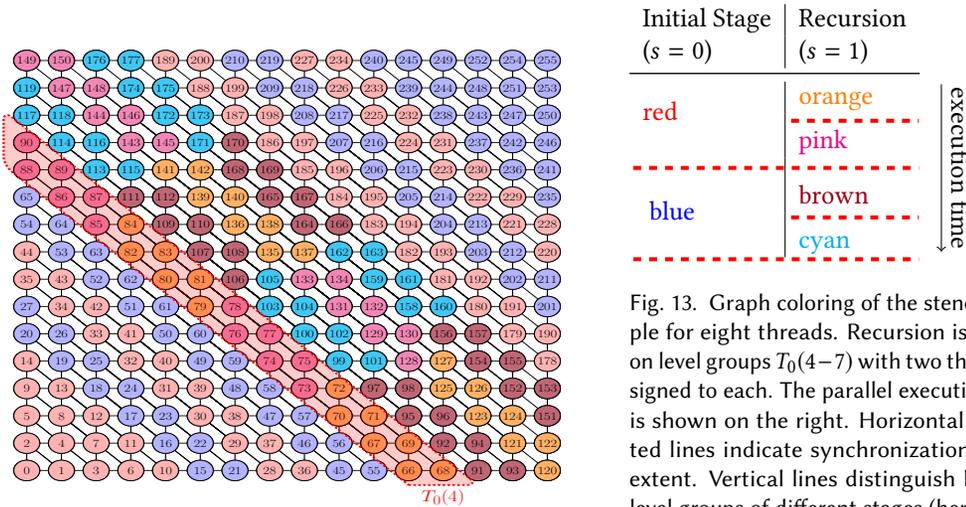
\begin{figure}[t]
       	\begin{minipage}[c]{0.6\textwidth}
       		\includegraphics[height=0.3\textheight,width=0.89\textwidth]{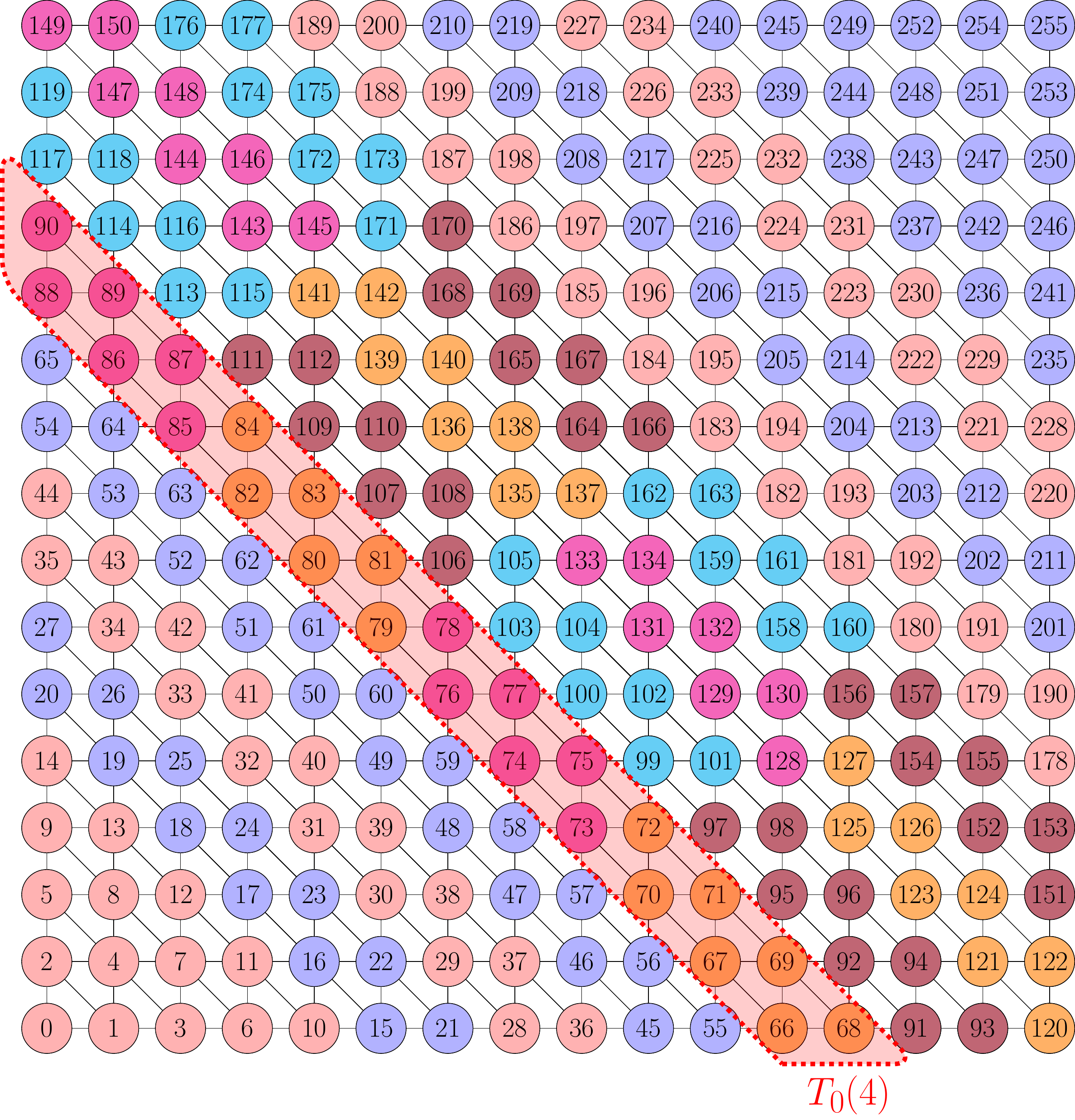}
       	\end{minipage}\hfill
       	\begin{minipage}[c]{0.4\textwidth}
	       	\begin{tabular}{l|l}
	       		{Initial Stage} & {Recursion}\\
	       		{($s=0$)} & {($s=1$)}\\
	       		\midrule
	       	   \multirow{2}{*}{\textcolor{red}{red}} & {\textcolor{amber}{orange}}\\
	       	   \rule{0pt}{3ex}
	       		& {\textcolor{magenta}{pink}}\\
				\rule{0pt}{4ex} 
	       	   \multirow{2}{*}{\textcolor{blue}{blue}} & {\textcolor{carmine}{brown}}\\
	       	   \rule{0pt}{3ex}
	       	   & {\textcolor{cyan}{cyan}}\\
	       	\end{tabular}
	       	\begin{tikzpicture}[overlay]
		       	\draw[-, dashed, red, line width=1.5] (-1.8,0.22) -- (-0.1,0.22);
		       	\draw[-, dashed, red, line width=1.5] (-3.9,-0.4) -- (-0.1,-0.4);
		      	\draw[-, dashed, red, line width=1.5] (-1.8,-1.05) -- (-0.1,-1.05);
		       	\draw[-, dashed, red, line width=1.5] (-3.9,-1.6) -- (-0.1,-1.6);
		       	\draw[->] (0.2,0.7) -- (0.2,-1.5);
		       	\node[rotate=-90] at (0.45,-0.4) {execution time};
	       	\end{tikzpicture}
       		\caption{Graph coloring of the \stex for eight
                  threads. Recursion is applied on \levelGroups $T_0(4-7)$ with
                  two threads assigned to each. The parallel execution order is
                  shown on the right.  Horizontal red dotted lines indicate
                  synchronization and its extent. Vertical lines distinguish
                  between \levelGroups of different stages (here $T_0$ and
                  $T_1$) which can run in parallel.
		}
       		\label{fig:rec_2d-7pt_graph}
       	\end{minipage}
       \end{figure}
     
          
\subsubsection{Level group construction and global load balancing} \label{subsec:subgraph_selection}

The recursive refinement of \levelGroups allows us to tackle load imbalance
problems and limited degree of parallelism as we are no longer restricted by the
one thread per \levelGroup constraint.  Instead, we have the opportunity to form
\levelGroups and assign appropriate thread counts to them such that the load per
thread approaches the optimal value, \ie the total workload divided by the number of
threads available. Pairs of adjacent level groups having different colors within a stage,
\ie $T_s(i)$ and $T_s(i+1)$ with $i=0,2,4,...$,
are typically handled by the same threads, so we assign an equal number of threads
to them. 
We then apply recursion to the \levelGroups with more than one
thread assigned. Starting with the original graph as the base \levelGroup
($T_{-1}(0)$) to which all available threads
$\acrshort{nthreads}(T_{-1}(0))=\acrshort{nthreads}$ and all vertices
$\acrshort{nrows}(T_{-1}(0))=\acrshort{nrows}^\mathrm{total}$ are assigned, we perform
the following steps to form \levelGroups $T_s(:)$ at stage $s \ge 0$ to which we
assign $\acrshort{nthreads}(T_{s}(:))$ threads. To illustrate the procedure we
use the $16 \times 16$ \stex and construct a coloring scheme for eight threads
(see \Cref{fig:rec_2d-7pt_graph}).
\begin{enumerate}
\item Assign weights to all levels at stage ($s$) of the recursion. Assuming
  that $L_s(i) \subset T_{s-1}(j)$, its weight is defined by
	\begin{align*}
		w(L_s(i)) &=
                \frac{\acrshort{nrows}(L_s(i))}{\frac{\acrshort{nrows}(T_{s-1}(j))}{\acrshort{nthreads}(T_{s-1}(j))}}
                = \frac{\acrshort{nrows}(L_s(i))}{\acrshort{nrows}(T_{s-1}(j))}
                \acrshort{nthreads}(T_{s-1}(j)).\\
	\end{align*}
	
For a given \levelGroup ($T_{s-1}(j)$) that has to be split up
($\acrshort{nthreads}(T_{s-1}(j)) > 1$), the weight describes the fraction of
the optimal load per thread,
$\frac{\acrshort{nrows}(T_{s-1}(j))}{\acrshort{nthreads}(T_{s-1}(j))}$, in the
specific level ($L_s(i)$).

Requesting $\acrshort{nthreads}(T_{-1}(0))=8$ threads for the
$\acrshort{nrows}(T_{-1}(0)) = 256$ vertices of the \stex in
\Cref{fig:rec_2d-7pt_graph} produces the following weights for the initial
($s=0$) levels:
	\begin{align*}
		\{w(L_0(0)), w(L_0(1)), w(L_0(2)), ...\} &= \Big{\{} \frac{1}{256} \times 8 , \frac{2}{256} \times 8 , \frac{3}{256} \times 8 , ...\Big{\}}.
	\end{align*}
	
\item The above definition implies that if the weight is close to a natural
  number $b$, the corresponding workload is near optimal for operation with $b$
  threads. Thus, starting with $L_s(0)$ we aggregate successive levels until
  their combined weight forms a number $a$ close to a natural number
  $b$. Distance-k coloring is ensured by enforcing it to aggregate \atleast $2
  \times k$ \levels, \ie for \DTWO coloring at least four levels (two for red
  and two for blue). Closeness to the natural number is quantified by a parameter
  $\epsilon$ defined as
	\begin{align*}
		\epsilon =  1 - \abs(a-b), & \text{ where } b= \max(1,[a])\\
				& \text{and } [a] \text{ is the nearest integer to $a$},
	\end{align*}
	and controlled by the criterion
	\begin{align*}
	\epsilon &> \epsilon_s, \text{where the $\epsilon_s \in  [0.5,1)$ are user defined parameters.} 	
	\end{align*}		   
The choice of this parameter may be different for every stage of recursion.
Once we find a collection of successive levels satisfying this criterion, the
natural number $b$ is fixed. We try to further increase the number of levels to
test if there exists a number $a'>a$ which is closer to $b$ leading to an
$\epsilon$ value closer to one. We finally choose the set of levels with the
best $\epsilon$ value and define them to form $T_s(0)$ and $T_s(1)$ which are to
be executed by $\acrshort{nthreads}(T_s(0))=\acrshort{nthreads}(T_s(1))=b$
threads.  In \Cref{fig:rec_2d-7pt_graph} we choose $\epsilon_s = 0.6$, which
selects the first seven levels to form $T_0(0)$ and $T_0(1)$.  As their combined
weight is $\frac{28}{32}=0.875$, one thread will execute these two \levelGroups.

\item We continue with subsequent pairs of \levelGroups ($T_s(i), T_s(i+1);
  i=2,4 ...$) by applying this procedure starting with the very next
  \level. Finally, once all the levels have been touched, a total of
  $\acrshort{nthreads}(T_{s-1}(j))$ threads have been assigned to the levels
  $L_s(i) \subset T_{s-1}(j)$. For example, for $T_0(4)$ and $T_0(5)$ in
  \Cref{fig:rec_2d-7pt_graph} two threads satisfy the criterion as the total
  weight of the four levels included is $\frac{54}{32}=1.69$.
	
\item The distribution between adjacent red and blue \levelGroups which are
  assigned to the same thread(s) as well as the final global load balancing is
  performed using a slight modification of the scheme presented in \Cref{subsec:LB}
  (shown at the beginning of \Cref{alg:LB} in \cref{Sec:algo}):
Now the calculation of mean and variance must consider the number of threads
($\acrshort{nthreads}(T_{s}(j))$) assigned to each \levelGroup. 
The worker array now has to be replaced by the number of
threads assigned to each \levelGroup ($\acrshort{nthreads}(T_{s}(j))$). The
algorithm then tries to minimize the variance of the number of vertices per
thread in \levelGroups. Ideally, after
this step the load per thread in each \levelGroup should approach the optimal
value given above.
\end{enumerate}
Once the \levelGroup of stage $s$ has been formed, the recursion and the above
procedure are separately applied to all new \levelGroups with more than one
thread assigned. This continues until every \levelGroup is assigned to one
thread. The depth of the recursion is determined by the parameter $\epsilon_s$
and depends on the matrix structure as well as degree of parallelism requested.

For our \stex in \Cref{fig:rec_2d-7pt_graph} the inner four \levelGroups of
stage $s=0$ required one stage of recursion. This led to 16 \levelGroups at
stage $s=1$, as we require four new \levelGroups per recursion to schedule two
threads.  In terms of parallel computation, first the red vertices will be
computed in parallel with the orange ones using four threads for both
colors. Once the orange vertices are done, each pair of threads assigned to
$T_0(4)$ and $T_0(6)$ synchronize locally (\ie within $T_0(4)$ and $T_0(6)$
separately). Then the pink vertices are computed followed by a global
synchronization of all threads. The scheme continues with the blue vertices and
the brown/cyan ones, which represent the two blue \levelGroups to which
recursion has been applied (see table in \Cref{fig:rec_2d-7pt_graph}).

\begin{figure}[t]
		 \includegraphics[width=\textwidth, height=0.2\textheight]{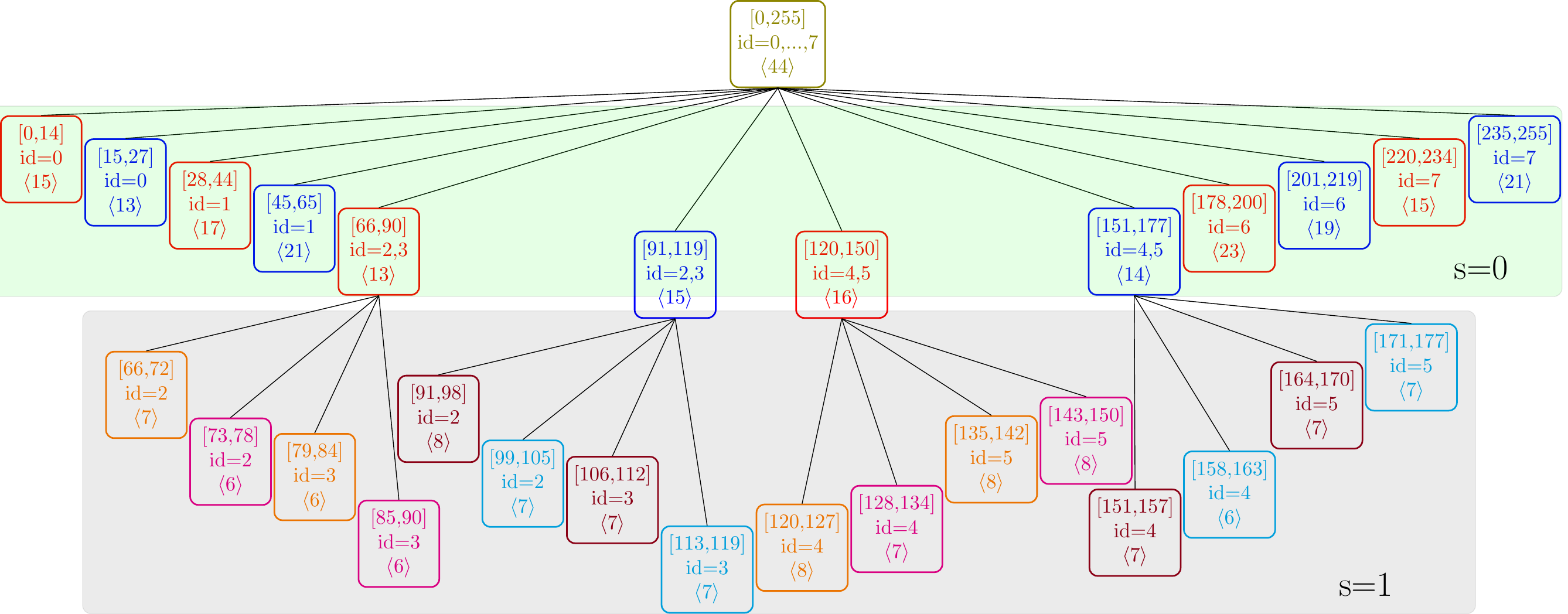}
	 	\caption{The internal tree structure of \acrshort{RACE}
                  representing the \stex for domain size $16 \times 16$ and
                  eight threads. The range $[\ldots]$ specified in each leaf
                  represents the vertices belonging to each \levelGroup and the
                  id refers to the thread id assigned to each \levelGroup
                  assuming compact pinning . The last entry
                  $\langle\acrshort{nrowsEff}\rangle$ gives the effective row
                  count introduced in \Cref{Sec:param_study}. }
	 	\label{fig:rec_2d-7pt_tree}
	 \end{figure}

The recursive nature of our scheme can be best described by a tree data
structure, where every node represents one \levelGroup and the maximum depth is
equivalent to the maximum level (stage) of recursion. The data structure for the
colored graph in \Cref{fig:rec_2d-7pt_graph} and its thread assignments are
shown in \Cref{fig:rec_2d-7pt_tree}. The root node represents our baseline
\levelGroup $T_{-1}(0)$ comprising all 256 vertices and all eight threads
(having unique $id=0,\ldots,7$). The first level of child nodes gives the
initial ($s=0$) distribution, with each node storing the information of a
\levelGroup including its color. Threads are mapped consecutively to the
\levelGroups. The red $T_0(4)$ \levelGroup, which consists of
vertices $66,\ldots,90$ (omitting the superscript for level numbers), is executed
by threads with $id=2,3$.  Applying recursion to $T_0(4)$, this node spawns four
new child nodes at stage $s =1$, \ie \levelGroups $T_1(0,\ldots,3) \subset
T_0(4)$, to be executed by the two threads. Synchronization only happens between
threads having the same parent node after executing the same color. Note that
actual computations are only performed on the leaf nodes of the final tree.

%% file: parameter_study.tex


The \acrshort{RACE} method has a set of input
parameters $\{\epsilon_s; s=0,1,\ldots\}$ which control the
assignment of threads to adjacent level groups.
To determine useful settings, we analyze the interaction
between these input parameters, the number of threads used, and the
parallel efficiency of the generated workload distribution.

As the internal tree structure contains all information about the
final workload distribution, we can use it to identify the critical
path in terms of workload and thus the parallel
efficiency. To this end we introduce the \effRow for every node
(or \levelGroup) $\acrshort{nrowsEff}(T_s(i))$, which is a measure for
the absolute runtime to calculate the
corresponding \levelGroup. For \levelGroups that are not further
refined (leaf nodes) this value is their actual workload, \ie the
number of rows assigned to them ($\acrshort{nrowsEff}(T_0(0)) = 15$
in \Cref{fig:rec_2d-7pt_tree}). For an inner node, the \effRow is the
sum of the maximum workload (\ie the maximum \effRow value) across each of
the two colors of its child nodes:
\begin{align*}
\acrshort{nrowsEff}(T_s(i)) &= \max\left(\acrshort{nrowsEff}(T_{s+1}(j) \subseteq T_s(i))\right) + \max\left(\acrshort{nrowsEff}(T_{s+1}(j+1) \subseteq T_s(i))\right)\\
 & \text{for } j=0,2,\ldots
\end{align*}
Such a definition is based on the idea that nodes at a given stage $s$
have to synchronize with each other and have to wait for their
siblings with the largest workload in each sweep (color). Propagating
this information upwards on the tree until we reach the root node
constructs the critical path in terms of longest runtime taking into
account single thread workloads, dependencies, and
synchronizations. Thus, the final value in the root node
$\acrshort{nrowsEff}(T_{-1}(0))$ can be considered as the effective
maximum workload of a single thread. Dividing the globally optimal
workload per thread,
${\acrshort{nrows}^\mathrm{total}}/{\acrshort{nthreads}}$, by
this number gives the parallel efficiency ($\eta$) of our workload
distribution:
\begin{align*}
	\eta &= \frac{ \acrshort{nrows}^\mathrm{total}} {\acrshort{nrowsEff}(T_{-1}(0)) \times \acrshort{nthreads}}. 
\end{align*}
For the tree presented in \Cref{fig:rec_2d-7pt_tree}, the parallel
efficiency is limited to $\eta=\frac{256}{44 \times 8 } = 0.73$ on
eight threads, \ie the maximum parallel speedup is $5.8$.

\subsection{Parameter analysis and selection}
\label{subsec:param_analysis}
The parallel efficiency \acrshort{eta} as defined above can be calculated for
any given matrix, number of threads $\acrshort{nthreads}$, and choice of
$\{\epsilon_s; s=0,1,\ldots\}$; it reflects the quality of parallelism generated
by \acrshort{RACE} for the problem at hand.  This way we can understand the
interaction between these parameters and identify useful choices for
the $\epsilon_s$. Of course, running a specific kernel such
as \acrshort{SymmSpMV} on actual hardware will add further hardware and software
constraints such as attainable memory bandwidth or cost of synchronization.

\begin{figure}[t]
	\centering
	\subfloat[$\eta$ versus \acrshort{nthreads} for \emph{inline\_1} matrix, $\epsilon_1 = 0.5$ ]{\label{fig:inline-a}\includegraphics[width=0.45\textwidth , height=0.2\textheight]{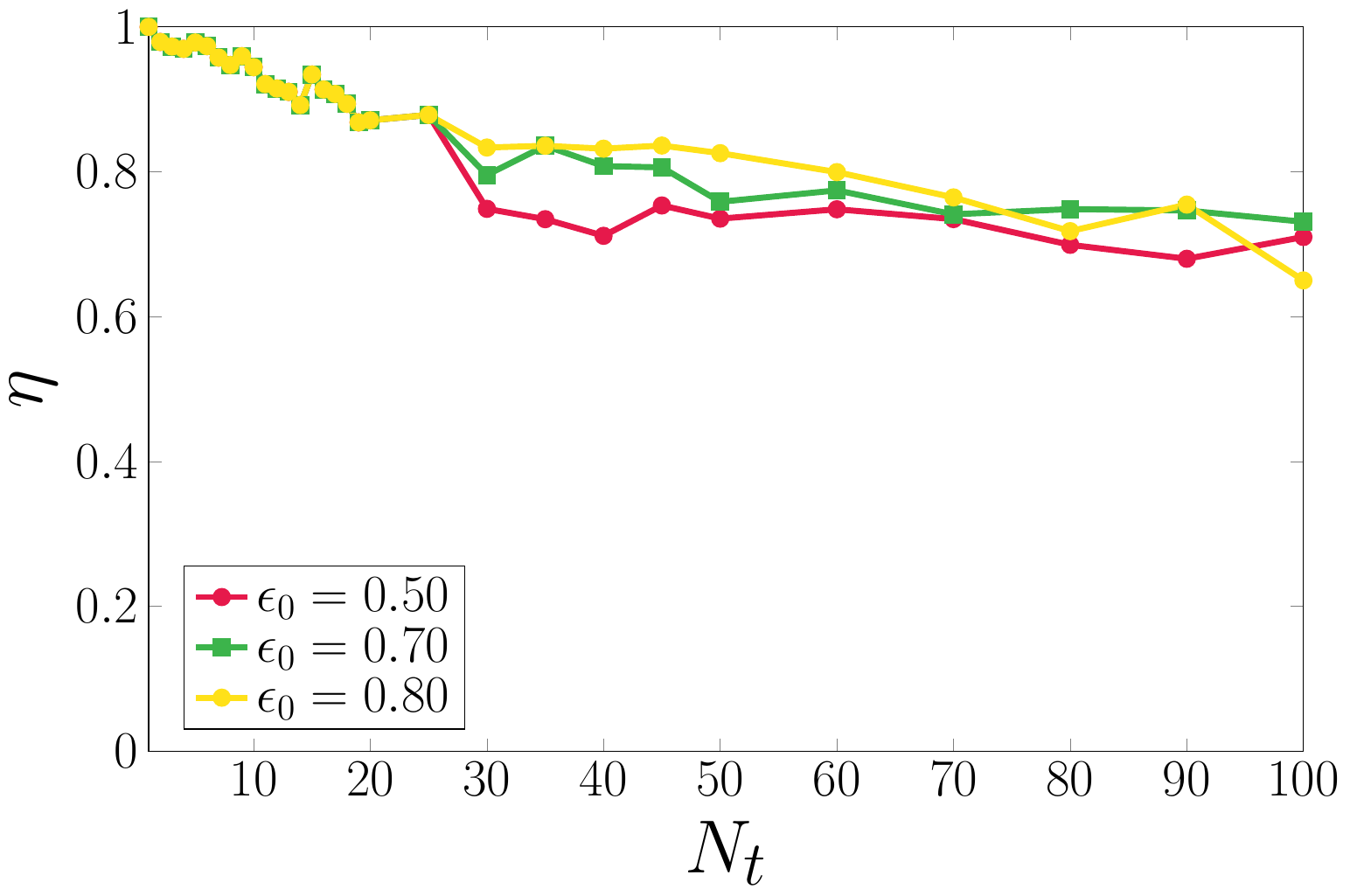}}
	\hspace{1.5em}
	\subfloat[\acrshort{nthreads}=25]{\label{fig:inline-b}\includegraphics[width=0.45\textwidth , height=0.2\textheight]{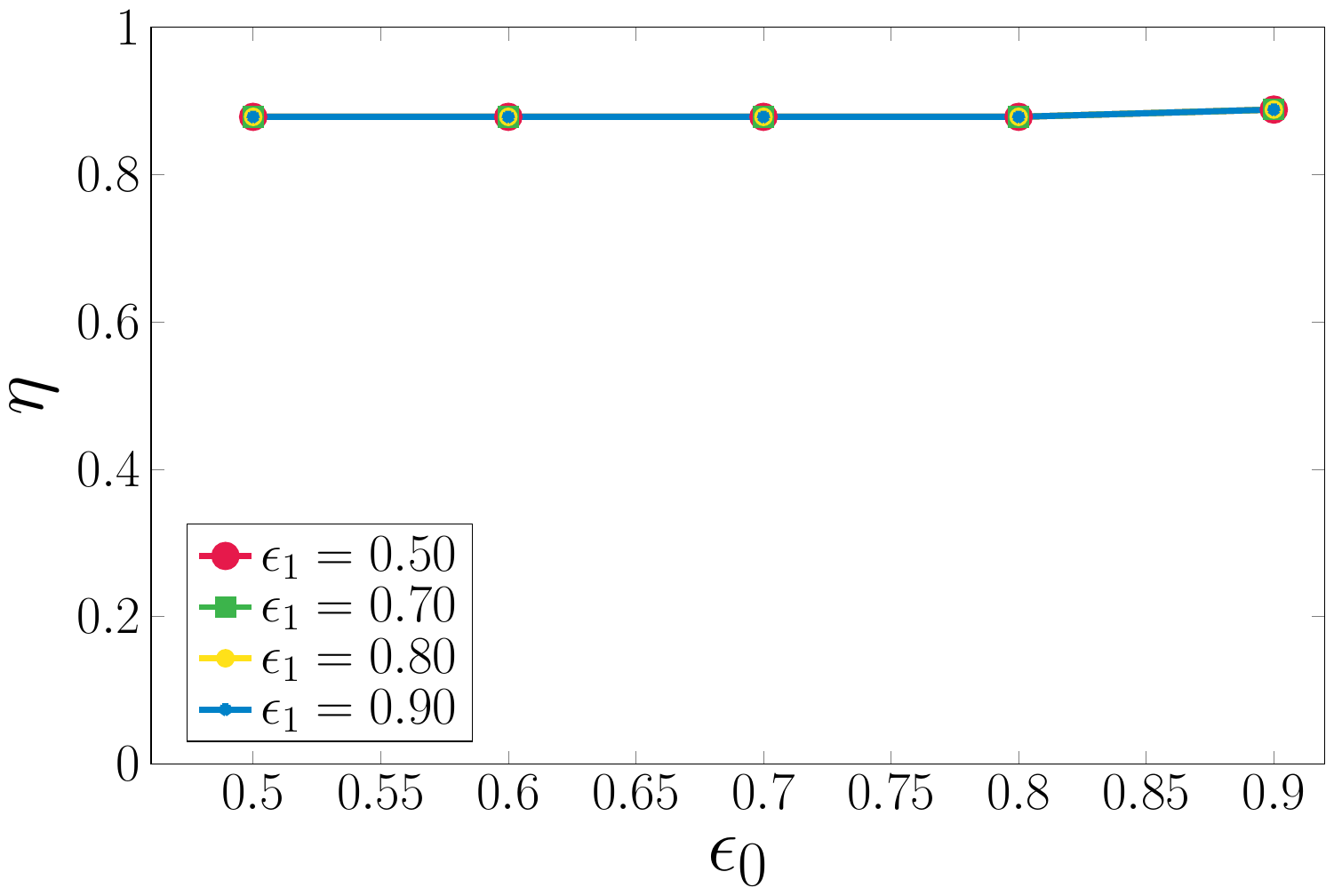}}
	
	\subfloat[\acrshort{nthreads}=45 ]{\label{fig:inline-c}\includegraphics[width=0.45\textwidth , height=0.2\textheight]{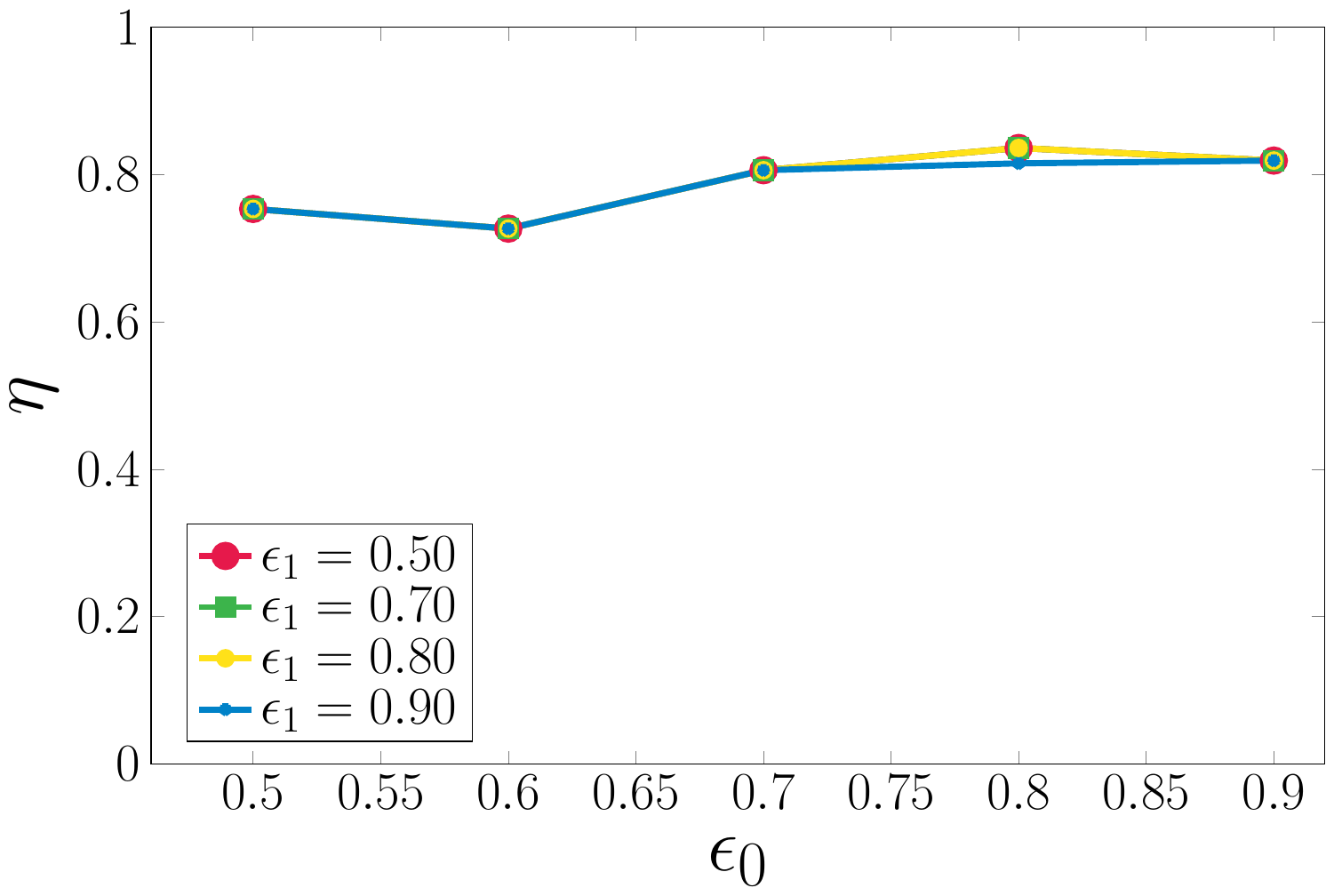}}
	\hspace{1.5em}
	\subfloat[\acrshort{nthreads}=100 ]{\label{fig:inline-d}\includegraphics[width=0.45\textwidth , height=0.2\textheight]{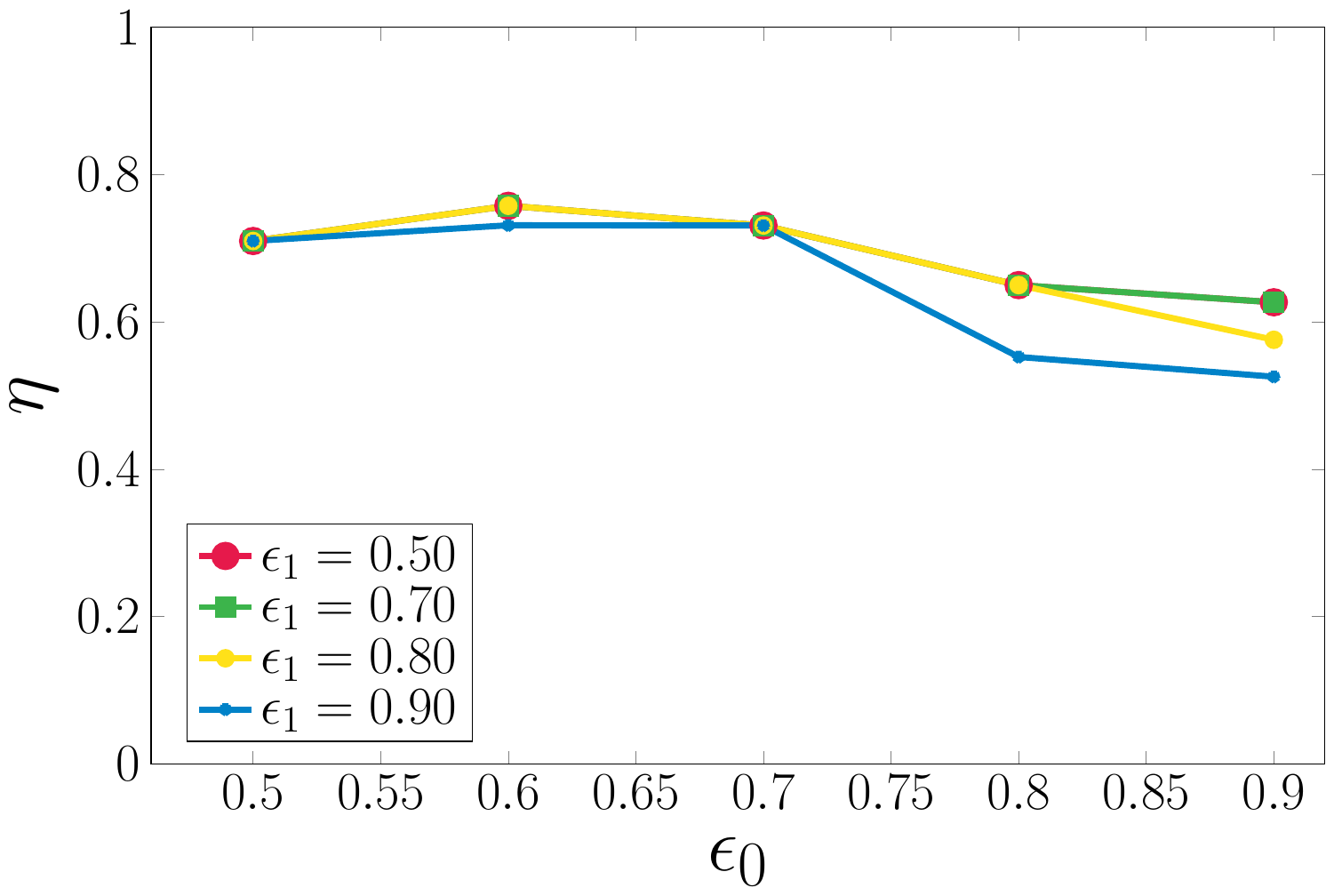}}
	\caption{Parameter study on the \emph{inline\_1} matrix. In \Cref{fig:inline-b,fig:inline-c,fig:inline-d} each of the lines in the plot are iso-$\epsilon_1$ and impact of $\eta$ with respect to $\epsilon_0$ is shown. $\epsilon_s$ for $s>1$ is fixed to $0.5$.}
	\label{fig:inline_param_study}
\end{figure}

\begin{figure}[t]
   	\centering
   	\includegraphics[width=0.9\textwidth]{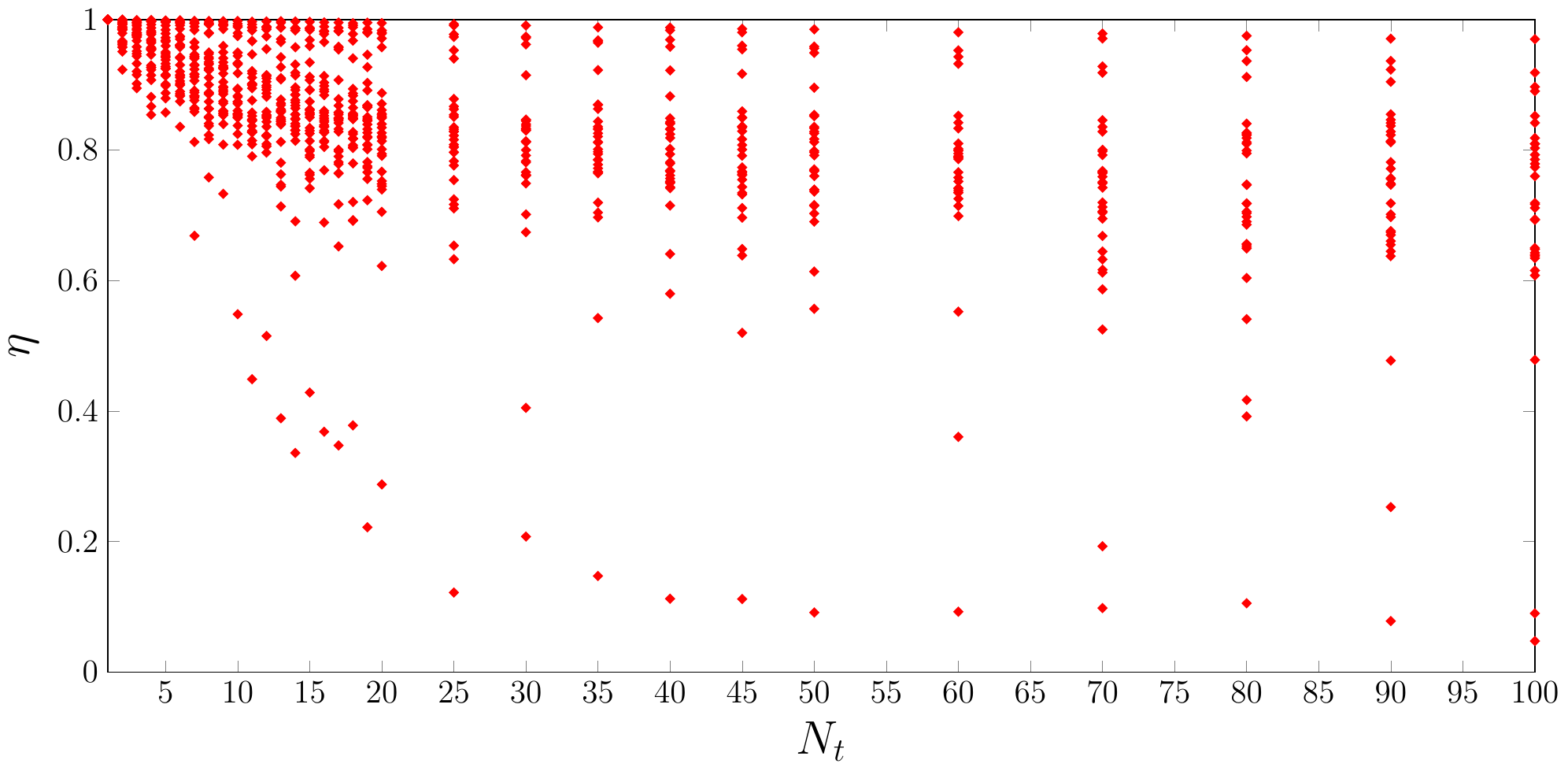}
   	\caption{Parallel efficiency $\eta$ versus \acrshort{nthreads} for all test matrices with $\epsilon_{0,1} = 0.8$ and $\epsilon_{s>1} = 0.5$.}
  	\label{fig:param_all_mtx_stat}
   \end{figure}

As a first step we can limit the parameter space by simple corner case
analysis. Setting all parameters close to one requests high-quality load
balancing but may prevent our balancing scheme from terminating. In the extreme
case of $\{\epsilon_s=1; s=0,1,\ldots\}$ the scheme may generate only
two \levelGroups (one of each color) in each recursion, assign all threads to
them, and may further attempt to refine them in the same way.  The lowest
possible value of $\epsilon_s$ is the maximum deviation of a real number from
its nearest integer, which is $0.5$. A range of [$0.5$,$0.9$] for the
$\epsilon_s$ is therefore used in the following. For a basic analysis we have
selected the \emph{inline\_1} matrix (see \Cref{tab:test_mtx}) as it has a
rather small amount of parallelism and allows us to cover basic
scenarios. In \Cref{fig:inline_param_study} we demonstrate the impact of
different choices for $\epsilon_0$ and $\epsilon_1$ on the parallel efficiency
for thread counts up to 100, which is a useful limit for modern CPU-based
compute nodes.
For $s > 1$ we always set the minimum value of $\epsilon_s=0.5$. The limited
parallelism can be clearly observed in \Cref{fig:inline-a}, with efficiency
steadily decreasing with increasing thread count. At
$\epsilon_1=0.5$ there is only a minor impact of the parameter
$\epsilon_0$. In \Cref{fig:inline-b,fig:inline-c,fig:inline-d} the interplay
between these two parameters is analyzed at different thread counts in more
detail. We find that up to intermediate parallelism ($\acrshort{nthreads}=50$)
the exact choice has only a minor impact on the parallel efficiency (see $y$-axis
scaling). For larger parallelism the interplay becomes more intricate,
where too large values of $\epsilon_{0,1}$ may lead to stronger imbalance. Based
on this evaluation, we choose $\epsilon_{0,1}=0.8$ and $\epsilon_s=0.5$ for $s>1$
for all subsequent performance measurements. The quality of this choice in terms of
parallel efficiency for all matrices is presented
in \Cref{fig:param_all_mtx_stat}. Here we plot the $\eta$ value for all the
matrices over a large thread count. We find that our parameter setting achieves
parallel efficiencies of 80\% or higher for a substantial fraction of the
matrices up to intermediate thread counts.
Representing the upper (lower) values in \Cref{fig:param_all_mtx_stat} is the
best (worst) case matrix \emph{Graphene-4096} (\emph{crankseg\_1}), exhibiting
almost perfect (very low) parallel efficiency at intermediate to high thread
counts.

\begin{figure}[t]
	\centering
	\subfloat[\emph{crankseg\_1}]{\label{fig:crankseg_param}\includegraphics[width=0.24\textwidth]{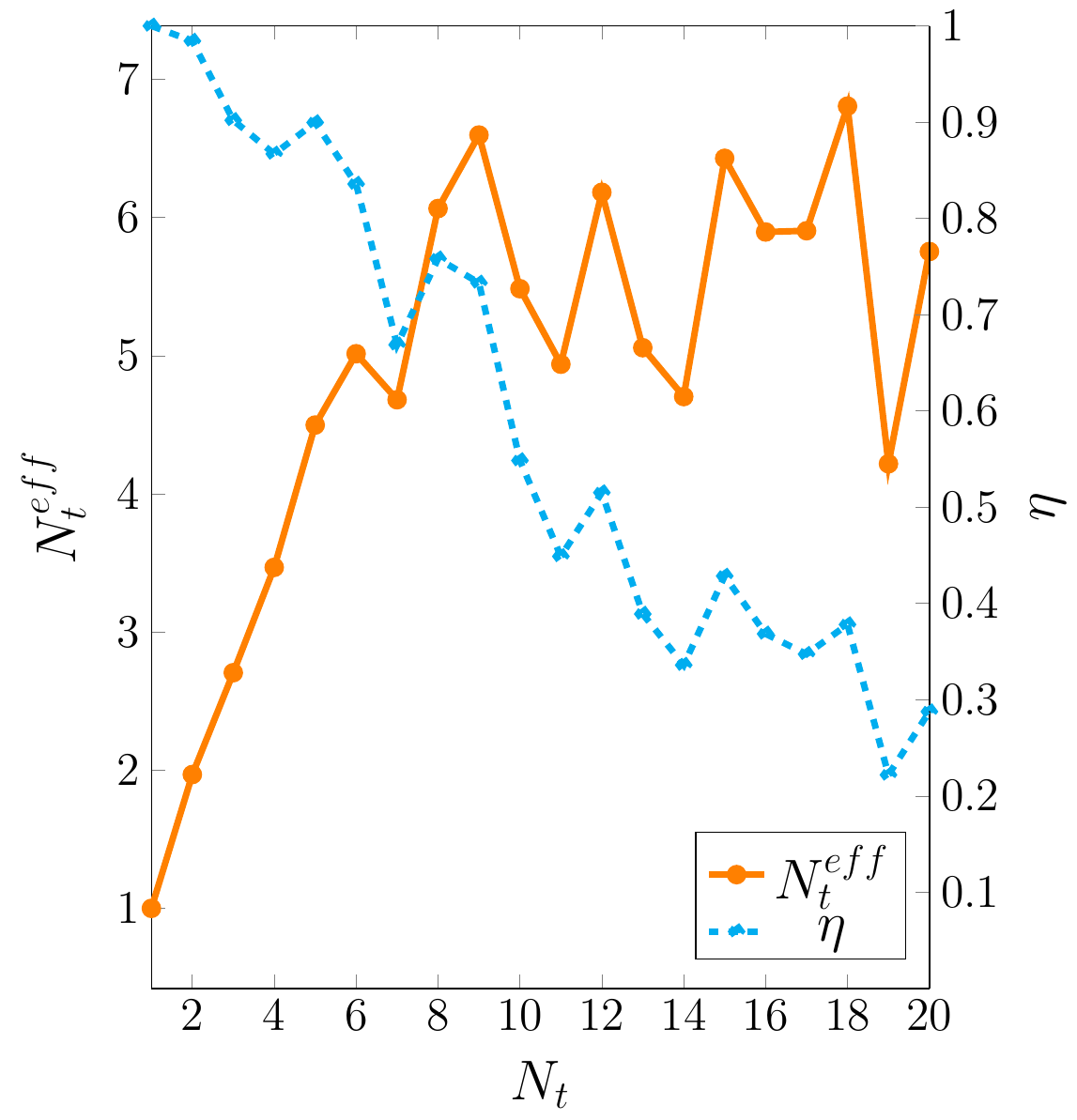}}
	\subfloat[\emph{inline\_1}]{\label{fig:inline_param}\includegraphics[width=0.24\textwidth]{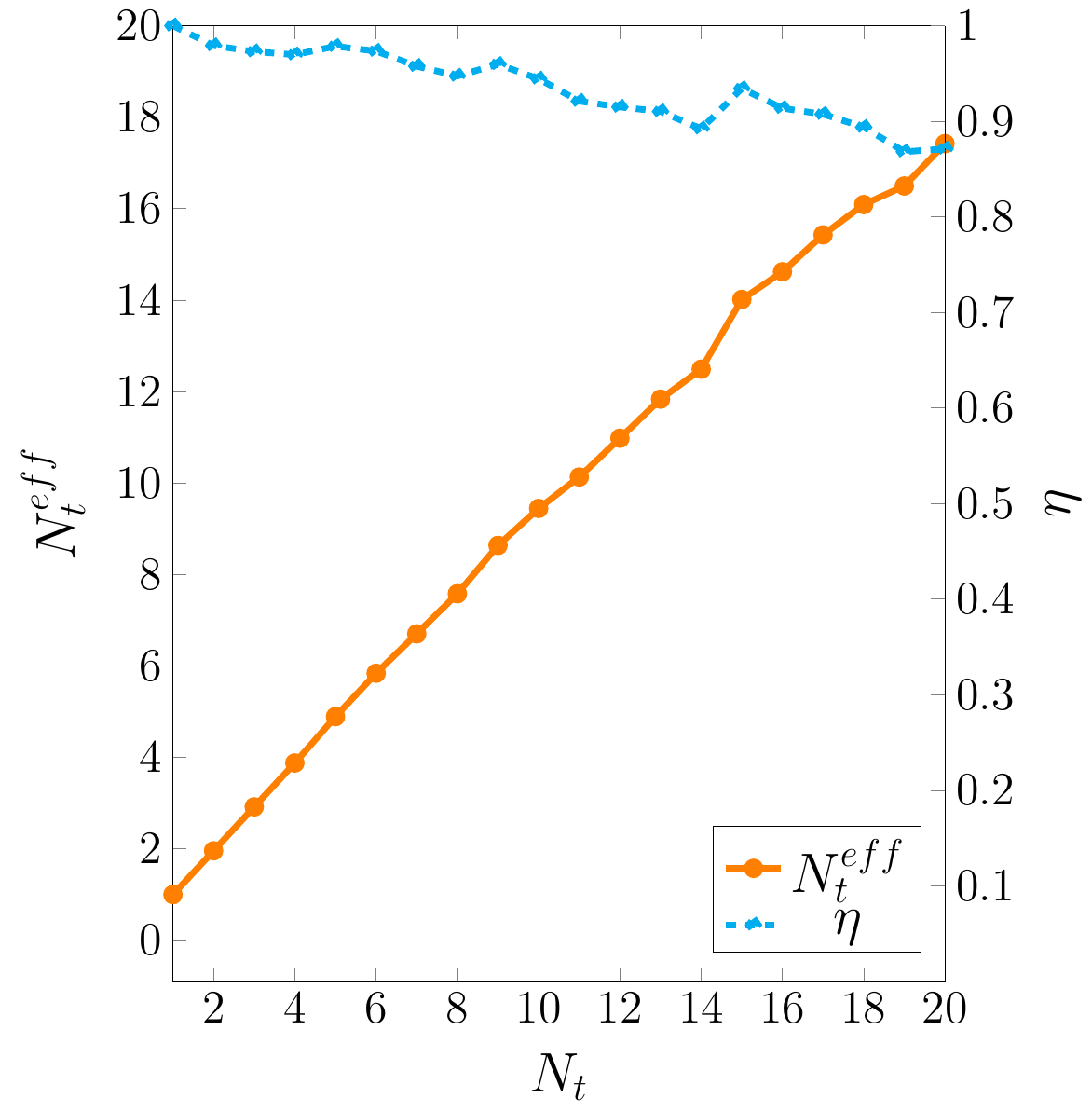}}	
	\subfloat[\emph{parabolic\_fem}]{\label{fig:parabolic_fem_param}\includegraphics[width=0.24\textwidth]{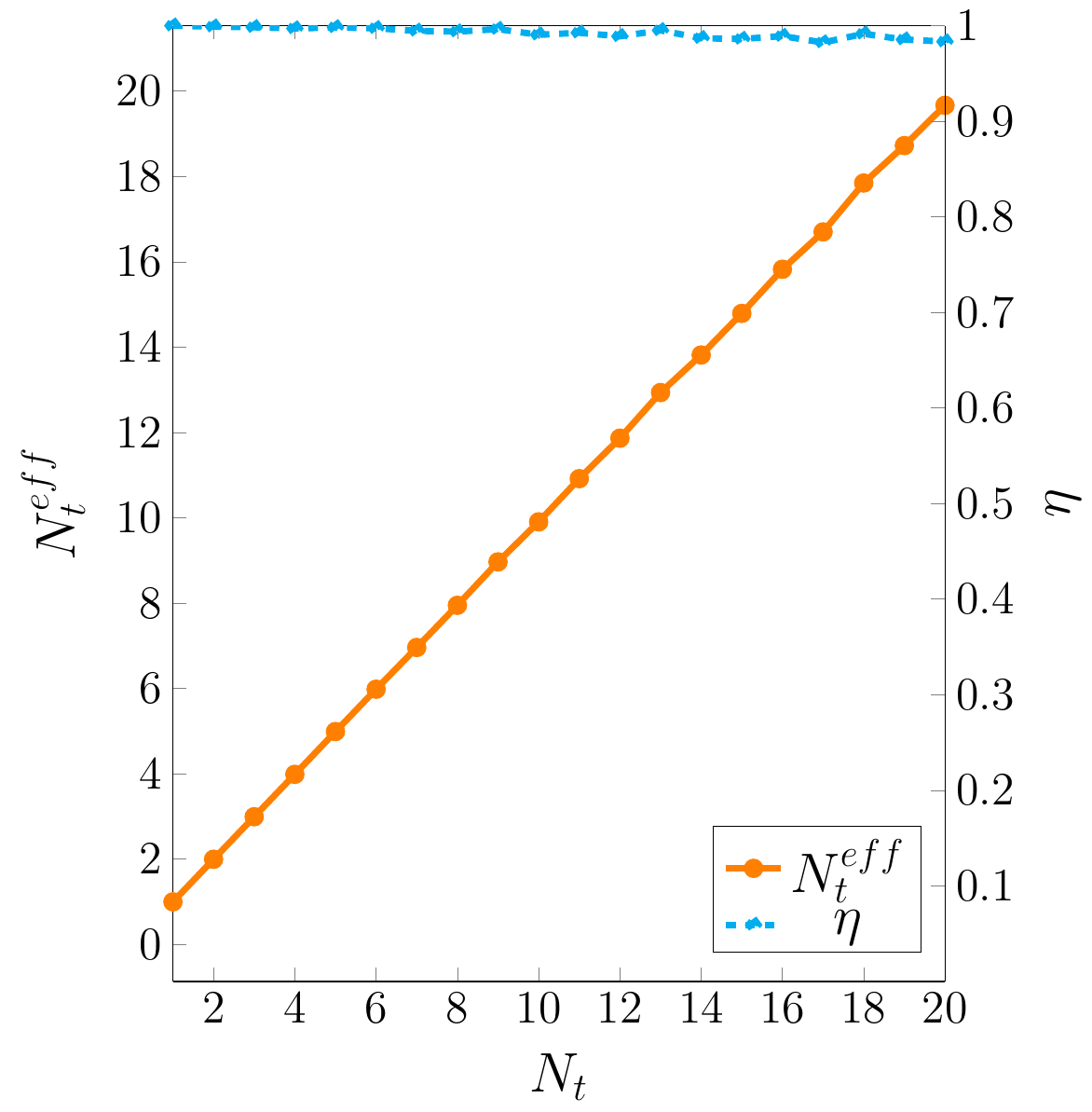}}
	\subfloat[\emph{Graphene-4096}]{\label{fig:Graphene-4096_param}\includegraphics[width=0.24\textwidth]{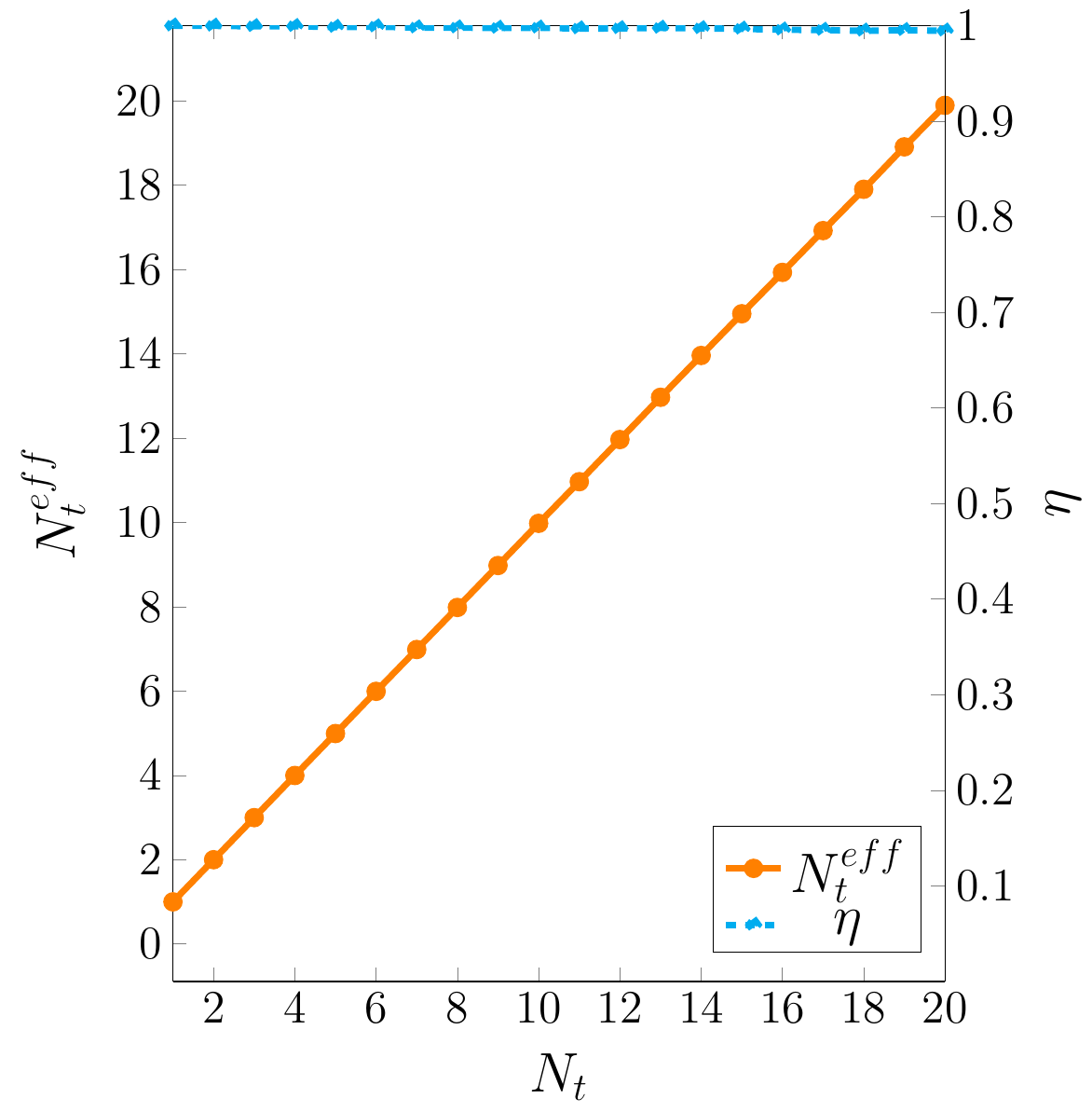}}	
	\caption{\acrshort{threadEff} and $\eta$ versus \acrshort{nthreads} for
	the four corner case matrices, with the same settings used in experiment
	runs. \acrshort{threadEff} is defined as
	$\eta\times\acrshort{nthreads}$.}
	\label{fig:corner_cases_param}
\end{figure}

Finally, we evaluate the scalability of RACE using these two corner cases and
the \emph{inline\_1} matrix as well as the \emph{parabolic\_fem} matrix, which
is small enough to fit into the cache.  In \Cref{fig:corner_cases_param} we
mimic scaling tests on one Skylake processor with up to 20 cores (\ie
threads) and plot the parallel efficiency $\eta$ as well as the maximum number
of threads which can be ``perfectly'' used \acrshort{threadEff} (\ie
$\acrshort{threadEff} = \eta\times\acrshort{nthreads}$).  The unfavorable structure
of the \emph{crankseg\_1} matrix puts strict limits on parallelism even for low
thread counts.  The combination of small matrix size with a rather dense
population (see \Cref{table:bench_matrices}) leads to large inner levels when
constructing the graph, triggering strong load imbalance if using more than six
threads. A search for better $\epsilon_s$ slightly changes the characteristic
scaling but not the maximum parallelism that can be extracted. For
the \emph{inline\_1} matrix we find a weak but steady decrease of the parallel
efficiency, which is in good agreement with the discussion
of \Cref{fig:inline_param_study}. The other two matrices scale very well in the
range of thread counts considered.

\begin{figure}[t]
	\centering
	\subfloat[\emph{crankseg\_1}]{\label{fig:crankseg_scaling}\includegraphics[width=0.23\textwidth , height=0.18\textheight]{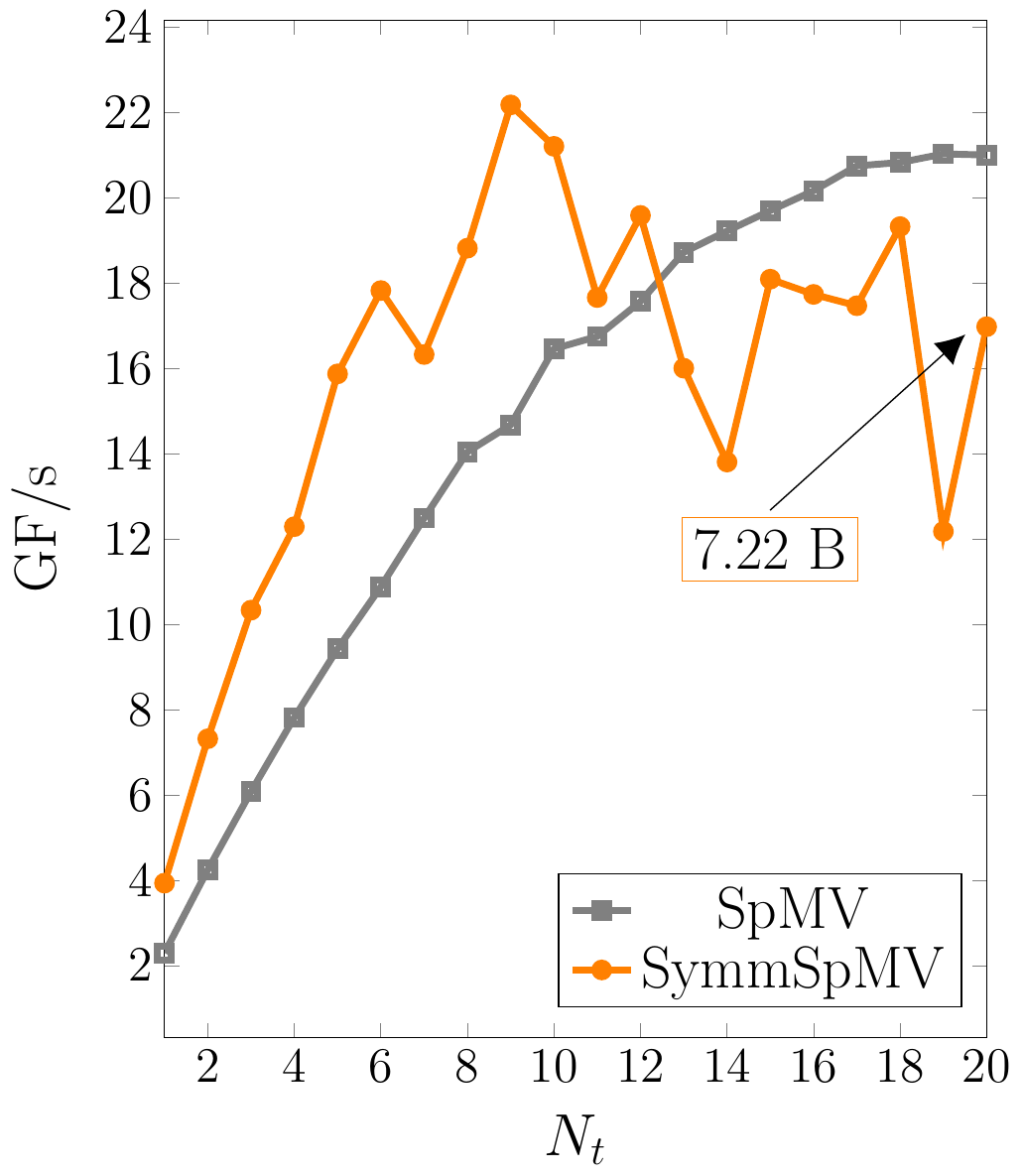}}
	\subfloat[\emph{inline\_1}]{\label{fig:inline_scaling}\includegraphics[width=0.23\textwidth , height=0.18\textheight]{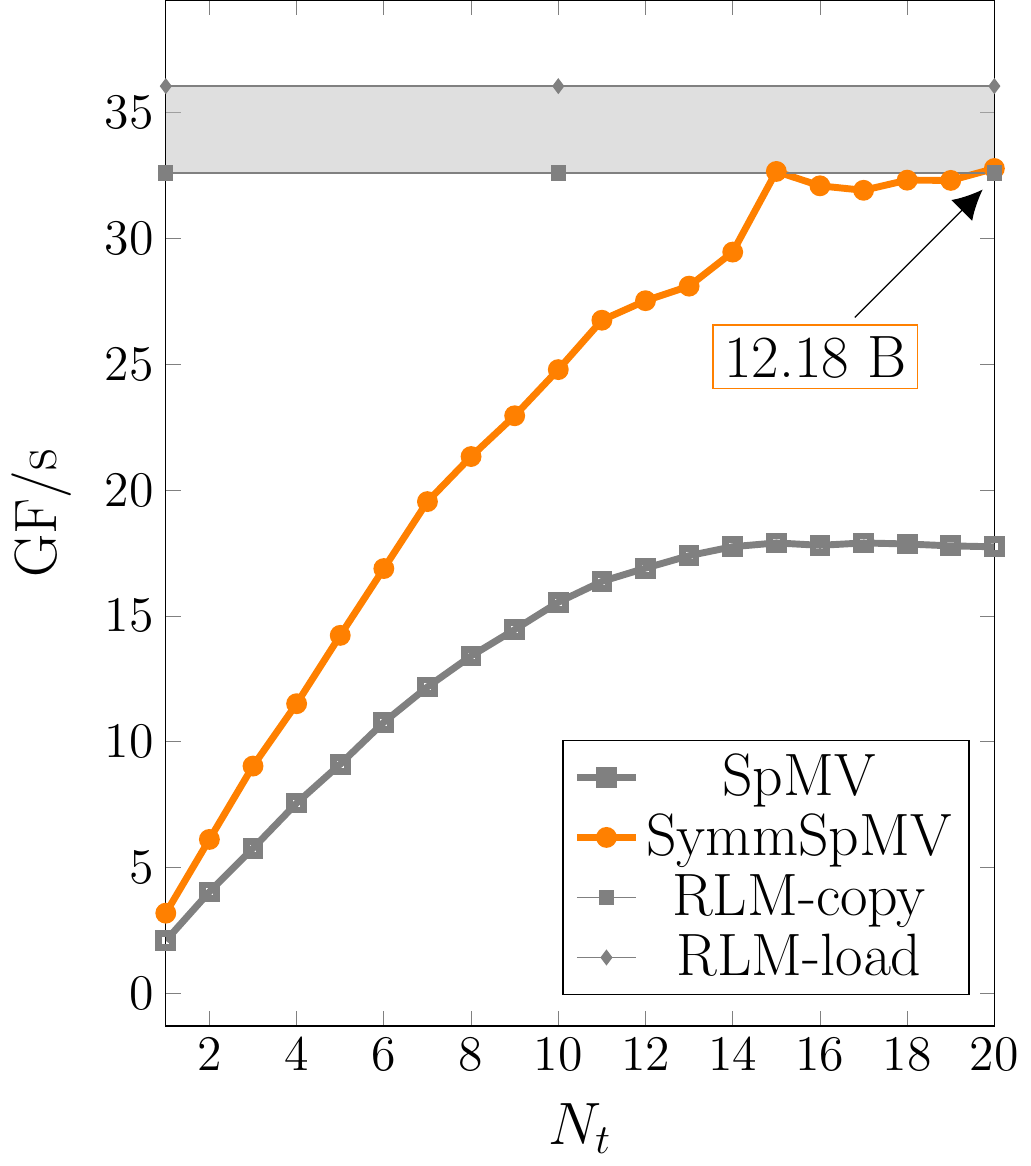}}	
	\subfloat[\emph{parabolic\_fem}]{\label{fig:parabolic_fem_scaling}\includegraphics[width=0.23\textwidth , height=0.18\textheight]{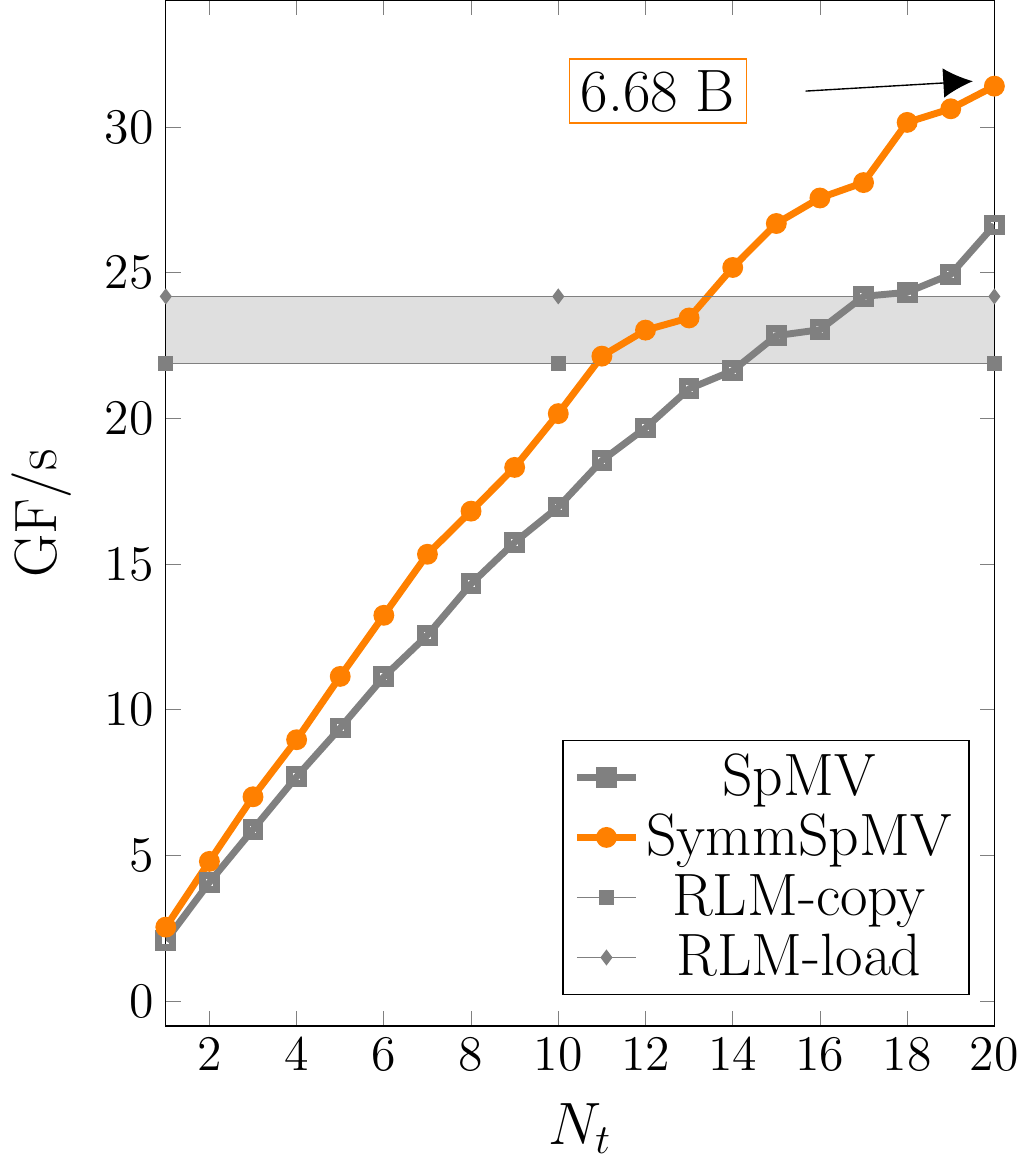}}
	\subfloat[\emph{Graphene-4096}]{\label{fig:Graphene-4096_scaling} \includegraphics[width=0.23\textwidth , height=0.18\textheight]{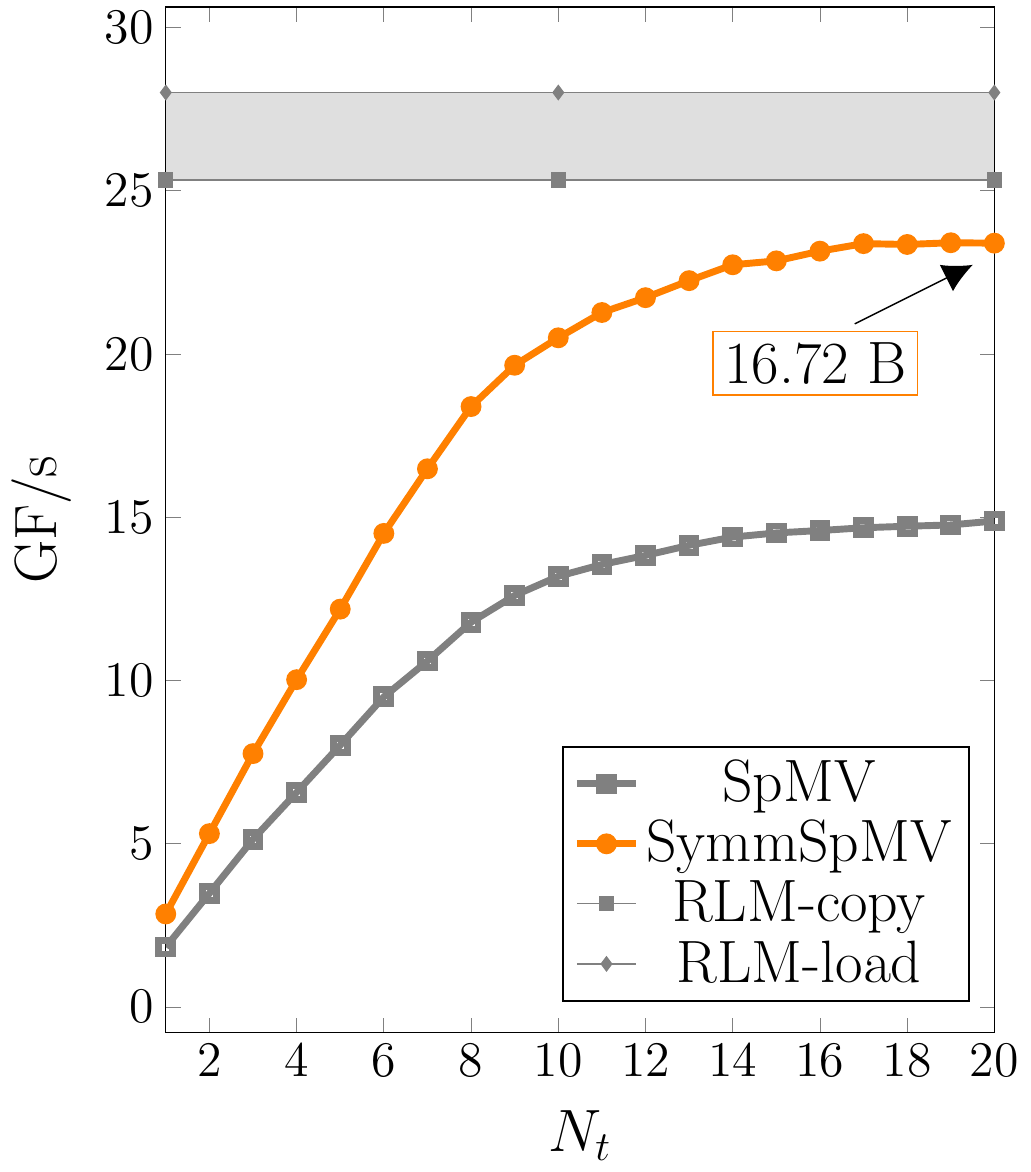}}	
	\caption{Parallel performance measurements of \acrshort{SymmSpMV}
	with \acrshort{RACE} on one \SKX socket for the four
	corner case matrices. The performance of the
	basic \acrshort{SpMV} kernel is presented for reference. For the
	matrices \Cref{fig:inline_scaling,fig:parabolic_fem_scaling,fig:Graphene-4096_scaling}
	the maximum roof{}line performance limits \Cref{eq:upper_performance} are
	given using the computational intensity \Cref{eq:SymmSpMV_intensity} for
	the two extreme cases of load-only memory bandwidth (RLM-load) and copy
	memory bandwidth (RLM-copy). The measured full socket main memory data
	traffic per nonzero entry of the symmetric matrix (in \BYTE) for the \acrshort{SymmSpMV}
	 operation is also shown, where values below 12 \BYTE indicate caching of the matrix entries.}
	\label{fig:corner_cases_scaling}
\end{figure}

The corresponding performance measurements for the \acrshort{SymmSpMV} kernel
(see \Cref{sect:SymmSpmv}) on a single \SKX processor chip with 20 cores are
shown in \Cref{fig:corner_cases_scaling}.\footnote{For the benchmarking setup
see \Cref{Sec:expt}.}
For the \emph{crankseg\_1} matrix (see \Cref{fig:crankseg_scaling}) we recover
the limited scaling due to load imbalance as theoretically predicted. A
performance maximum is at nine cores, where the maximum \acrshort{SpMV}
performance can be slightly exceeded. However, based on the roof{}line performance
model given by \Cref{eq:SymmSpMV_intensity,eq:upper_performance} together with
the matrix parameters from \Cref{table:bench_matrices}, a theoretical speedup of
approximately two as compared to \acrshort{SpMV} can be expected for the full
processor chip under best conditions.
Indeed, in case of the \emph{inline\_1} and \emph{Graphene-4096}
matrices, performance
scales almost linearly until the main memory bandwidth bottleneck is hit. The
saturated performance is in good agreement with the roof{}line limits.
Note  that even though the \emph{inline\_1} matrix does not exhibit
perfect theoretical efficiency ($\eta$ $\approx 0.85$ at
$\acrshort{nthreads}=20$), it still generates sufficient parallelism to achieve
main memory saturation: The memory bottleneck can mitigate
a limited load imbalance.

The peculiar performance behavior of
\emph{parabolic\_fem}
(see \Cref{fig:parabolic_fem_param,fig:parabolic_fem_scaling}) is due to 
its smallness ($\approx 23$ \MB), which lets it fit into the caches of the
Skylake processor (\acrshort{LLC} size $= 28$ \MB). Thus, performance is not
limited by the main memory bandwidth constraint and the roof{}line model limits do
not apply.

We have demonstrated that a simple choice for the only set of RACE input
parameters $\{\epsilon_s; s=0,1,\ldots\}$ can extract sufficient parallelism for
most matrices considered in this study. Moreover, the parallel efficiency as
calculated by RACE in combination with the roof{}line performance model is a good
indication for scalability and maximum performance of the actual computations.

%% file: experiment_and_results.tex

We evaluate the performance of the \acrshort{SymmSpMV} based on parallelization and reordering performed by \acrshort{RACE} and compare it with the two MC approaches introduced above and the \acrshort{MKL}. 
As a yardstick for baseline performance we choose the general \acrshort{SpMV} kernel and use the performance model introduced in \Cref{Sec:test_kernels} to quantify the quality of our absolute performance numbers.
As the deviations between different measurement runs are less than 5\%, we do  not show the 
error bar in our performance measurements.

\subsection{Experimental Setup}

All matrix data are encoded in the CRS format. For the \acrshort{SymmSpMV}  only the nonzeros of the upper triangular matrix are stored. In the case of RACE and the coloring approaches every thread executes the \acrshort{SymmSpMV} kernel \Cref{alg:SymmSpMV} with appropriate outer loop boundary settings depending on the color (MC, ABMC) or \levelGroups (\acrshort{RACE}) to be computed. \Inorder to ensure vectorization of the inner loop in \Cref{alg:SymmSpMV} we use the SIMD pragma \texttt{\#pragma simd reduction(+:tmp) vectorlength(VECWIDTH)}. Here \texttt{VECWIDTH} is the maximum vector width supported by the architecture, \ie \texttt{VECWIDTH = 4 (8)} for \IVB (\SKX).

The  \acrshort{MKL} offers two choices for the two sparse matrix kernels under consideration: First, CRS based data structures are provided and are used in the subroutines (\texttt{mkl\_cspblas\_dcsrgemv} for \acrshort{SpMV}  and  \texttt{mkl\_cspblas\_dcsrsymv} for \acrshort{SymmSpMV}) without any modification (MKL). This mode of operation is deprecated from \acrshort{MKL}.v.18. Instead, the inspector-executor mode (MKL-IE) is recommended to be used. Here, the user initially provides the matrix along with hints (\eg symmetry) and operations to be carried out to the inspector routine (\texttt{mkl\_sparse\_set\_mv\_hint}). Then an optimization routine (\texttt{mkl\_sparse\_optimize}) is called where the matrix is  preprocessed based on the inspector information to achieve best performance and highest parallelism for the problem at hand. The subroutine \texttt{mkl\_sparse\_d\_mv} is then used to do the \acrshort{SpMV} or \acrshort{SymmSpMV} operations on this optimized matrix structure. This approach does not provide any insight into which kernel or data structure is actually used ``under the hood.''

In the performance measurements the kernels are executed multiple times \inorder to ensure reasonable measurement times and average out potential performance fluctuations. Doing successive invocations of the kernel on the same two vectors, however, may lead to unrealistic caching of these vectors if the number of rows is small enough. Thus, we use two ring buffers (at least of 50 MB each) holding separate vectors of size \acrshort{nrows}. After each kernel invocation we switch to the next vector in the two buffers. This way we mimic the typical structure of iterative sparse solvers where between successive matrix-vector operations other data intensive kernels are executed, \eg several Level 1 BLAS routines or preconditioning steps. We run over these two buffers 100 iterations (times) and report the mean performance.

For all methods and libraries the input matrices have been preprocessed with \acrshort{RCM} bandwidth reduction using the \SPMP library \cite{SpMP}. This provides the same or better performance on all matrices as compared to the original ordering. If not otherwise noted we use the full processor chip and assign one thread to each core. As we focus on a single chip and \acrfull{SNC} is not enabled on \SKX, no NUMA data placement effects impact our results.

\subsection{Results}
Before we evaluate the performance across the full set of matrices presented in \Cref{table:bench_matrices} we return to the analysis of the \acrshort{SymmSpMV} performance and data traffic for the Spin-26 matrix  which we have presented in \Cref{Sec:motivation} for the established coloring approaches. 
\subsubsection{Analysis of \acrshort{SymmSpMV} kernel using RACE for the Spin-26 matrix}
\label{Sec:Spin26full}
\begin{figure}[t]
 	\centering
 	\subfloat[SymmSpMV]{\label{fig:motivation_symm_spmv_w_RACE}\includegraphics[width=0.248\textwidth, height=0.20\textheight]{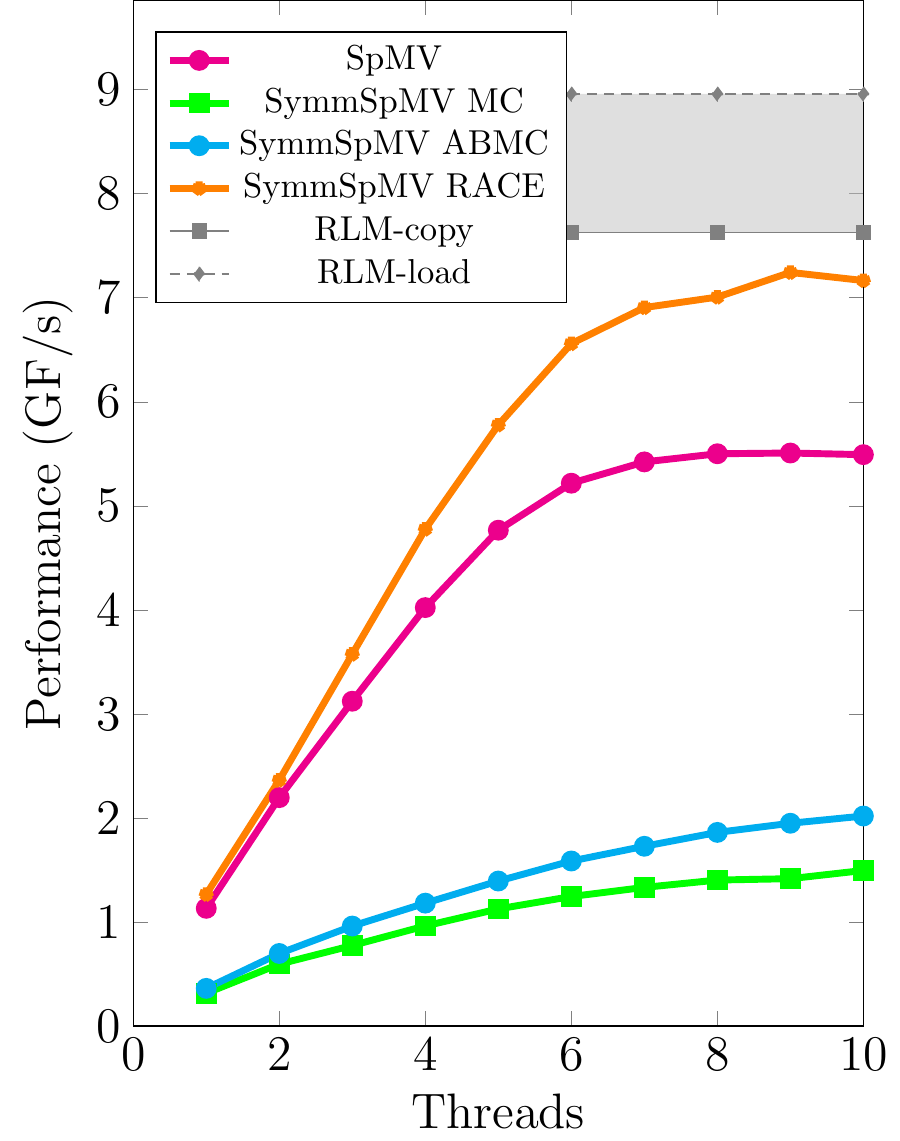}}
 	\subfloat[Data traffic]{\label{fig:motivation_data_w_RACE}\includegraphics[width=0.248\textwidth, height=0.20\textheight]{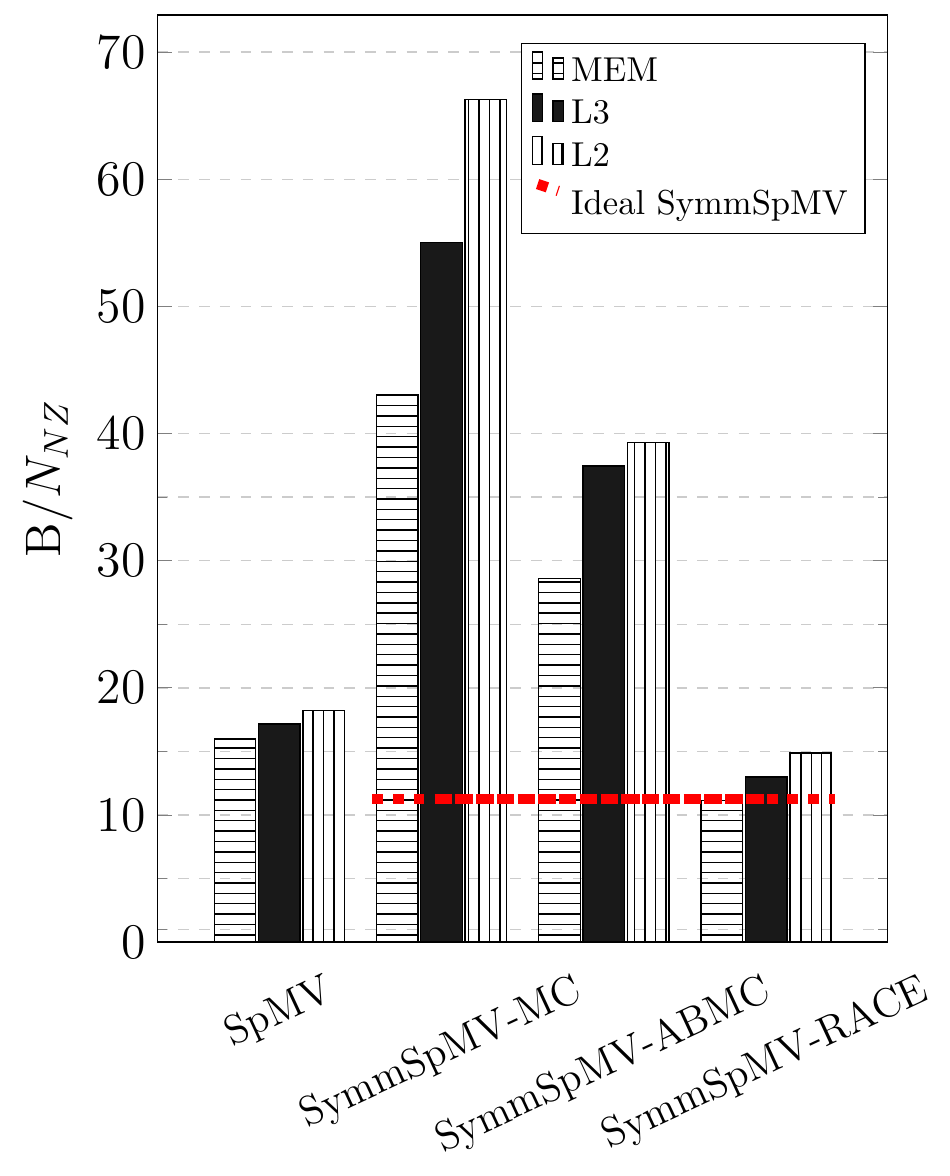}}
  	\subfloat[SymmSpMV]{\label{fig:motivation_symm_spmv_w_RACE_skx}\includegraphics[width=0.248\textwidth, height=0.20\textheight]{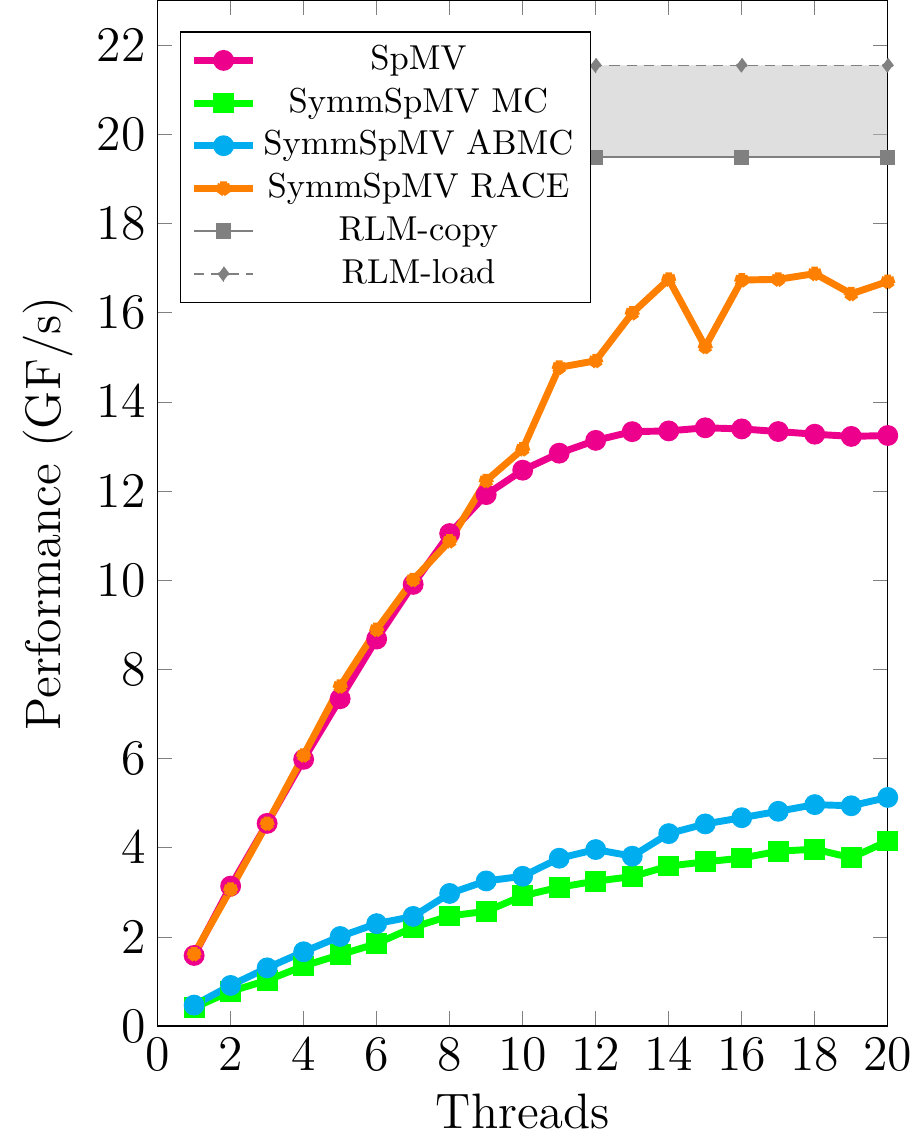}}
  	\subfloat[Data traffic]{\label{fig:motivation_data_w_RACE_skx}\includegraphics[width=0.248\textwidth, height=0.20\textheight]{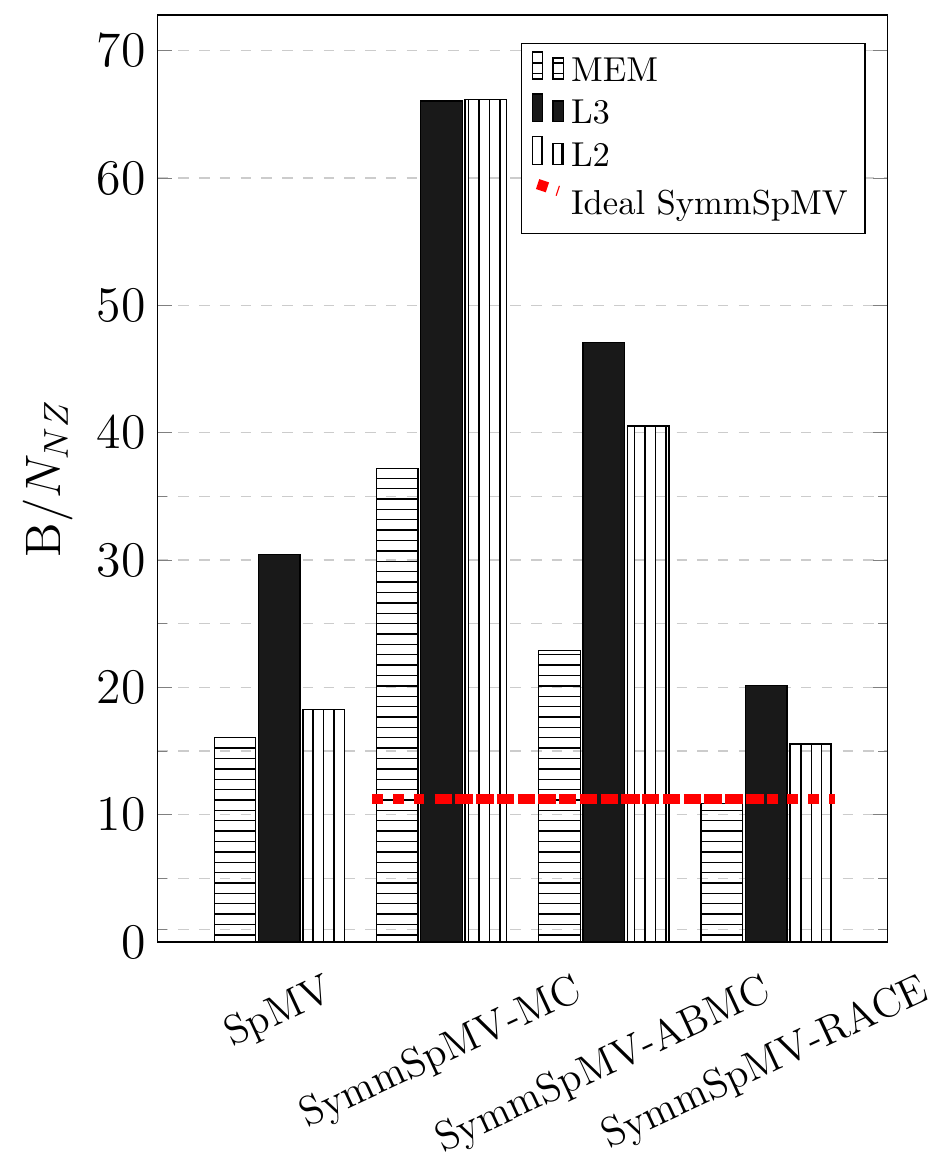}}
 	\caption{Performance (\Cref{fig:motivation_symm_spmv_w_RACE}) and data traffic (\Cref{fig:motivation_data_w_RACE}) analysis for \acrshort{SymmSpMV} kernel with Spin-26 matrix using \acrshort{MC}, \acrshort{ABMC}, and \acrshort{RACE} on a single socket of \IVB. The corresponding measurements for a single socket of \SKX is shown in \Cref{fig:motivation_symm_spmv_w_RACE_skx,fig:motivation_data_w_RACE_skx}. The roofline performance model (using copy and load-only bandwidth) and the performance of the \acrshort{SpMV} kernel is plotted for reference in scaling plots \Cref{fig:motivation_symm_spmv_w_RACE,fig:motivation_symm_spmv_w_RACE_skx}. The average data traffic per nonzero entry ($\acrshort{NNZR}$) of the full matrix as measured with \LIKWID for all cache levels and main memory is shown together with the minimal value for main memory access (horizontal dashed line) in \Cref{fig:motivation_data_w_RACE,fig:motivation_data_w_RACE_skx}.}
 	\label{fig:motivation_w_RACE}
 \end{figure}

The shortcomings in terms of performance and excessive data transfer for parallelization of \acrshort{SymmSpMV} using MC and ABMC have been demonstrated in \Cref{fig:motivation}. We extend this evaluation by comparison with the \acrshort{RACE} results in \Cref{fig:motivation_w_RACE}.
The figures clearly demonstrate the ability of \acrshort{RACE} to ensure high data locality in the parallel \acrshort{SymmSpMV} kernel. The actual main memory traffic achieved is inline with the minimum traffic for that matrix (see discussion in \Cref{Sec:motivation}) and a factor of up to 4$\times$ lower than the coloring approaches. Correspondingly, \acrshort{RACE} \acrshort{SymmSpMV} performance is at least $3.3\times$ higher than its best competitor and $25\%$ better than the \acrshort{SpMV} kernel on both architectures. It achieves more than 84\% of the roofline performance limit based on the copy main memory performance. Note that the indirect update of the LHS vector will generate a store instruction for every inner loop iteration (see \Cref{alg:SymmSpMV}), while the \acrshort{SpMV} kernel only does a final store at the end of the inner loop iteration. In combination with the low number of nonzeros per row (\acrshort{NNZR}) of the Spin-26 matrix, the ``copy'' induced limit poses a realistic upper performance bound.  
\subsubsection{Analyzing absolute performance of RACE}
\begin{figure}[t]
	\centering
	\subfloat[\IVB]{\label{fig:spmv_vs_symm_spmv_ivy}\includegraphics[width=0.95\textwidth, height=0.3\textheight]{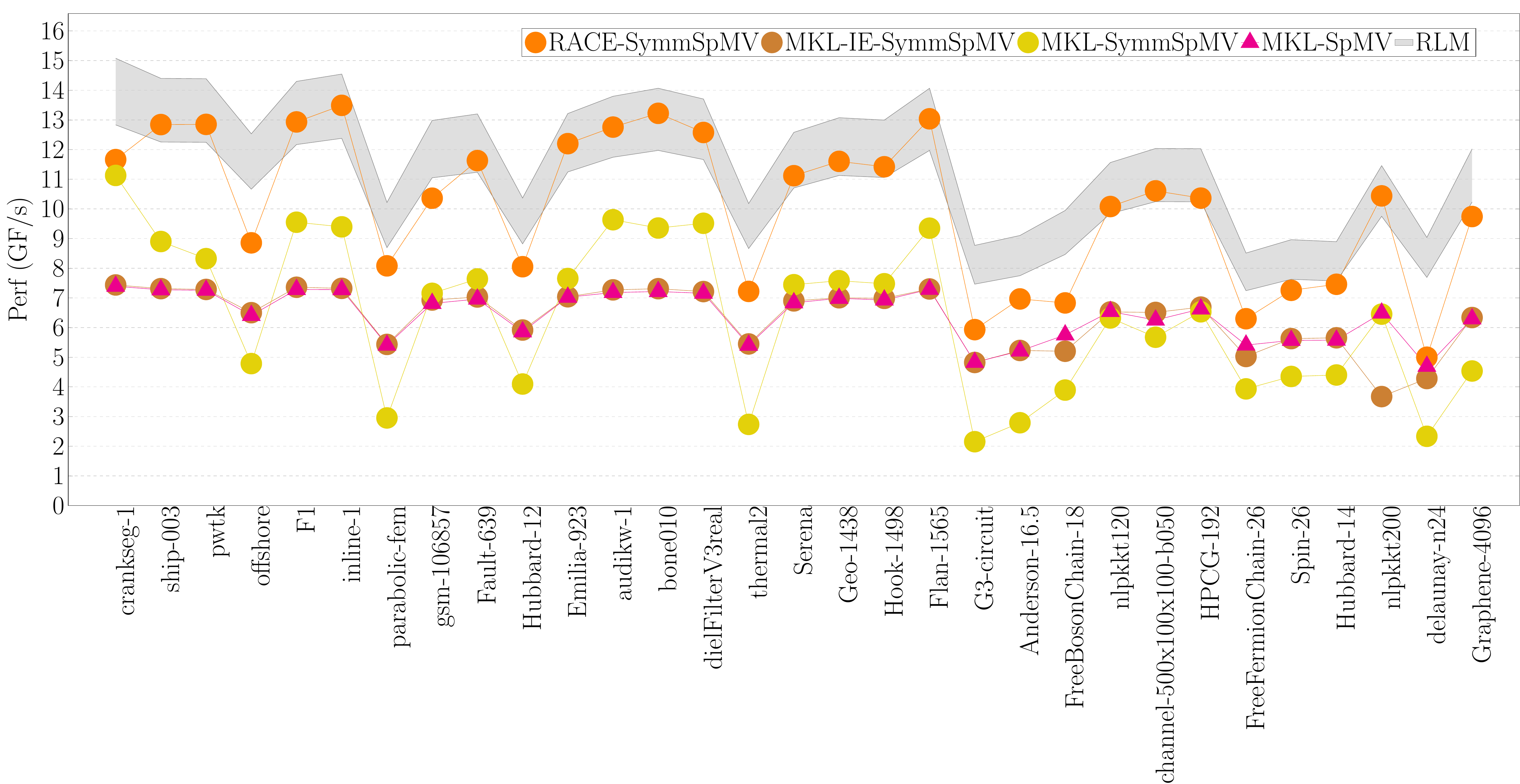}}
	\hspace{1em}
	\subfloat[\SKX]{\label{fig:spmv_vs_symm_spmv_skx}\includegraphics[width=0.95\textwidth, height=0.3\textheight]{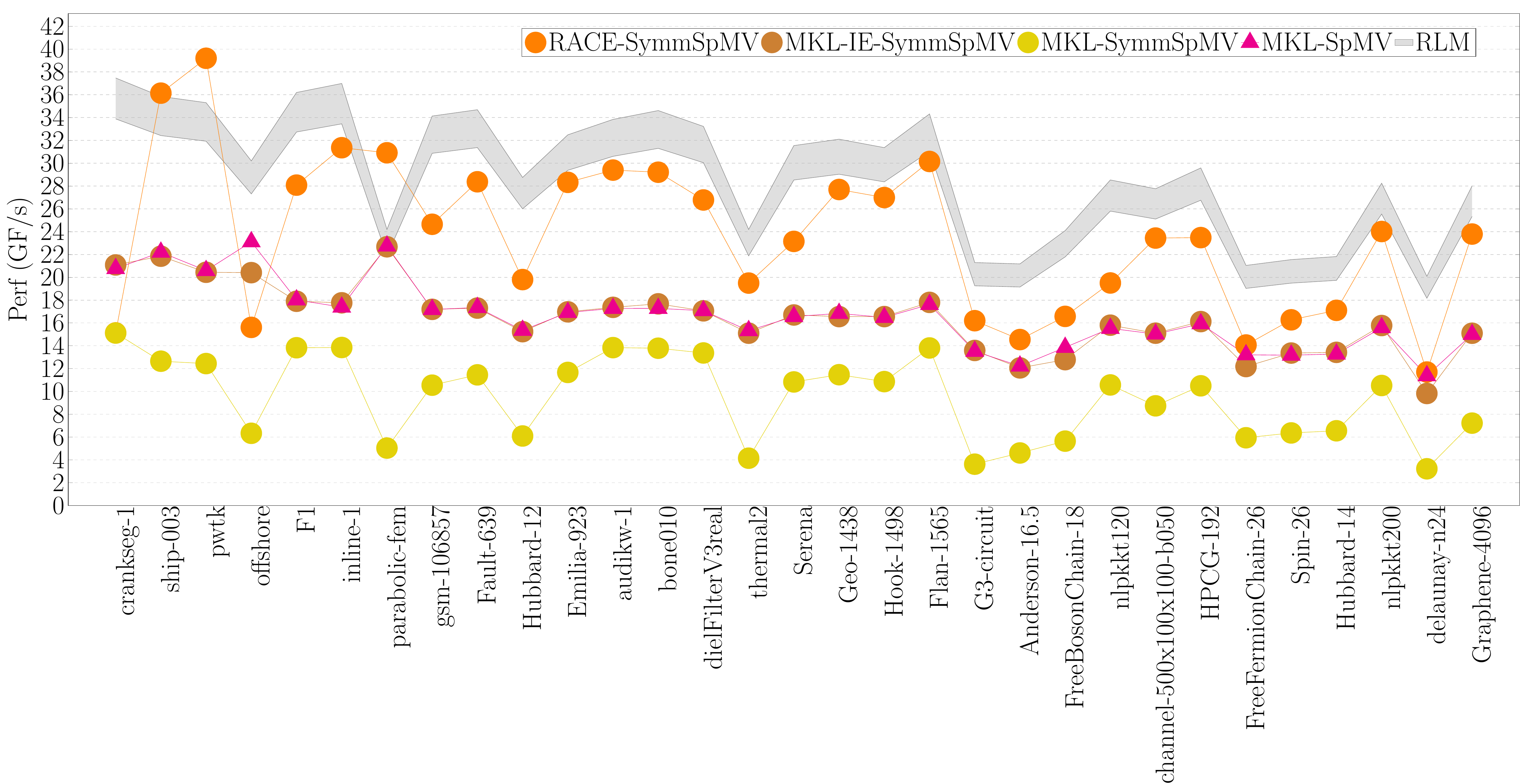}}
	\caption{Performance of \acrshort{SymmSpMV} executed with \acrshort{RACE} compared to the performance model and \acrshort{MKL} implementations. \acrshort{SpMV} performance obtained using \acrshort{MKL} library is also shown for reference. The model prediction is derived for  bandwidths in the range of load and copy bandwidth, and using the measured $\alpha_{\acrshort{SpMV}}$ shown in \Cref{table:alpha_values}.}
	\label{fig:SpMV_vs_SymmSpMV}
\end{figure}

We now extend our \acrshort{RACE}  performance investigation to the full set of test matrices presented in~\Cref{table:bench_matrices}. In \Cref{fig:spmv_vs_symm_spmv_ivy,fig:spmv_vs_symm_spmv_skx} the performance results for the full \IVB processor chip (10 cores) and the full \SKX processor chip (20 cores) are presented along with the upper roofline limits and the performance of the baseline \acrshort{SpMV} kernel using \acrshort{MKL}.  
The matrices are arranged along the abscissa according to the ascending number of rows (\acrshort{nrows}), \ie increasing size of the two vectors involved in \acrshort{SymmSpMV}. Overall \acrshort{RACE}  performance comes close to or matches our performance model for many test cases on both architectures. 
A comparison of the architectures shows that the corner case matrices \texttt{crankseg\_1} and \texttt{parabolic\_fem} have a strikingly different behavior for \acrshort{RACE}. For \texttt{crankseg\_1} this is caused by the limited amount of parallelism in its structures. Here we refer to the discussion of \Cref{fig:crankseg_param} where best performance and highest parallelism (\acrshort{threadEff}) were achieved at approximately 10 cores. Using only 9 cores on \SKX lifts the \acrshort{SymmSpMV} performance of  \texttt{crankseg\_1} slightly above the \acrshort{SpMV} level of the MKL. The \texttt{parabolic\_fem} has been chosen to fit into the \acrshort{LLC} of the \SKX architecture to provide a corner case where scalability is not intrinsically limited by main memory bandwidth (see  \Cref{fig:parabolic_fem_param}) and thus our roofline performance limit does not apply for this matrix on \SKX. However, on \IVB the matrix data set just exceeds the LLC and the performance is inline with our model. 

On both architectures a characteristic drop in performance levels is encountered around the \texttt{Flan\_1565} and \texttt{G3\_circuit} matrices, where the aggregate size of the two vectors (25 MB) approaches the available LLC sizes. For smaller matrices we have a higher chance that the vectors stay in the cache during the \acrshort{SymmSpMV}, \ie the vectors must only be transferred once between main memory and the processor for every kernel invocation. 
For larger matrices (\ie larger $N_r$) the reuse of vector data during a single \acrshort{SymmSpMV} kernel decreases and vector entries may be accessed several times from the main memory. This is reflected by the increase in measured $\alpha_{\acrshort{SpMV}}$ (assumed $\alpha_{\acrshort{SymmSpMV}}$) values for matrices with index 20 and higher in \Cref{table:alpha_values}. 

In short, \acrshort{RACE} has an average speedup of 1.4$\times$ and 1.5$\times$ compared to \acrshort{SpMV} on the \SKX and \IVB architectures, respectively. On \SKX \acrshort{RACE} \acrshort{SymmSpMV} attains on an average 87\% and 80\% of the roofline performance limits predicted using the copy and load bandwidth, respectively, while on \IVB we are 91\% and 83\% close to the respective performance models.

The MKL implementations of \acrshort{SymmSpMV} deserves a special consideration in this context.
Therefore,  in \Cref{fig:SpMV_vs_SymmSpMV} we also compare  our approach with the two \acrshort{MKL} options described above. For the MKL-IE variant we specify exploiting the symmetry of the matrix when calling the inspector routine. 
On the \IVB architecture, \acrshort{RACE} always provides superior performance levels and the best performing Intel variant depends on the underlying matrix. On the \SKX, however, MKL-IE always outperforms the deprecated MKL routine and is superior to \acrshort{RACE} for two matrices (\texttt{crankseg-1,offshore}). These are the same matrices where \acrshort{RACE} is slower than the MKL \acrshort{SpMV} kernel (see \Cref{fig:spmv_vs_symm_spmv_skx}). It can be clearly seen that the MKL-IE data for \acrshort{SymmSpMV} 
 are identical with the MKL \acrshort{SpMV} numbers presented in \Cref{fig:SpMV_vs_SymmSpMV}, \ie the inspector calls the baseline \acrshort{SpMV} kernel and uses the full matrix, though it knows about the symmetry of the matrix. One reason for that strategy might be that the parallelization approach used in the deprecated MKL implementation for \acrshort{SymmSpMV} is not scalable which would explain the fact that MKL is worse than MKL-IE for all cases on \SKX.  As neither the algorithm used to parallelize the \acrshort{SymmSpMV} nor its low level code implementation is known, we refrain from a deep analysis of the Intel performance behavior. 
In summary we find that \acrshort{RACE} is on average 1.4$\times$ faster than the best Intel variant and can achieve speedups of up to 2$\times$. Note 
	that on \SKX the best MKL variant is always MKL-IE, 
	which has almost twice the memory footprint 
	compared to the  \acrshort{SymmSpMV} with \acrshort{RACE}.

\subsubsection{Single core performance}
\begin{figure}[t]
 	\centering
 	\subfloat[\SKX]{\includegraphics[width=0.95\textwidth, height=0.3\textheight]{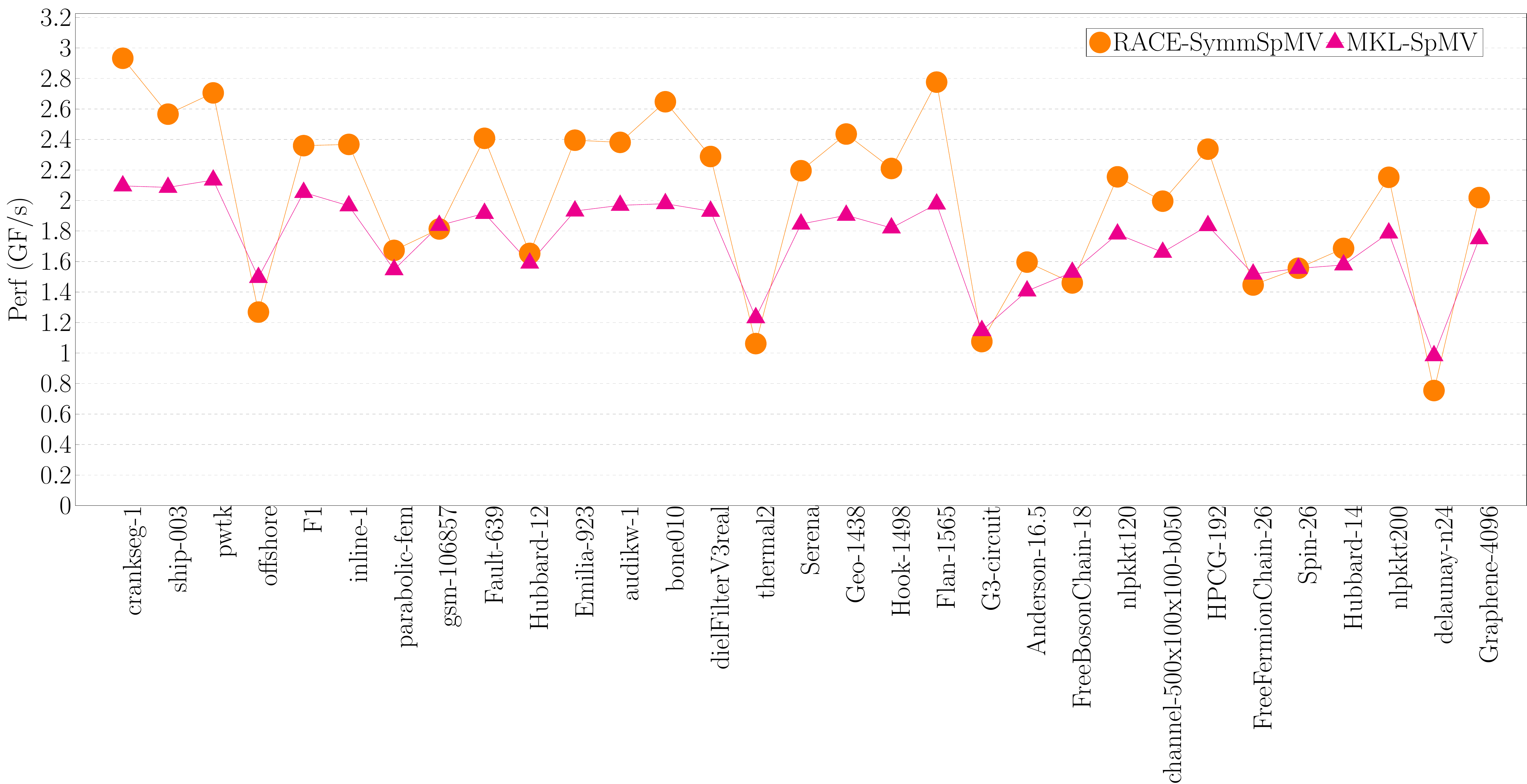}}
 	\caption{Single core performance of \acrshort{SymmSpMV} executed with \acrshort{RACE} compared to \acrshort{SpMV} performance using \acrshort{MKL}.}
 	\label{fig:SpMV_vs_SymmSpMV_single_core}
 \end{figure}

\begin{figure}[t]
  	\centering
  	\begin{minipage}[c]{0.4\textwidth}
  		\includegraphics[width=\textwidth]{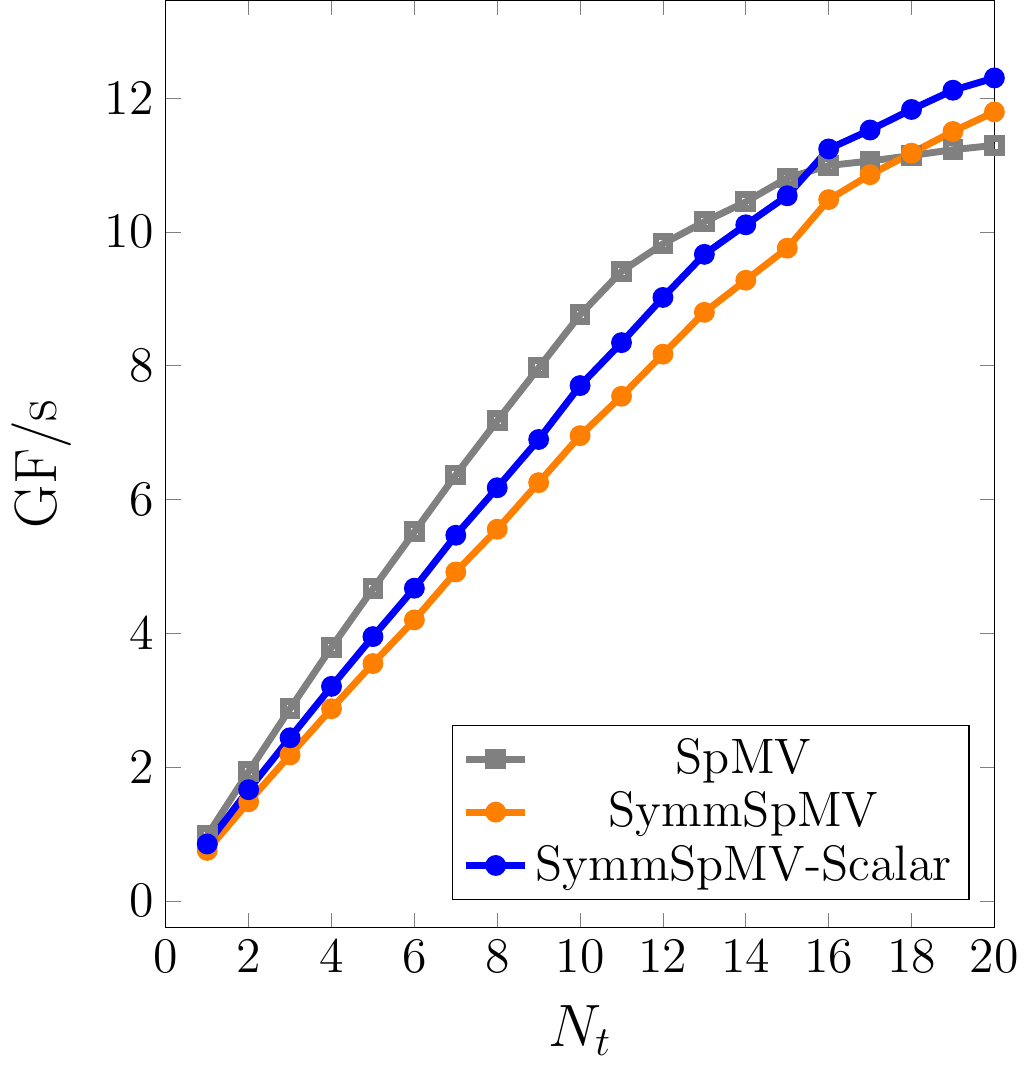}
  	\end{minipage}\hfill
  	\begin{minipage}[c]{0.55\textwidth}
  		\caption{Parallel performance  of \acrshort{SymmSpMV} (with \acrshort{RACE})  and \acrshort{SpMV} (with \acrshort{MKL}) for the \texttt{delaunay\_n24} matrix on one socket of \SKX. To disable vectorization (SymmSpMV-Scalar) we set \texttt{VECWIDTH = 1} when compiling the \acrshort{SymmSpMV} kernel.}
  		\label{fig:scaling_delaunay}
  	\end{minipage}
  \end{figure}

Although single core performance is often considered not to be crucial for 
the full chip \acrshort{SymmSpMV} performance, we demonstrate that it is
vital to explain some of the performance behaviors. For example the drop in 
\acrshort{SymmSpMV} performance
 for matrices like  \texttt{Hubbard-12} and \texttt{delaunay\_n24} strongly correlates with
 the lower performance of the baseline \acrshort{SpMV} (see \Cref{fig:SpMV_vs_SymmSpMV}).
 These matrices are characterized by a rather low $\acrshort{NNZR}$ and a 
 larger $\alpha_{\acrshort{SpMV}}$ value. Note that  $\alpha_{\acrshort{SpMV}}$ 
 measured for the \acrshort{SpMV}  kernel mainly accounts for the RHS vector traffic 
  and the actual $\alpha_{\acrshort{SymmSpMV}}$ may even be higher as \acrshort{SymmSpMV}
  requires two vectors to stay in cache concurrently. 
 Moreover, for these matrices the inner loop lengths are typically very short (approximately $\acrshort{NNZR}/2$ on average) and consequently the SIMD vectorization performed by the 
 compiler may become inefficient. This leads to lower single core performance
  as shown in \Cref{fig:SpMV_vs_SymmSpMV_single_core} for the \SKX architecture, 
  where bad performance of \acrshort{SymmSpMV}  and \acrshort{SpMV} can often be 
  correlated with a small \acrshort{NNZR} value. 
 For several matrices these combined effects overcompensate the reduced matrix data traffic of the \acrshort{SymmSpMV} leading to worse single core performance than running \acrshort{SpMV} with the full matrix. Using the \texttt{delaunay\_n24} matrix as a representative for this class of matrices we demonstrate the basic challenge for \acrshort{SymmSpMV} to exploit its basic performance advantage over \acrshort{SpMV} in \Cref{fig:scaling_delaunay}.
Starting with an approximately 25\% lower single core performance (0.75 GF/s versus 0.98 GF/s) but having a 50\% higher roofline performance limit (approximately 18 GF/s; see \Cref{fig:spmv_vs_symm_spmv_skx}) than the \acrshort{SpMV}, the \acrshort{SymmSpMV} is not able to saturate the main memory bandwidth of the \SKX on its 20 cores. As speculated above, the single core performance is limited by inefficient SIMD vectorization of the extremely short inner loop and switching back to scalar code does improve performance by 15\% (see \Cref{fig:scaling_delaunay}). As we are still substantially off the bandwidth limit we see this benefit over the full chip. Using chips with larger core counts would allow for further improving the \acrshort{SymmSpMV} performance of this matrix. The same arguments hold for the \texttt{offshore} matrix but here the effect compared to \acrshort{SpMV} performance is even more pronounced on \SKX.  Here the full matrix can at least partially be held in the large aggregate cache between successive kernel invocations and its performance is not limited by the main memory bandwidth. In terms of caching effects we have also further identified at least partial caching of the matrix for \texttt{ship-003} and \texttt{pwtk} test cases by analyzing the overall data traffic in the kernel invocations. This is inline with their higher performance levels presented in \Cref{fig:spmv_vs_symm_spmv_skx}. 

\subsubsection{Comparing \acrshort{RACE} with \acrshort{MC} and \acrshort{ABMC}}
\begin{figure}[t]
	\centering
	\subfloat[\IVB]{\label{fig:symm_spmv_ivy}\includegraphics[width=0.95\textwidth, height=0.3\textheight]{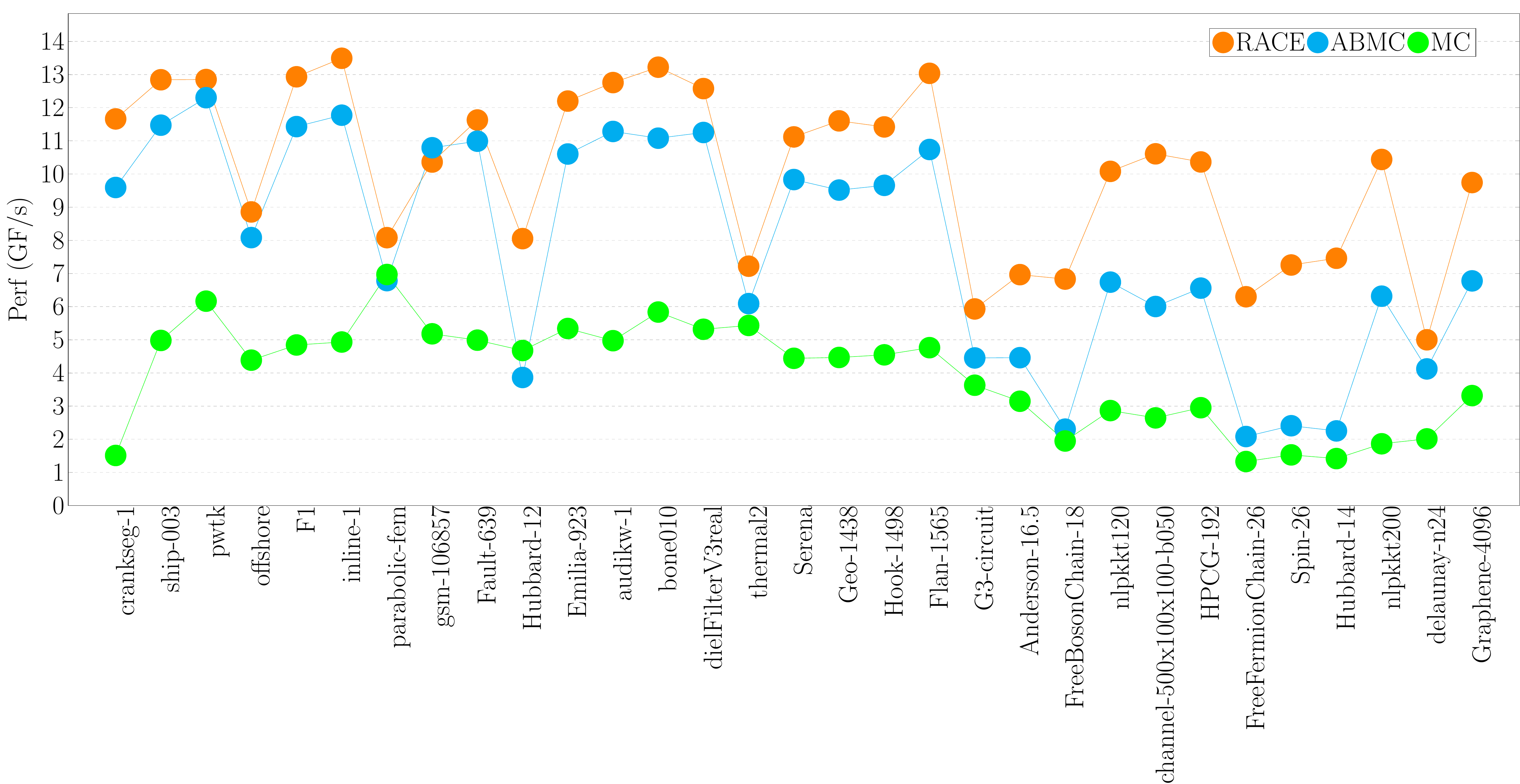}}
	\hspace{1em}
	\subfloat[\SKX]{\label{fig:symm_spmv_skx}\includegraphics[width=0.95\textwidth, height=0.3\textheight]{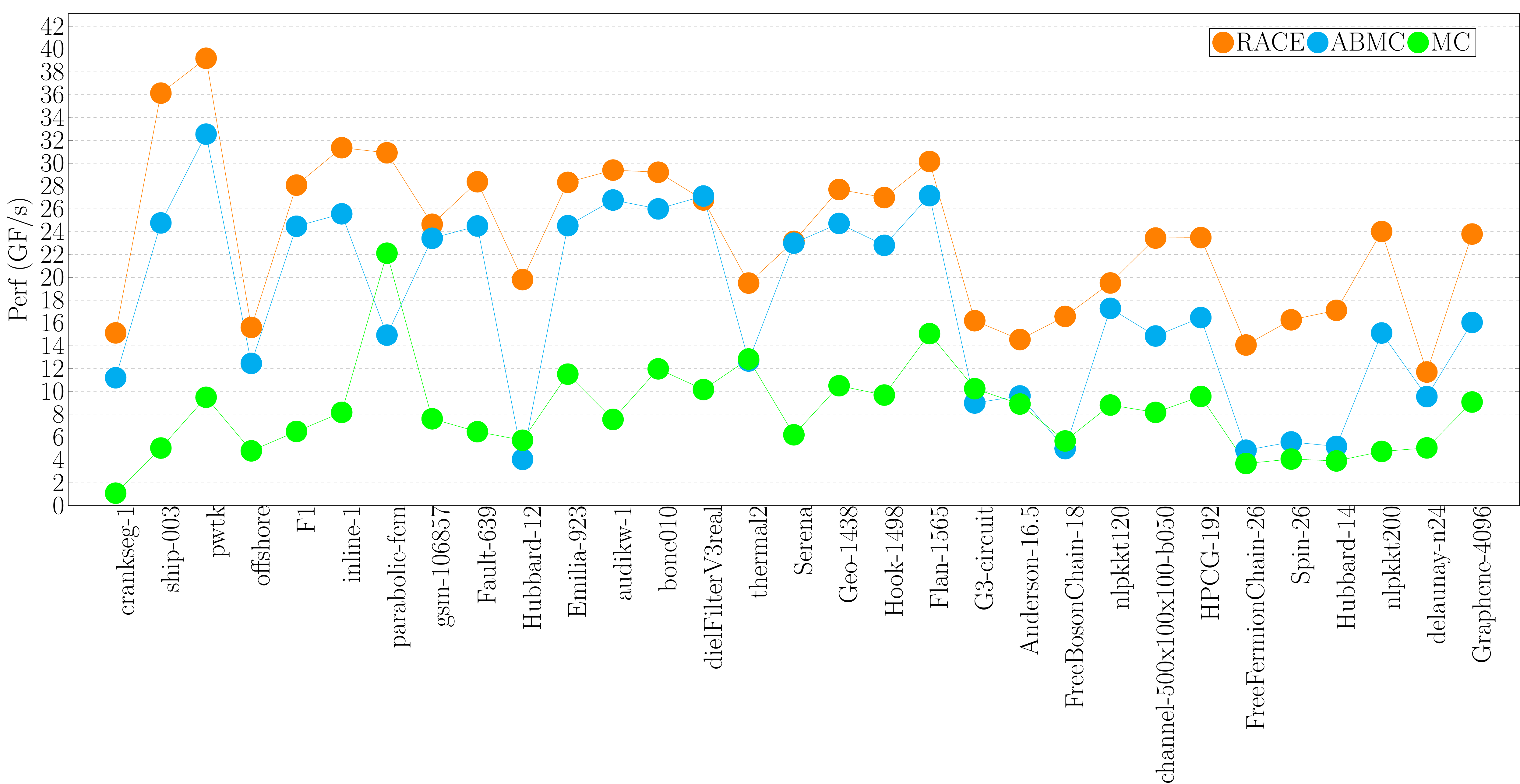}}
	\caption{Comparison of \acrshort{SymmSpMV} performance between \acrshort{RACE} and coloring variants \acrshort{MC} and \acrshort{ABMC}. Matrices are arranged in increasing number of rows (\acrshort{nrows}).}
	\label{fig:symm_spmv}
\end{figure}
Having well understood the performance characteristics of  \acrshort{SymmSpMV} with \acrshort{RACE} we finally compare this with the performance achieved by the two coloring methods in \Cref{fig:symm_spmv}. Here the underlying algorithm as well as implementation are known and are closely related to our approach. 
%
Overall the \acrshort{MC} is not competitive and provides low performance levels for almost all the matrices on both architectures. The \acrshort{ABMC} shows similar performance characteristics as \acrshort{RACE} until the two vectors involved in \acrshort{SymmSpMV} approach the size of the caches (cf. discussion of \Cref{fig:SpMV_vs_SymmSpMV}). For matrices with sufficiently small $N_r$ (left in the diagram) the method can achieve between 70\% and 90\% of \acrshort{RACE} performance on most cases. For matrices in the right part of the diagram with their higher $N_r$ and $\alpha_{SpMV}$ values, the \acrshort{ABMC} falls substantially behind \acrshort{RACE}. Here, the strict orientation of the \acrshort{RACE} design towards data locality in the vector accesses delivers its full power. See also the data transfer discussion in \Cref{Sec:Spin26full} for the \texttt{Spin-26} matrix.
In total there are only three cases where \acrshort{ABMC} performance is on a par with or slightly above the \acrshort{RACE} measurement and the average speedup of \acrshort{RACE} is $1.5\times$ and $1.65 \times$ for \IVB and \SKX, respectively.
Note that all three methods use the same baseline kernels and thus performance differences between the methods do not arise from different low level code but from the ability to generate appropriate degrees of parallelism and to maintain data locality.

%% file: conclusion.tex
In this paper we have developed \acrshort{RACE}, a coloring algorithm and open-source library
implementation for exploiting parallelism in algorithms with inherent dependencies.
\acrshort{RACE} generates hardware-efficient \DK colorings of undirected graphs and puts 
emphasis on data access locality, load balancing, and 
parallelism that is adapted to the number of cores of the underlying architecture.  We
demonstrated these benefits by applying \acrshort{RACE} to \acrfull{SymmSpMV} on modern
multicore architectures and compared its performance against
standard multicoloring, algebraic block multicoloring, and \acrshort{MKL}
implementations. Average and maximum speedups of 1.4 and 2, respectively,
could be observed across a representative set of 31 matrices on
two modern Intel processors. 
Our entire experimental and performance analysis process was backed by the
Roof{}line performance model, corroborating the optimality of
the \acrshort{RACE} approach in terms of resource utilization and shedding some new
light on the challenges of the \acrshort{SymmSpMV} kernel on modern hardware.
We demonstrated that \acrshort{RACE} runs very close to the Roof{}line limit for
most of the 31 test cases. Outliers were analyzed and discussed in detail.

Similar to other coloring algorithms, the \acrshort{RACE} method is not
limited to the \acrshort{SymmSpMV} kernel and can be used to efficiently
parallelize solvers and kernels having general \DK dependencies. Moreover, due
to the level-based formulation of \acrshort{RACE}, the framework has an added
advantage that allows us to address other classes of problems. Future work
with \acrshort{RACE} will involve variants of linear solvers and kernel
operations like in-place matrix powers and polynomials, which are of high
interest in the scientific community.